\input harvmac
\input epsf

\noblackbox \def\ev#1{\langle #1 \rangle}

\Title
 {\vbox{
 \baselineskip12pt
 \hbox{HUTP-01/A045}
\hbox{WIS/19/01-SEPT-DPP}
\hbox{UCSD-PTH-01-16}
\hbox{ILL-(TH)-01-9}
 \hbox{hep-th/0110028}\hbox{}\hbox{}
}}
{\vbox{
 \centerline{A Geometric Unification of Dualities}
 }}
 \medskip

\centerline{F. Cachazo$^1$, B. Fiol$^2$, K. Intriligator$^3$ , S. Katz$^{4,5}$
and
C. Vafa$^1$}
\vskip .5cm

\centerline{$^1$ Jefferson Physical Laboratory, Harvard University,
Cambridge, MA 02138, USA}
\medskip
\centerline{$^2$ Department of Particle Physics, The Weizmann
 Institute of Science, Rehovot, 76100, Israel}
\medskip
\centerline{$^3$ UCSD Physics Department, 9500 Gilman Drive,
La Jolla, CA 92093}
\medskip
\centerline{$^4$ Departments of Mathematics and Physics}
\centerline{ University
 of Illinois at Urbana-Champaign, Urbana, IL 61801 USA}
\centerline{$^5$ Department of Mathematics, Oklahoma State University,
Stillwater, OK 74078 USA}
\bigskip

\vskip .1in\centerline{\bf Abstract}
We study the dynamics of a large class of ${\cal N}=1$ quiver theories,
geometrically realized by type IIB D-brane probes wrapping cycles of
local Calabi-Yau threefolds. These include ${\cal N}=2$ (affine)
A-D-E quiver theories deformed by superpotential terms, as well
as chiral ${\cal N}=1$ quiver theories obtained in the presence of vanishing
4-cycles inside a Calabi-Yau.
We consider
the various possible geometric transitions of the 3-fold
and show that they correspond to Seiberg-like dualities
(represented by Weyl reflections in the A-D-E case or `mutations' of bundles
in the case of vanishing 4-cycles)
or large $N$ dualities involving gaugino condensates (generalized
conifold transitions).  Also duality cascades are naturally
realized in these classes of theories, and are related to the affine Weyl
group symmetry in the A-D-E case.

 \smallskip
\Date{October 2001}

\def\co{\Lambda_0}
\def\v#1{{\vec{#1}}}
\def\vrho{{\vec\rho}}
\def\fund{  \> {\vcenter  {\vbox
              {\hrule height.6pt
               \hbox {\vrule width.6pt  height5pt
                      \kern5pt
                      \vrule width.6pt  height5pt }
               \hrule height.6pt}
                         }
                   } \>
           }

\batchmode
  \font\bbbfont=msbm10
\errorstopmode
\newif\ifamsf\amsftrue
\ifx\bbbfont\nullfont
  \amsffalse
\fi
\ifamsf
\def\IR{\hbox{\bbbfont R}}
\def\IC{\hbox{\bbbfont C}}

\def\IZ{\hbox{\bbbfont Z}}
\def\IF{\hbox{\bbbfont F}}
\def\IP{\hbox{\bbbfont P}}
\else
\def\IR{\relax{\rm I\kern-.18em R}}
\def\IZ{\relax\ifmmode\hbox{Z\kern-.4em Z}\else{Z\kern-.4em Z}\fi}
\def\IF{\relax{\rm I\kern-.18em F}}
\def\IP{\relax{\rm I\kern-.18em P}}
\fi
\def\semidirect{\mathbin{\hbox{\hskip2pt\vrule height 4.1pt depth -.3pt
width .25pt \hskip-2pt$\times$}}}
\def\N{{\cal N}}

\lref\kentaro{K. Hori, H. Ooguri and Y. Oz, ``Strong Coupling Dynamics of
Four-Dimensional N=1 Gauge
Theories from M Theory Fivebrane'', Adv.Theor.Math.Phys. {\bf 1} (1998) 1-52 [hep-th/9706082].}

\lref\diagom{D.-E. Diaconescu and J. Gomis, ``Fractional branes and
boundary states in orbifold theories'', [hep-th/9906242].}

\lref\dfr{M. R. Douglas, B. Fiol and C. R\"omelsberger, ``The spectrum
of BPS branes on a noncompact Calabi-Yau'', [hep-th/0003263].}

\lref\dfrbrane{M. Douglas, B. Fiol and C. Römelsberger, ``Stability and BPS
branes,'' [hep-th/0002037].}

\lref\ItzyksonFU{
C.~Itzykson,
``Simple Integrable Systems And Lie Algebras,''
Int.\ J.\ Mod.\ Phys.\ A {\bf 1}, 65 (1986).
}

\lref\CKV{F.~Cachazo, S.~Katz, and C.~Vafa, ``Geometric Transitions and N=1
Quiver Theories,'' [hep-th/0108120]}

\lref\km{S. Katz and D. Morrison, ``Gorenstein Threefold Singularities with
Small Resolutions via Invariant Theory for Weyl Groups'', J. Algebraic
Geometry {\bf 1} (1992) 449-530.}

\lref\agav{M. Aganagic and C. Vafa, ``Mirror Symmetry and a $G_2$ Flop,''
[hep-th/0105225].}

\lref\aget{M. Aganagic, A. Karch, D. Lust, A. Miemiec,
``Mirror Symmetries for Brane Configurations and Branes at Singularities,''
 Nucl.Phys. {\bf B569} (2000) 277-302 [hep-th/9903093].}

\lref\flop{K. Dasgupta, K. Oh, R. Tatar,
``Open/Closed String Dualities and Seiberg Duality
from Geometric Transitions in M-theory,''
[hep-th/0106040].}
%
\lref\ElitzurHC{
S.~Elitzur, A.~Giveon, D.~Kutasov, E.~Rabinovici and A.~Schwimmer,
``Brane dynamics and N = 1 supersymmetric gauge theory,''
Nucl.\ Phys.\ B {\bf 505}, 202 (1997)
[hep-th/9704104].
}
\lref\KutasovSS{
D.~Kutasov, A.~Schwimmer and N.~Seiberg,
``Chiral Rings, Singularity Theory and Electric-Magnetic Duality,''
Nucl.\ Phys.\ B {\bf 459}, 455 (1996)
[hep-th/9510222].
}

\lref\SeibergPQ{
N.~Seiberg,
``Electric - magnetic duality in supersymmetric nonAbelian gauge theories,''
Nucl.\ Phys.\ B {\bf 435}, 129 (1995)
[hep-th/9411149].
}

\lref\KutasovNP{
D.~Kutasov and A.~Schwimmer,
``On duality in supersymmetric Yang-Mills theory,''
Phys.\ Lett.\ B {\bf 354}, 315 (1995)
[hep-th/9505004].
}

\lref\ArgyresEH{
P.~Argyres, M.~Ronen Plesser and N.~Seiberg,
``The Moduli Space of N=2 SUSY {QCD} and Duality in N=1 SUSY {QCD},''
Nucl.\ Phys.\ B {\bf 471}, 159 (1996)
[hep-th/9603042].
}

\lref\LeighEP{
R.~Leigh and M.~Strassler,
``Exactly marginal operators and duality in
four-dimensional N=1 supersymmetric gauge theory,''
Nucl.\ Phys.\ B {\bf 447}, 95 (1995)
[hep-th/9503121].
}

\lref\KutasovVE{
D.~Kutasov,
``A Comment on duality in N=1 supersymmetric nonAbelian gauge theories,''
Phys.\ Lett.\ B {\bf 351}, 230 (1995)
[hep-th/9503086].
}
\lref\GubserIA{
S.~Gubser, N.~Nekrasov and S.~Shatashvili,
``Generalized conifolds and four dimensional N = 1 superconformal  theories,''
JHEP {\bf 9905}, 003 (1999)
[hep-th/9811230].
}

\lref\AharonyNE{
O.~Aharony, J.~Sonnenschein and S.~Yankielowicz,
``Flows and duality symmetries in N=1 supersymmetric gauge theories,''
Nucl.\ Phys.\ B {\bf 449}, 509 (1995)
[hep-th/9504113].
}

\lref\DouglasSW{ M.~Douglas and G.~Moore,
``D-branes, Quivers, and ALE Instantons,''
hep-th/9603167.
}

\lref\GubserIA{
S.~Gubser, N.~Nekrasov and S.~Shatashvili,
``Generalized conifolds and four dimensional N = 1 superconformal  theories,''
JHEP {\bf 9905}, 003 (1999)
[hep-th/9811230].
}

\lref\NSVZ{M. Shifman and A. Vainshtein,
Nucl. Phys. {\bf B277} (1986) 456; Nucl. Phys.
 {\bf B539} (1991) 571; V. Novikov,
M. Shifman, A. Vainshtein, and V. Zakharov, Nucl. Phys. {\bf B229} (1983)
381.}

\lref\BlumMM{
J.D. ~Blum and K.~Intriligator,
``New phases of string theory and 6d RG fixed points
via branes at  orbifold singularities,''
Nucl.\ Phys.\ B {\bf 506}, 199 (1997)
[hep-th/9705044].
}

\lref\IntriligatorDH{
K.~Intriligator,
``New string theories in six dimensions via branes at orbifold
singularities,''
Adv.\ Theor.\ Math.\ Phys.\  {\bf 1}, 271 (1998)
[hep-th/9708117].
}

\lref\WittenZH{
E.~Witten,
``Some comments on string dynamics,''
[hep-th/9507121].
}
\lref\WittenSC{
E.~Witten,
``Solutions of four-dimensional field theories via M-theory,''
Nucl.\ Phys.\ B {\bf 500}, 3 (1997)
[hep-th/9703166].
}

\lref\AnselmiYS{
D.~Anselmi, J.~Erlich, D.~Freedman and A.~Johansen,
``Positivity constraints on anomalies in supersymmetric gauge theories,''
Phys.\ Rev.\ D {\bf 57}, 7570 (1998)
[hep-th/9711035].
}
\lref\CarlinoUK{
G.~Carlino, K.~Konishi and H.~Murayama,
``Dynamical symmetry breaking in supersymmetric SU(n(c)) and USp(2n(c))
gauge theories,''
Nucl.\ Phys.\ B {\bf 590}, 37 (2000)
[hep-th/0005076].
}

\lref\strom{A. Strominger, ``Massless Black Holes and Conifolds in String
Theory,'' Nucl.Phys. {\bf B451} (1995) 96-108 [hep-th/9504090].}

\lref\strogmo{B. Greene, D. Morrison and A. Strominger,``Black Hole
Condensation and the Unification of String Vacua,''
Nucl.Phys. {\bf B451} (1995) 109-120  [ hep-th/9504145].}

\lref\beret{M. Bershadsky, A Johansen, T. Pantev, V. Sadov and C. Vafa,
`` F-theory, Geometric Engineering and N=1 Dualities,''
Nucl.Phys.{\bf B505} (1997)  153-164 [hep-th/9612052]; C. Vafa and
B. Zwiebach, ``N=1 Dualities of SO and USp Gauge Theories
and T-Duality of String Theory,'' Nucl.Phys. {\bf B506}
(1997) 143-156 [hep-th/9701015].}

\lref\ov{H. Ooguri and C. Vafa,``Geometry of N=1 Dualities in Four
Dimensions,'' Nucl.Phys. {\bf B500} (1997) 62-74 [hep-th/9702180].}

\lref\reva{O. Aharony, S.S. Gubser, J. Maldacena, H. Ooguri and
Y. Oz,``Large N Field Theories, String Theory and Gravity,''
 Phys.Rept. {\bf 323} (2000) 183-386 [hep-th/9905111 ].}

\lref\gopva{R. Gopakumar and C. Vafa,``On the Gauge Theory/Geometry
Correspondence,'' Adv.Theor.Math.Phys. {\bf 3} (1999) 1415-1443
[hep-th/9811131].}

\lref\gopvaii{R. Gopakumar and C. Vafa, ``Branes and Fundamental
Groups,'' Adv. Theor. Math. Phys. {\bf 2} (1998) 399-411 [hep-th/9712077].}

\lref\vaaug{C. Vafa, ``Superstrings and Topological Strings at Large N,''
[hep-th/0008142].}

\lref\ach{B.S. Acharya,``On Realising N=1 Super Yang-Mills in M theory,''
[hep-th/0011089].}

\lref\amv{M. Atiyah, J. Maldacena and C. Vafa,''An M-theory Flop as a Large
N Duality,'' [hep-th/0011256].}

\lref\aw{M. Atiyah and E. Witten,`` M-Theory Dynamics On A Manifold Of $G_2$
Holonomy,'' [hep-th/0107177].}

\lref\stkl{I. Klebanov and M. Strassler,``Supergravity and a Confining
Gauge Theory: Duality Cascades and $\chi$SB-Resolution of Naked
Singularities,''  JHEP {\bf 0008} (2000) 052 [hep-th/0007191].}

\lref\civ{F. Cachazo, K. Intriligator and C. Vafa,``A Large N Duality via a
Geometric Transition,'' Nucl.Phys. {\bf B603} (2001) 3-41 [hep-th/0103067].}

\lref\kutet{S. Elitzur, A. Giveon, D. Kutasov, E. Rabinovici and
A. Schwimmer,``Brane Dynamics and N=1 Supersymmetric Gauge Theory,''
 Nucl.Phys. B505 (1997) 202-250 [hep-th/9704104].}

\lref\kmv{S. Katz, P. Mayr and C. Vafa, ``Mirror symmetry and Exact
Solution of 4D N=2 Gauge Theories I,'' Adv.Theor.Math.Phys. 1
(1998) 53-114 [hep-th/9706110].}

\lref\hiv{K. Hori, A. Iqbal and C. Vafa, ``D-Branes And Mirror
Symmetry,'' [hep-th/0005247].}

\lref\klebwi{I. Klebanov and E. Witten, ``Superconformal Field Theory on
Threebranes at a Calabi-Yau Singularity,''
Nucl.Phys. {\bf B536} (1998) 199-218 [hep-th/9807080].}

\lref\anpi{I. Antoniadis and B. Pioline,``Higgs branch, HyperKahler
quotient and duality in SUSY N=2 Yang-Mills theories,''
Int.J.Mod.Phys. {\bf A12} (1997) 4907-4932  [hep-th/9607058].}

\lref\wim{E. Witten, `` Solutions Of Four-Dimensional Field Theories Via M
Theory,''  Nucl.Phys. {\bf B500} (1997) 3-42 [hep-th/9703166].}

\lref\lnv{A. Lawrence, N. Nekrasov and C. Vafa, ``On Conformal Theories in
Four Dimensions,'' Nucl.Phys. {\bf B533} (1998) 199-209 [hep-th/9803015].}

\lref\mn{J. Maldacena and C. Nunez,``Towards the large N limit of pure N=1
super Yang Mills,'' Phys.Rev.Lett. {\bf 86} (2001) 588-591 [hep-th/0008001].}

\lref\witli{E. Witten, ``Phases of $N=2$ Theories In Two Dimensions,''
 Nucl.Phys. {\bf B403} (1993) 159-222 [hep-th/9301042].}

\lref\horv{K. Hori and C. Vafa, ``Mirror Symmetry,'' [hep-th/0002222].}

\lref\klmvw{ A. Klemm, W. Lerche, P. Mayr, C.Vafa and N. Warner,``Self-Dual
Strings and N=2 Supersymmetric Field Theory,''
Nucl.Phys. {\bf B477} (1996) 746-766 [hep-th/9604034].}

\lref\SYZ{A. Strominger, S.T. Yau and E. Zaslow, ``Mirror Symmetry is
T-Duality,'' Nucl.Phys. {\bf B479} (1996) 243-259 [hep-th/9606040].}

\lref\cfiv{S. Cecotti, P. Fendley, K. Intriligator and C. Vafa, ``A New
Supersymmetric Index,'' Nucl.Phys. B386 (1992) 405-452 [hep-th/9204102].}

\lref\modgr{M. Douglas, B. Greene and D. Morrison, ``Orbifold Resolution by
D-Branes,'' Nucl.Phys. {\bf B506} (1997) 84-106 [hep-th/9704151].}

\lref\seith{P. Seidel and R. Thomas, ``Braid group actions on derived
categories of coherent sheaves'', Duke Math. Jour. {\bf 108} (2001) 37--108
[math.AG/0001043].}

\lref\cecov{S. Cecotti and C. Vafa, ``On Classification of N=2
Supersymmetric Theories,'' Commun.Math.Phys. {\bf 158} (1993)
569-644 [hep-th/9211097].}

\lref\zas{E. Zaslow,``Solitons and Helices: The Search for a Math-Physics
Bridge,'' Commun.Math.Phys. {\bf 175} (1996) 337-376  [hep-th/9408133].}

\lref\hanse{B. Feng, A. Hanany, Y. He, A. Uranga, ``Toric Duality as
Seiberg Duality and Brane Diamonds,'' [hep-th/0109063].}

\lref\plb{C. Beasley and M. Plesser,``Toric Duality Is Seiberg Duality,''
[hep-th/0109053].}

\lref\dm{M. Douglas and G. Moore, ``D-branes, Quivers, and ALE
Instantons,'' [hep-th/9603167].}

\lref\ks{S. Kachru, E. Silverstein,``4d Conformal Field Theories and
Strings on Orbifolds,'' Phys.Rev.Lett. {\bf 80} (1998) 4855-4858
[hep-th/9802183].}

\lref\morpl{D. Morrison and M. Plesser, ``Non-Spherical Horizons, I,''
Adv.Theor.Math.Phys. {\bf 3} (1999) 1-81 [hep-th/9810201].}

\lref\av{B. Acharya and C. Vafa,``On Domain Walls of N=1 Supersymmetric
Yang-Mills in Four Dimensions,'' [hep-th/0103011].}

\lref\alt{K. Altmann, ``The versal deformation of an isolated toric
Gorenstein singularity'', Inv. Math. {\bf 128} (1997) 443--479
[alg-geom/9403004].}

\lref\gros{M. Gross, ``Deforming Calabi-Yau Threefolds,''
[alg-geom/9506022].}

\lref\ih{A. Hanany and A. Iqbal, ``Quiver Theories from D6-branes via
Mirror Symmetry,'' [hep-th/0108137].}

\lref\tayva{T. Taylor and C. Vafa, ``RR Flux on Calabi-Yau and Partial
Supersymmetry Breaking,''
Phys.Lett. {\bf B474} (2000) 130-137 [hep-th/9912152].}

\lref\mayr{P. Mayr, `` On Supersymmetry Breaking in String Theory and its
Realization in Brane Worlds,''
Nucl.Phys. {\bf B593} (2001) 99-126 [hep-th/0003198].}

\lref\polsh{S. Giddings, S. Kachru and J. Polchinski, ``Hierarchies from
Fluxes in String Compactifications,'' hep-th/0105097.}

\lref\PolchinskiDY{
J.~Polchinski,
``Scale And Conformal Invariance In Quantum Field Theory,''
Nucl.\ Phys.\ B {\bf 303}, 226 (1988).
}

\lref\pless{C. Beasley, B. Greene, C. Lazaroiu and M. Plesser,``D3-branes
on partial resolutions of abelian quotient singularities of Calabi-Yau
threefolds,'' Nucl.Phys. {\bf B566} (2000) 599-640 [hep-th/9907186]}

\lref\hananpa{ B. Feng, A. Hanany, Y. He,``D-Brane Gauge Theories from
Toric Singularities and Toric Duality,'' Nucl.Phys. {\bf B595} (2001) 165-200}

\lref\quar{ G. Gibbons and P. Townsend, ``A Bogomol`nyi equation for
intersecting domain walls,'' Phys.Rev.Lett. {\bf 83} (1999) 1727-1730
[hep-th/9905196] \semi S. Carroll, S. Hellerman, M. Trodden, ``Domain Wall
Junctions are 1/4-BPS States,'' Phys.Rev. {\bf D61} (2000) 065001
[hep-th/9905217] \semi P. Townsend, ``PhreMology--calibrating M-branes,''
Class.Quant.Grav. {\bf 17} (2000) 1267-1276 [hep-th/9911154]
\semi S. Nam and K. Olsen, ``Domain Wall Junctions in
Supersymmetric Field Theories in D=4,'' JHEP {\bf 0008} (2000) 001
 [hep-th/0002176].}

\lref\morep{J. Edelstein, K. Oh, and
R. Tatar, ``Orientifold, Geometric Transition and Large N Duality for SO/Sp
Gauge Theories,'' JHEP {\bf 0105} (2001) 009  [hep-th/0104037]\semi
K. Dasgupta, K. Oh, and R. Tatar, ``Geometric Transition, Large N Dualities
and MQCD Dynamics,'' [hep-th/0105066] \semi
K. Dasgupta, K. Oh, and R. Tatar,
``Open/Closed String Dualities and
Seiberg Duality from Geometric Transitions in
    M-theory,'' [hep-th/0106040].}

\lref\bohe{B. Feng, A. Hanany, Y. He, `` Phase Structure of D-brane Gauge
Theories and Toric Duality,'' JHEP {\bf 0108} (2001) 040 [hep-th/0104259] }

\lref\gvw{S. Gukov, C. Vafa and E. Witten, `` CFT's From Calabi-Yau
Four-folds,'' Nucl.Phys. B584 (2000) 69-108 [hep-th/9906070];  A. Shapere
and C. Vafa, ``BPS Structure of Argyres-Douglas Superconformal
Theories,'' [hep-th/9910182].}

\lref\arn{V. Arnold, S. Gusein-Zade and A. Varchenko, ``Singularities of
Differentiable Maps,'' Vol. II, Monographs in Mathematics, vol. 82,
Birkh$\ddot{\rm a}$use, Boston, Basel, Stuttgart, 1985.}


\newsec{Introduction}

A deeper understanding of string theory on background geometries
with some vanishing cycles has played
a key role in various aspects of string dualities.  An early example
of this was in the context of the physical
interpretation of the conifold \strom\
singularity and its possible transitions \strogmo.
Geometric transitions have also played an important role
in deriving field theoretic dualities from string theory.  In particular
by considering spacetime filling
D-branes wrapped around cycles of Calabi-Yau 3-folds,
Seiberg's duality was derived in this way in the context of type IIB
\beret\
and type IIA \ov\ string theories.

Geometric transitions have also played a key role in large $N$
dualities.  The AdS/CFT correspondence \reva\ can be viewed as an
example of such a transition \gopva, where before the transition
(small `t Hooft parameter) there are D-branes wrapped around cycles,
and after the geometric transition (large `t Hooft parameter) these
cycles which supported the D-branes have disappeared, and have been
replaced with flux through a dual cycle.  The large $N$ duality of
Chern-Simons with topological strings
\gopva\ is an example of this kind.  The geometric transition duality
was embedded in type IIA superstring \vaaug, with D6 branes wrapping
an $S^3$ on one side of the transition and fluxes through a dual $S^2$
on the other side; this leads to a large $N$ duality for ${\cal N}=1$
Yang-Mills theory in 4 dimensions.  This duality was lifted up to
M-theory \refs{
\ach, \amv}\ where it was interpreted as a purely geometric transition.
Since, as argued in \amv\ and further elaborated in \refs{\agav, \aw}\ the
transition in quantum geometry is smooth in the M-theory lift, this
leads to a derivation of the geometric transition duality.

The type IIB mirror of these large $N$ dualities has also been studied
\refs{\stkl, \civ} (see also
\morep).  One
aim of this paper is to generalize these constructions and show that
the Seiberg-like dualities and large $N$ dualities/gaugino
condensation can be viewed in a unified way as geometric transitions
in the same setup.  We consider a wide variety of 4d, $\N =1$
supersymmetric gauge theories, which can be constructed via branes
which partially wrap cycles of a (non-compact) Calabi-Yau 3-fold $X$.
In type IIB, one can consider general combinations of $D3,D5,D7$
branes, wrapped over various cycles and filling the 4 dimensional
spacetime.  This generically leads to a theory with gauge group
$\prod_i U(N_i)$ with some matter in the bifundamental
representations, and some superpotential terms (depending on the
complex structure of $X$).  Changing the Kahler parameters of the
underlying CY 3-fold translates to changing the coupling constants of
the gauge theory (and sometimes also to FI terms).

We find, as in \refs{\beret, \ov}\ that changing the Kahler parameters
of $X$ (or, in the type IIA mirror with wrapped D6 branes, changing
the complex parameters) changes the description of the gauge theory.
As we pass through transitions in the 3-fold geometry, corresponding
to blowing up different Kahler classes, we find dual gauge theory
descriptions of the same underlying theory.  These are transitions
where some 2- or 4-cycles shrink and other 2- or 4-cycles grow.  On
the other hand, for a class of these theories which eventually
confine, with gaugino condensates, we find transitions of the type
where the 2-cycle or 4-cycle has shrunk, disappeared, and instead a
number of finite size $S^3$'s have emerged, with fluxes through them.
The description in terms of the blown up $S^3$'s is better at large
$N$ (in the IR), where the size of the $S^3$'s, which corresponds
to the gaugino condensation, is large.

By using the holographic picture, and following the geometric
transitions, we can smoothly follow the field theory dualities and
dynamics along the renormalization group flow.  In the UV, which
corresponds to far distance to the geometry, we have a description
which is best given in terms of finite size 2-cycles and
4-cycles. This is the weak coupling limit.  The renormalization group
flow to the IR corresponds in the geometry to going towards the tip of
the cone (or more precisely towards the ``tips'' of the cone).  In
doing so, the description changes: some 2-cycles or 4-cycles shrink,
and others emerge, corresponding to Seiberg-like dualities in the
field theory.  Eventually the gauge theory flows to e.g. a RG fixed
point, a free-magnetic phase, or confinement with gaugino condensation.
This is seen by following the geometry towards the tips of the cone.
E.g. deep in the IR, or in the very large $N$ limit, the description
might be best in terms of the blown up $S^3$'s; this is where the
gauge theory confines and gaugino condensation has taken place.

In this way we have a unified geometric picture, where both
kinds of dualities can be seen in the same RG trajectory,
depending on where in
the geometry we are.  This unification
sharpens the picture of Seiberg duality given in
\refs{\beret, \ov} (a similar comment applies to the brane construction of
\kutet): Rather than just seeing that
two gauge theories are connected by changing
the moduli of the theory, which by itself is not a complete derivation
of duality\foot{For example, by similar changes of the moduli one can
relate $\N =2$ $U(N_c)$, with $N_f$ flavors, to $\N =2$ $U(N_f-N_c)$
with $N_f$ flavors.  But here this duality misses part of the story.
The original $U(N_c)$ theory does indeed contain
the free-magnetic $U(N_f-N_c)$ theory in its spectrum, but
this description is only good on part of the Higgs branch, and it also
must be augmented with an extra $U(1)^{2N_c-N_f}$ where this
Higgs branch part intersects the Coulomb branch \ArgyresEH.}
) we can use the geometry to follow the RG trajectory, and see
which description is best, at which scale, as we flow to the IR.

We consider two classes of local 3-folds in type IIB.  One type (i)
involves certain Calabi-Yau threefolds which only has compact 2-cycles
and no compact 4-cycles.  The other type (ii) involves Calabi-Yau's
which have compact 2- and 4-cycles.  For type (i) we consider $X$ to
have the geometry of an A-D-E 2-fold geometry fibered over a plane,
with some blown up 2-cycles $S_i^2$'s in one to one correspondence
with the simple roots of A-D-E.  We can wrap D5 branes over these
2-cycles, which fill the directions transverse to $X$.  In addition,
one could also include $N_0$ additional D3 branes transverse to $X$.
The 3-folds $X$ which we consider can thus be labelled (up to
deformations) as $X(k,G)$ with $G$ the A-D-E group and $k$ an integer
which labels the data about how the holomorphic 2-cycles of the A-D-E
are fibered over the plane \CKV .\foot{ More generally we can consider
one $k$ for each simple root of A-D-E, but this can also be obtained,
by deformations, from the case we consider.}  For $N_0=0$, the gauge
groups obtained via wrapping various numbers $N_i$ D5 branes over the
various $S^2_i$ of $X(k,G)$ are quiver gauge theories with gauge group
$\prod _{i=1}^r U(N_i)$, with the quiver diagram the $G$ Dynkin
diagram and $r=$rank$(G)$, and the matter in hypermultiplets dictated
by the links of the Dynkin diagram.  The theory arises from the
corresponding $\N =2$ quiver theory, broken to $\N =1$ by the
additional superpotentials for the adjoint superfields $\phi _i$ in
the $\N =2$ $U(N_i)$ vector multiplet
\eqn\wadded{W_i={g_i\over k+1}\Tr \phi _i^{k+1}+\hbox{lower order}.}
The precise form of the superpotential is dictated by the fibration data.
Adding $N_0$ D3 branes, the quiver gauge theory becomes $\prod _{i=0}^r
U({\widehat{N}}_i)$, based on the affine $\widehat G$ Dynkin diagram, with
\eqn\NandM{{\widehat N}_i=N_0d_i+N_i,}
for $i\not =0$
with $d_i$ the Dynkin indices.  We also set ${\widehat N}_0=N_0$.

The inequivalent blowups for $\N=1$ A-D-E quiver theories are given
by the action of the Weyl group.  As we will discuss, a Weyl reflection on
a node is related to a Seiberg-like duality on the corresponding
gauge group.  A similar statement applies to the affine case.
The duality cascade of
\stkl, for example, corresponds to the affine $\widehat A_1$ case
of $X(k=1,G={\widehat A_1})$.  This will be
generalized here to the arbitrary affine case.  The generalized duality
cascade is related to the affine Weyl group,
which is the semi-direct product of the Weyl group and translation by the
root lattice; the translation is responsible for the cascading reduction
of the D3 branes as we flow to the IR.

For the type (ii) case, with compact 4-cycles in addition to the
two-cycles, we consider local threefolds which have a toric
realizations, as in the examples studied in \kmv.  We can then
consider wrapping general classes of D3,D5 and D7 branes.  In this
case, it is more convenient to use the mirror IIA picture of the
manifold and branes, as it does not suffer from quantum corrections.
Using the appropriate mirror symmetry in the context of branes
\hiv, we write down the corresponding quiver theory, as well as the
corresponding Seiberg-like dualities.  The dualities involve changes
of the classical parameters in the type IIA mirror.  We specialize to
the Calabi-Yau threefolds involving delPezzo and their
transitions. Certain aspects of this case have been noted recently in
\refs{\plb, \hanse}.

The organization of this paper is as follows:  In section 2 we give
an overview of the ${\cal N}=1$ A-D-E quiver theories and the results
we will find for them in this paper.  In section 3 we give the
description of classical aspects of the A-D-E quiver gauge theories under
consideration.  In section 4 we discuss some aspects of the quantum
dynamics of the gauge couplings and their running.  In section 5
we discuss gaugino condensation in the non-affine A-D-E ${\cal N}=1$ quiver
theories.  In section 6
we consider the geometric engineering of these theories and their large
$N$ dual, involving the leading quantum corrections and
the geometric realization of gaugino condensates.
In section 7 we discuss Seiberg-like dualities
for the A-D-E  quiver theories anticipated
from geometry.  In section 8 we discuss the gauge
theoretic interpretation of these dualities.
In section 9 we consider the gauge theory
dynamics of the $A_2$ quiver in more detail, as a typical
situation where the Seiberg-like duality is relevant. In section 10
we discuss  dynamical aspects
of the affine quiver theory and its relation to the
non-affine case.  We also note the connection
of RG cascades in this class of theories
with affine Weyl reflection.  In section 11 we discuss examples of
${\cal N}=1$ superconformal A-D-E quiver theories.  In section
12 we setup the geometric engineering of the type (ii) local
threefolds, as well as dualities predicted by geometry.  In section
13 we specialize to a class of examples and illustrate how the
gaugino condensation takes place in these chiral theories and what
geometric transition they correspond to.

\newsec{Basic structure of the type (i) ${\cal N}=1$ quiver theories
and their large $N$ duals}

The class of type (i) theories which we consider are fibrations
of a A-D-E twofold geometry over a plane.  The corresponding
field theory is that of an $\N =2$ $A$-$D$-$E$ or affine
$\widehat A$-$\widehat D$-$\widehat E$ quiver theory, deformed to $\N=1$
by superpotential terms $W_i(\phi_i)$, with $\phi _i$
the adjoint field in the $\N=2$ $U(N_i)$ vector multiplet.
The choice of $W_i$'s are encoded in the fibration data.
For simplicity we consider the case where all the superpotentials
are polynomials of degree $k+1$.

The case $X(k=1, G=A_1)$, for example, corresponds to the small
resolution of the conifold, in which the $S^2$ is blown up.  Wrapping
$N$ D5 branes on the $S^2$ leads to $\N =1$ $U(N)$ pure Yang-Mills.
It was argued in \vaaug\ that for large $N$, or in the IR, the theory
is better described by the geometric conifold transition:
$X\rightarrow \widetilde X$, where $\widetilde X$ is the deformed conifold,
with its blown up $S^3$ having RR flux.  The generalization to
$X(k,A_1)$ for arbitrary $k$ was discussed in \civ: the worldvolume
theory is $\N =2$ $U(N)$ gauge theory, broken to $\N =1$ by a
superpotential as in \wadded.  Geometrically this means that instead
of having holomorphic $S^2$'s over the whole plane (corresponding to
vev of $\Phi$) they only appear at $k$ points. Let us label these
$S^2$'s by $S^2_p$ where $p=1,...,k$.  One can distribute the $N$ D5
branes by wrapping them on any of the $S^2_p$'s, leading to a Higgsing
$U(N)\rightarrow \prod_{p=1}^k U(M_p)$.  The geometric transition duality of
\civ\ involves $X(k,A_1)\rightarrow \widetilde X(k,A_1)$ in which
every $S^2_p$ is blown down and replaced with a blown up $S_p^3$
having RR flux. The geometric transition duality yields a new field
theory duality, in which the original $U(N)$ theory is dual to a
$\N=2$ $U(1)^k$ theory, which is broken to $\N =1$ by a particular
superpotential (which can be regarded as electric and magnetic FI
terms).  This duality was shown to be a powerful tool for obtaining
exact results about these supersymmetric field theories
\civ.

The case of $N_0$ D3 branes transverse to $X(k=1,G)$, without wrapped D5s,
was discussed in \GubserIA.  The gauge group is as in \NandM, with all
$N_i=0$, and these theories flow to $\N =1$ superconformal field theories.
These superconformal field theories have a holographic dual description
in terms of IIB string theory on $AdS_5\times M_5(1,G)$ which was
discussed in \GubserIA ,
generalizing the work \klebwi\ ccorresponding to $G=A_1$.
We can now add wrapped D5s (sometimes referred to as adding
fractional D3 branes), which breaks the conformal invariance.  As will
be discussed, this theory undergoes a RG cascade generalizing that of
\stkl, which is the case coming from $X(k=1,G=A_1)$.

The geometry of the general $X(k,G)$ and the classical gauge theories
associated with arbitrary wrapped D5s, and arbitrary transverse D3s,
was obtained in \CKV.  It was shown there that the basic aspects
of the geometry and geometric transition duality matches with what one
expects for the field theory in terms of gaugino condensates.  One
major aim of the present work is to analyze the dynamics of these
gauge theories in detail, and verify that the geometry properly
predicts the correct gauge theory dynamics.  We will see that the
associated field theory dualities are geometrically realized via two
different possible geometric dual operations:
\eqn\dualops{\eqalign{(A):\ S_i^2&\rightarrow \sum _j A_{ij}S_j^2\cr
(B):\ S_p^2&\rightarrow S_p^3.}}
The operation $(B)$ is the geometric transition duality, which occurs
when a $U(N)$ gauge theory confines, with gaugino condensation.  The size
of the $S^3_p$ is related to the gaugino condensate \refs{\vaaug, \civ}.

The operations $(A)$ on the other hand, correspond to Seiberg-type
dualities \SeibergPQ\ which, from essentially the same viewpoint,
was discussed in \refs{\beret, \ov}
(see also the related work \refs{\aget, \flop}).  As we will discuss, these are related to
the $G$ Weyl group for $N_0=0$ or, for general $N_0$, to the $\widehat
G$ affine Weyl group.  Each Seiberg-like duality corresponds to a
Weyl reflection about a simple root, with the reflections about each
of the simple roots generating the full Weyl group.
In particular, all of the $A_{ij}$ in \dualops\ are given by Weyl
reflections about each simple root $\vec e_{i_0}$ as
\eqn\weylti{\vec e_i\rightarrow {\vec e_i}\hskip0.003in '=\vec e_i -(\vec e_i \cdot \vec
e_{i_0}) \vec e_{i_0}\equiv \sum _j A_{ij}\vec e_j}
(including the affine root $\vec e_0$
and its Weyl translation in the $\widehat G$ case).   The rank of the
gauge group is determined by D-brane charge conservation
(as in \refs{\beret, \ov}):
$$\sum N_i \vec e_i=\sum N_i' {\vec e_i}\hskip0.003in '$$
which implies that
$$N_i'=(A^{-T})_{ij} N_j$$

The Weyl symmetry $(A)$ acts on the $U(N_i)$ coupling constants
and on the superpotentials as
\eqn\weylact{g _i^{-2}\rightarrow \sum _j A_{ij}g _j^{-2}, \qquad
W_i(\phi _i)\rightarrow \sum _j A_{ij}W_j(\phi _i).}
The action of \weylact\ on $g_i^{-2}$
follows from the fact that this is identified with the quantum volume
of $S^2_i$.  When $S^2_i$ shrinks the $1/g_i^2\rightarrow 0$ (i.e. the
theory is strongly coupled) and if we continue it past that it become
negative.  However we know that another $S^2$ has emerged whose volume
is $-1/g_i^2$ which now is positive.  This is the dual gauge theory.
{}From this point of view the duality can be viewed as an attempt to
make the $1/g_i^2$'s positive. In the field
theory, dimensional transmutation can occur, with the running $g_i$
written in terms of dynamical scales $\Lambda _i$;
the action of \weylact\ on $g_i^{-2}$
then becomes a statement about matching the dynamical scales $\Lambda
_i$ of the dual theories.    The duality is inherited from that
duality of the corresponding ${\cal N}=2$ theory with $W_i=0$, which in
the field theory setup was noted in \ArgyresEH (see also \anpi),
corresponding to $U(N_c)$ theory with $N_f$ flavors getting related to
$U(N_f-N_c)$ with $N_f$ flavors.  Breaking to $\N =1$ by $W\sim \Tr\phi^2$
was considered in \ArgyresEH\ and the case of more general
$W\sim \Tr\phi ^{k+1}$
was considered e.g. in \kutet\ via NS brane constructions.

The two transitions $(A)$ and $(B)$ combine in a beautiful way in the
geometric dual description.  Geometrically, if we start far from the
tip of the cone, the geometry has a description in terms of $S^2_i$'s,
which change in size as we come closer to the tip of the cone (which
is the geometric realization of moving towards the IR).  Sometimes an
$S^2$ shrinks and a dual $S^2$ grows, an $(A)$ type transition, which
is interpreted as Seiberg-like duality.  Sometimes an $S^2$ shrinks
and an $S^3$ grows, a $(B)$ type transition, and this corresponds to
the occurrence of confinement and gaugino condensation.  The nice
thing about this picture is that not only can we ``derive''
Seiberg-like dualities by connecting branes wrapping cycles of
Calabi-Yau, as in \refs{\beret, \ov}, but in fact we are able to see how
they occur in a dynamical sense, i.e. following the RG trajectory and
seeing that they become equivalent.  This picture works equally well
for $G$ as well as for the affine ${\widehat G}$ type quiver theories.
The application of duality is particularly striking in the affine
case, as one may have to undergo infinitely many applications of
duality as we go from the UV to the IR.  In particular the RG cascade
of \stkl\ corresponds to the $\widehat A_1$ Weyl group.  Upon flowing
to the IR, one undergoes a series of $(A)$ type transitions until
eventually the theory confines and undergoes the $(B)$ type
transition.

Consider the $G=A$-$D$-$E$ quiver theories.  Using the action \weylact\ of the
Weyl group on the coupling constants, one can represent the coupling
constants as a $r=$rank$(G)$ vector $\vec x$ such that
\eqn\chamber{{1\over g_i^2(\vec x)}=\vec e_i \cdot \vec x >0.}
The space of $\vec x$ satisfying this condition is a $G$ Weyl chamber,
a fundamental domain for the action of the Weyl group on ${\IR^r /W}$
where $W$ is the Weyl group.  The Weyl chamber is a conical wedge,
which has $r$ codimension one boundaries, given by $g_i^{-2}=0$ for
any $i=1\dots r$.  The RG flow corresponds to moving $\vec x$ inside
the fundamental domain along a straight line, until it hits a boundary
where one of the $1/g_i^2=0$.  After this, if the ranks of the dual
theories are all positive, there is a reflection off the boundary,
with incident angle equal to the reflection angle; this corresponds to
a Seiberg-like duality.  If the rank of the dual theory is not
positive, there is no reflection and this is indicative of the (B)
type transition in \dualops, corresponding to confinement and gaugino
condensation.

The above picture also applies to the affine quiver theories, with the
couplings for the non-affine nodes still labeled by $\vec x$ exactly
as in \chamber, for $i=1\dots r$. The
only new feature is the existence of the extra affine node, $i=0$,
whose gauge coupling is given by
\eqn\giparm{g_0^{-2}(\vec x)={1\over g_s}-\sum_{i=1}^r
 d_i\vec e_i  \cdot \vec x.}
The $d_i$ in \giparm\ are the Dynkin indices, and the extending
simple root is $\vec e_0=-\sum _{i=1}^r d_i \vec e_i$, which can
be written as $\sum _{i=0}^r d_i\vec e_i=0$ with $d_0\equiv 1$.
There is now the further restriction on the space of allowed
$\vec x $ that the RHS of
\giparm\ is also non-negative.  This gives an additional codimension one
boundary, cutting the Weyl chamber wedge to a finite sized box; this
space of allowed $x$ is the $\widehat G$ Coxeter box.  It can be
viewed as the fundamental domain of the affine Weyl group action on
$\IR^r\over {\widehat W}$ where $\widehat W$ is the affine Weyl group.
Equivalently it can be viewed as the fundamental domain of the Cartan
torus by the Weyl group action, $T^r/W$, noting that $\widehat W
=W\semidirect T$ where $T$ is the translation group of the root
lattice.  Note that a linear combination $\sum _{i=0}^r d_i
g_i^{-2}(x)=1/g_s$ is actually independent of $\vec x$.  This is the
gauge coupling of a diagonal $U(N_0)$.  Including the theta angles the
complex version of this statement is also true: the complex gauge
coupling
\eqn\taudiag{\tau _D\equiv \sum _{i=0}^r d_i \tau _i=\tau _{IIB},}
with $\tau _{IIB}$ the IIB string coupling.

A special case of the above discussion for the $\N =1$ affine
$\widehat G$ quiver theories is the case $N_i=0$ wrapped D5s, with
$N_0\neq 0$ transverse D3s.  This case leads to a $\N =1$
superconformal field theory with a $r+1$ complex dimensional moduli
space of gauge couplings $\tau _i$, which are the complexification of
the couplings in \chamber\ and \giparm.  The Weyl reflection dualities
maps the theory back to itself, except for changing the coupling
constants.  This is part of the S-duality group of these theories.
The remaining S-duality is the usual $SL(2,\IZ )$ action on the
diagonal gauge coupling \taudiag.  So a fundamental domain of the
moduli space of the $\N =1$ superconformal field theories is given by
(the complexification of )\chamber\ and \giparm, with $\vec x$ in the
$\widehat G$ Coxeter box, along with the $SL(2,\IZ )$ fundamental
domain for $\tau _D$.

This same picture holds for the special case where the deforming
$W_i(\phi _i)$ vanish, leading to $\N =2$ rather than $\N =1$
superconformal field theories.  The Coxeter box structure for the
moduli space was found in the related case of D5 branes at a $\IC
^2/\Gamma _G$ singularity \refs{\DouglasSW, \BlumMM}\ in
\IntriligatorDH.  The moduli space for the $\N =2$ superconformal
$\widehat G$ theories, setting $W_i=0$, was studied in \wim\ for the
$\widehat A$-case and generalized to all the $\widehat A$-$\widehat
D$-$\widehat E$ quiver cases in
\kmv, where the moduli space of couplings was shown to be identified
with moduli of flat A-D-E connections on $T^2$.  The Coxeter box can
be identified with the moduli space of flat A-D-E connection on $S^1$ and
the description of the moduli space along the lines discussed here was
noted in \lnv.  Again, for both the $\N =2$ and the $\N =1$ superconformal
theories with $W_i(\phi _i)\neq 0$, the S-duality group
corresponds to $SL(2,\IZ )$ action on \taudiag, along with
the Weyl reflections on the $\tau _i$, as in \weylact.

%
%
%
%

\newsec{The classical quiver gauge theories}

\subsec{4d ${\cal N}=1$ A-D-E quiver theories}

The class of ${\cal N}=1$ quiver gauge theories we consider is a
deformation of ${\cal N}=2$ quiver gauge theories with gauge group
$\prod _i U(N_i)$, with $i$ running over the nodes of the quiver
diagram, and bi-fundamental hypermultiplets for the linked nodes $i$
and $j$; these hypermultiplets can be written as ${\cal N}=1$ chiral
superfields $Q_{ij}$ in the $(N_i, \overline N_j)$ and $Q_{ji}$ in the
$(\overline N_i, N_j)$ of $U(N_i)\times U(N_j)$.  The quiver diagrams
of interest here are the $G=A,D,E$, or affine $\widehat G= {\widehat
A},{\widehat D},{\widehat E}$, Dynkin diagrams; these are the most
general asymptotically free, or conformal respectively, ${\cal N}=2$
quiver gauge theories \kmv.  We consider deformations of these
theories to ${\cal N}=1$ supersymmetric theories by adding a
superpotential for the adjoint fields, $W_i(\phi_i)$, so the full
tree-level superpotential is
\eqn\Wnii{W=\sum _i [\Tr \sum _{j} s_{ij}Q_{ij}Q_{ji} \phi_i  -\Tr
W_i(\phi_i)]}
where $s_{ij}=-s_{ji}$ is the
intersection matrix of nodes $i$ and $j$, which is zero if the nodes
are not linked and $\pm 1$ if they are linked (nothing depends on the
choice for the $s_{ij}$ signs).  The first term
in \Wnii\ is that of the original undeformed $\N =2$ theory.

In the non-affine case, there is no restriction on $W_i (\phi_i)$.  In
the affine case, however, the geometric engineering of these quiver
theories \CKV\ leads to one restriction on the superpotentials:
$$\sum_{i=0}^rd_i W_i(x)=0$$

The equations of motion following from \Wnii\ are
\eqn\niieom{ \sum _j s_{ij}Q_{ij}Q_{ji}=\partial_i{W_i (\phi_i)},
 \qquad \phi _i Q_{ij}=Q_{ij}\phi _j,}
for every $Q_{ij}$.  The vacua are the solutions of these equations,
modulo complexified gauge transformations.  We now review the vacuum
structure, which was derived in \CKV.  For the case where
the quiver diagram is a non-affine $G=A,D,E$ Dynkin diagram, there
are various vacua which are given in terms of the positive roots
$\vrho _K\subset \Delta ^+$ of $G$; here $K=1,\dots , R_+$, with $2R_+
+r=|G|$, and the positive roots can be expanded in terms of the simple
roots $\vec e_i$ as
\eqn\vrhoexp{\vrho _K=\sum _{i=1}^r n^i_K\vec e_i,}
for appropriate $n_K^i\geq 0$.  This corresponds to the fact that the
associated geometry has 2-cycles $S^2_K$ corresponding to the positive
roots $\vrho _K$.

For each $\vrho_K$ there  are a number of irreducible branches of the
 supersymmetric theory, given by the roots $x$ the equation
\eqn\FIcond{W' _K(x )\equiv \sum _i n^i_K W'_i(x)=0.}
For simplicity we take all $W$'s to be polynomials of the same
degree $k+1$ (the more general case can
also be constructed geometrically \CKV).  Then,
for each positive root $\vec \rho _K$, the above equation has $k$
roots, which we label as
$x=a_{(p,K)}$, with $p=1,...,k$ and $K=1\dots R_+$.  There
is a susy vacuum for every choice of $M_{(p,K)}\geq 0$ such that
\eqn\NiK{N_i=\sum _{K=1}^{R_+}\sum_{p=1}^k M_{(p,K)} n^i_K.}
In these vacua $\phi_i$ has $n^i_K M_{(p,K)}$ eigenvalues given
by the root $a_{(p,K)}$ and the gauge group is Higgsed as
\eqn\genhiggs{\prod _{i=1}^r U(N_i)\rightarrow
\prod_{K=1}^{R_+}\prod_{p=1}^k U(M_{(p,K)}).}

For the case of affine quiver diagrams, the vacua are similarly
labeled by the positive affine roots \CKV.   We will consider the cases
where there are no pure 3-brane branches (this is the analogue
of the Coulomb branch of the ${\cal N}=4$).  In this case the Higgs branches
are also labeled by the positive roots of affine A-D-E, which are described
as follows:
Recall that the highest root
of $G$ is $\psi = \sum _{j=1}^r d_j e_j$, with $e_j$
the simple roots; the extending affine root $e_0$ is $e_0\equiv -\psi$, so
$\sum _{i=0}^r d_i e_i=0$, with $d_0\equiv 1$.  The extended Cartan
matrix is $C_{ij}=e_i\cdot e_j$ for all $i,j=0,\dots ,r$. For affine Lie
algebras one replaces $e_i\rightarrow \widehat e_i=(e_i,0)$ for $i=1,
\dots ,r$ and $e_0\rightarrow \widehat e_0=(-\psi, 1)$.  Note that
$\sum _{i=0}^r d_i \widehat e_i =\delta$, with $\delta=(0,1)$ which
we identify as the D3 brane charge direction (called the `imaginary
direction' for the affine algebra).  The positive roots of the affine
algebra are given by
$${\widehat \vrho}_{\widehat K}: \qquad (\Delta ,n^+), (\Delta^+,0)$$
where $n^+$ is a positive integer and $\Delta$ denotes all roots.
Each such vector can be written as positive combination of positive
affine roots:
$${\widehat \vrho}_{\widehat K}=\sum_{i=0}^r n^i_{\widehat K} {\widehat e_i}$$
For each such root, consider its projection to the root lattice
 which is either a positive root or its negative,
 given by $\pm \sum_{i=1}^r n^i_Ke_i$
as $K=1,..., R_+$.
For each such branch we consider solutions to
$$ W'({\widehat \vrho}_{\widehat K})=\pm \sum_i W'_i(x) n^i_K=0$$
which is exactly the equation we considered in the non-affine case
(the possible minus sign does not affect the solutions to the
above equation).  There are $k$ solutions for each branch, which
we label with $(p,{\widehat K})$.  Choose non-negative integers
$M_{(p,{\widehat K})}$ which label how many of each irreducible
branch we choose.  These should satisfy
$$\sum_{\widehat K} M_{(p,\widehat K )} n^i_{\widehat K}= {\hat N}_i$$
In this branch the gauge group is Higgsed to
\eqn\affhiggs{\prod _{i=0}^r U({\widehat N}_i)\rightarrow \prod _{\widehat K}
\prod_{p=1}^k
U(M_{(p,{\widehat K})}).}

\newsec{Aspects of the quantum dynamics: gauge couplings and their running}

In this section we discuss some aspects of the quantum dynamics of
the gauge couplings and their running as a function of scale.  We will
first consider the underlying ${\cal N}=2$ quiver theory and then
we will discuss aspects of the ${\cal N}=1$ deformed theory by adding
superpotential terms.

\subsec{$\N=2$ quiver theories}

First let us ignore the superpotentials $W_i(\phi_i)$ so
we have an $\N=2$ quiver theory.  This is also a good approximation
for the dynamics of the $\N=1$ quiver theory
for energy scales large enough compared to the superpotential deformations
(i.e. for scales $\mu$ large compared to the adjoint mass $W_i''(\mu)$).

The $N=2$ exact beta function for the coupling $\tau _i\equiv {\theta _i\over
2\pi}+4\pi i g_i^{-2}$ of $U(N_i)$ is
\eqn\niibeta{\beta _i\equiv -2\pi i \beta (\tau _i)=\sum _j C_{ij}N_j,}
with $C_{ij}=2\delta _{ij}-|s_{ij}|=\vec e_i\cdot \vec e_j$ the Cartan
matrix of the A-D-E diagram, or the extended Cartan matrix of the affine
$\widehat A$-$\widehat D$-$\widehat E$ diagram.  The sign of $\beta _i$ in
\niibeta\ is chosen so that the theory is asymptotically free if \niibeta\
gives $\beta _i>0$.  Note that this can be conveniently summarized
by a vector $\vec{\beta}$ whose projection on ${\vec e_i}$ gives
$\beta_i$
\eqn\betv{\beta _i = \vec e_i \cdot \vec \beta \qquad
\hbox{with}\qquad {\vec \beta}={\vec N} \equiv \sum _i N_i \vec e_i.}
In the affine $\widehat G$ case we include the affine node $i=0$ in
\betv.  In the affine case, since $\sum _{i=0}^r d_i \vec e_i=0$,
the $\widehat G$ affine quiver theory with ${\hat N}_i=N_0 d_i$
has $\vec \beta =0$; it's an $\N =2$
superconformal field theory for any $N_0$. This corresponds to $N_0$
D3 branes and no wrapped D5 branes.
There are $r+1$ complex moduli,
given by the $U({\hat N}_i)$ gauge couplings $\tau _i$ for $i=0,\dots ,r$.
More generally, for any ${\hat N}_i$, the beta functions \niibeta\ are
invariant under the shift
\eqn\nshift{{\hat N}_i\rightarrow {\hat N}_i+N_0d_i,} for any $N_0$.
Also, for any ${\hat N}_i$, the beta function for the coupling
\eqn\taudis{\tau _D\equiv \sum _{i=0}^r d_i \tau _i}
of a diagonally embedded $U(N)$ vanishes, as $\sum _i \beta _i d_i=0$.

In the construction of the affine $\widehat G$ quiver theories
\DouglasSW\ via D3 branes at $G$ type ALE singularities, $\tau _D$ is
the IIB string coupling, while the other $\tau _i$ are given by the
orbifold blowing up modes coming from the twisted sector NS or RR
fields, as in \lnv.  Thus $\tau _D$ must have the $SL(2,\IZ)$
S-duality of IIB string theory.  The other independent $\tau _i$ also
exhibit S-dualities, which correspond to $G$ Weyl reflections. As
already mentioned, this shows that the $r+1$ complex dimensional
moduli space of the $\widehat G$ quiver $\N =2$ superconformal
theories consists of the $SL(2,\IZ)$ fundamental domain for $\tau _D$,
along with the complexification of the $\widehat G$ Coxeter box for
the remaining linear combinations of the couplings $\tau _i$ of each
$U({\hat N}_i)$ gauge group factor.

\subsec{The ${\cal N}=1$ quiver theories}

Now consider the $\N =1$ A-D-E quiver theories, with superpotential as
in \Wnii, with $W_i(\phi _i)$ as in \wadded: $W_i \sim \Tr \phi
_i^{k+1}+ $lower order.  For $k=1$ the $\phi _i$ are massive and can
be integrated out; for $k>1$ the $\phi _i$ should be kept (unless one
adds the generic lower order terms in \wadded, in which case $\phi _i$
is again massive and can be integrated out at low energies).  Note
that, for $k>1$, the deformations of the superpotential appear to be
irrelevant ($k=2$ is marginally irrelevant), and thus divergent in the
UV limit.  In the UV one needs a cutoff to define the theory, but the
IR aspects we will discuss are universal and independent of the
cutoff.  The deforming operators are actually ``dangerously
irrelevant,'' much as in
\KutasovSS, in that they get large anomalous dimensions and they
control the IR dynamics.

The $\N =2\rightarrow \N =1$ superpotential deformation $W_i(\phi _i)$
does not change the (1-loop exact) holomorphic beta functions, so they are the
same as in
\niibeta:
\eqn\uninst{\beta _i =\sum _j C_{ij}N_j=\vec e_i \cdot \vec N,
\qquad\hbox{which gives}\qquad
e^{2\pi i \tau _i (\mu)}=\left({\Lambda _i\over \mu}\right) ^{\beta
_i}.}  The quantity appearing in \uninst\ is $e^{-S^i_{inst}}$, with
$S^i_{inst}$ the action for a $U(N_i)$ instanton.  The gauge coupling
running
\uninst\ and scales $\Lambda_i$ apply above the mass scale $\Delta \sim
W''$ where $\phi _i$ gets a mass
(this occurs for $k=1$ or, for higher $k$, if $W_i'(x)$ has
no coinciding roots).  Below the mass scale $\Delta$, the $\phi _i$ can
be integrated out and the holomorphic beta functions are
instead
\eqn\betalow{\beta ^{low}_i\equiv -2\pi i \beta (\tau _i)=
N_i+\sum _j C_{ij}N_j.}  Matching the running coupling $g_i$ at the
scale $\Delta$ gives that the low-energy theory has dynamical scale
$\Lambda _i^{low}$ given by $(\Lambda _i^{low})^{\beta
_i^{low}}=\Delta ^{N_i}\Lambda _i ^{\beta _i}$.  The more general
matching relations, associated with the different Higgsing branches,
will be discussed in detail the following section.

We refer to the above 1-loop beta functions as the ``holomorphic beta
functions'' since, as usual, they exactly give the running of the
coefficient of the $U(N_i)$ gauge kinetic term, when the
superpotentials are written in terms of the holomorphic (bare)
quantities. Also of interest are the ``physical beta functions,''
which are the ones of relevance for analyzing RG flows and
determining the existence of
RG fixed points.  The physical beta functions can be written
in terms of the anomalous dimensions \NSVZ, which for our theories
yields
\eqn\betae{\eqalign{\beta ^{phys}_i \equiv -2\pi i \beta ^{phys}(\tau _i )&=
3N_i-N_i(1-\gamma (\phi _i))-\sum _{j\neq i}|\vec e_i \cdot \vec e_j|
N_j(1-\gamma (Q_{ij}))\cr &=(1+\half
\gamma (\phi _i))\sum _j C_{ij}N_j+\half \sum _{j\neq i} |\vec e_i
\cdot \vec e_j|N_j\beta
({\lambda _{ij}}),}}
where $\lambda_{ij}$ is the coefficient of $Q_{ji}\Phi_i Q_{ij}$
in the superpotential (note that this can be scaled to one, by
rescaling $Q$'s).
Here $\gamma (\phi _i)$ is the anomalous dimension of $\phi _i$ and we
define
\eqn\betaij{\beta (\lambda _{ij})\equiv \gamma (\phi _i)+2\gamma (Q_{ij}),}
which is (proportional to) the beta function for the $\lambda _{ij}$.
  The expression \betae\ is essentially the NSVZ beta function,
though without the denominator factor of \NSVZ; this is because
we are using the holomorphic gauge kinetic terms, $\sim \int d^2 \theta
\tau W_\alpha W^\alpha$, and the canonical matter kinetic terms.

The beta function \betae\ applies above the possible scale $\Delta$
where $\phi _i$ could be integrated out.  Below such a scale, the
exact beta function is
\eqn\betael{\eqalign{\beta _i&=3N_i-\sum _{j\neq i} |\vec e_i \cdot
\vec e_j|N_j(1-\gamma (Q_{ij}))
\cr &={3\over 2}\sum _j C_{ij}N_j+\sum _{j\neq i} |\vec e_i \cdot
\vec e_j|\widehat \gamma (Q_{ij}),}}
where we define $\widehat \gamma (Q_{ij})=\gamma (Q_{ij})+\half$.  Integrating
out $\phi _i$ induces a quartic superpotential for the $Q_{ij}$, which
would be marginal if $\gamma (Q_{ij})=-\half$, corresponding to
$\widehat \gamma (Q_{ij})=0$.  Indeed, in this case the beta functions
\betael\ all vanish for the affine $\widehat G$ quiver theories with
$N_i=N_0d_i$.

\newsec{Gaugino condensation in the non-affine $\N =1$ quiver theories}

Let us now consider the dynamics of the $\N=1$ quiver theory
taking into account the fact that at scales lower than the relevant
scales for the superpotentials $W_i$ the theory gets higgsed to various
branches.  For simplicity let us assume that all the $W_i$'s become
relevant at the scale $\Delta$.  Thus for scales $\mu >>\Delta$
we effectively have an $\N=2$ quiver theory with the running of the
coupling constants we have noted.  Let us assume that at the scale
$\Delta$ all these couplings are still small, i.e. $1/g_i^2(\Delta)>>1$,
so that the classical analysis of the branches is reliable.
For scales below $\Delta$ the superpotential becomes relevant and
the theory is Higgsed (for generic $W_i$ to a product of pure $\N=1$
theories $\prod _{p=1}^k \prod _{K=1}^{R_+} U(M_{(p,K)})$ with some
additional massive fields.
The $SU(M_{(p,K)})$ factor in \genhiggs\ gets a
mass gap, with gaugino condensation and confinement.

The naive low-energy superpotential associated with the $\prod
SU(M_{p,K})$ gaugino
condensations can be written as
\eqn\Wgcn{W_{g.c.}=\sum _{p=1}^k
\sum _{K=1}^{R_+} S_{(p,K)} \left (\log \left({\Lambda _{(p,K)}^{3M_{(p,K)}}
\over S_{(p,K)}^{M_{(p,K)}}}\right)+M_{(p,K)}\right),}
where the $\Lambda _{(p,K)}$ are the scales
of the low energy $U(M_{(p,K)})$ gauge groups as found via naive threshold
matching, which we will discuss in what follows.
$S_{(p,K)}$ is the $SU(M_{(p,K)})$ glueball field $S_{(p,K)}\sim
\Tr _{SU(M_{(p,K)})} W_\alpha W^\alpha$, whose expectation value is the
$SU(M_{(p,K)})$ gaugino
condensate.  The $S_{(p,K)}$ in \Wgcn\ are massive and can be integrated out.

In general, the naive gaugino condensation superpotential is only a
leading approximation to a more non-trivial exact result.  This is
seen, among other examples, in the analysis of \civ, where the geometric
transition duality emerged as a powerful tool to obtain the exact
superpotential.  Though $W_{g.c.}$ is not exact, it does exactly give
the non-trivial monodromies of the superpotential.  Moreover
it should be a good approximation in the case where the
 $\N=2$ gauge couplings are weak at the scale $\Delta$ where the
Higgsing takes place.  The coupling constant of the $U(M_{(p,K)})$ theory
at the scale $\Delta$ where Higgsing takes place satisfies
$$g_{(p,K)}^{-2}(\Delta )=\sum_i n^i_K g_i^{-2}(\Delta ).$$
We still need to match the running gauge couplings $g_{(p,K)}$ across
the thresholds of various massive matter fields in order to relate the
low energy scales $\Lambda _{(p,K)}$ to the high energy scales
$\Lambda _i$ of the original quiver theory.  This is what we will now
do.

As discussed earlier, the possible eigenvalues of the adjoints $\phi _i$ are
the solutions $a_K^p$ , with $p=1\dots k$ and $K=1\dots R_+$, of
\FIcond.  E.g. for the case $k=1$, with $W_i=\half m_i \Tr \phi _i ^2$,
we have
\eqn\aKis{a_K={\sum _i n_K^i m_i a_i\over \sum _i n_K^im_i}.}
In general, $\phi _i$ can have $M_{(p,K)}n^i_K$ eigenvalues equal to
$a_{(p,K)}$, with $N_i=\sum _{(p,K)}n^i_KM_{(p,K)}$,
\eqn\phiig{\phi _i \rightarrow \oplus _{(p,K)} a_{(p,K)}
(1_{M_{(p,K)}})^{n^i_K},}
which breaks
\eqn\parthiggs{U(N_i)\rightarrow \prod _{p=1}^k
\prod _{K=1}^{R_+}U(M_{(p,K)})_i^{n_K^i}.}
Under this breaking
the $Q_{ij}$ decompose as
\eqn\qijdec{Q_{ij}\rightarrow \oplus _{(p,K)} \oplus _{(q,L)}
(M_{(p,K)},
\overline{M}_{(q,L)})^{n^i_Kn^j _L},}
with the bifundamental in \qijdec\ of mass $a_{(p,K)}-a_{(q,L)}$.

There is additional Higgsing, besides that of \phiig, due to non-zero
expectation values of components of the $Q_{ij}$; this Higgsing breaks
\eqn\umkd{\prod _i U(M_{(p,K)})_i^{n^i_K}\rightarrow U(M_{(p,K)})}
which is a diagonally embedded subgroup.  The $Q_{ij}$ expectation values
can be seen by plugging into the equations of motion \niieom,
\eqn\qvev{\sum _j s_{ij}Q_{ij}Q_{ji}=W_i'(\phi _i),}
which we should evaluate for the above eigenvalues $a_{(p,K)}$ of $\phi _i$.
In the end, the unbroken gauge group is
\eqn\higgsf{\prod _{i=1}^r U(N_i)\rightarrow \prod _{p=1}^k
\prod _{K=1}^{R_+}U(M_{(p,K)}), \qquad
\hbox{with}\quad N_i=\sum _{p=1}^k\sum _{K=1}^{R_+} n_K^i M_{(p,K)}.}
The naive superpotential \Wgcn\ arises from gaugino condensation
in the unbroken gauge group factors of \higgsf.

The original high energy $U(N_i)$ theory, with its adjoint included,
has beta functions as in \betv:
$\beta _i =\vec e_i \cdot \vec N$, with $\vec N\equiv \sum _i N_i \vec e_i$.
Using \higgsf\ we can also express these in terms of the ranks
$M_{(p,K)}$ of the low-energy gauge group:
\eqn\NMeqn{\beta _i =\vec e_i \cdot \vec N=\vec e_i \cdot \vec M \qquad
\hbox{where}\qquad \vec M\equiv \sum _{p=1}^k \sum _{K=1}^{R_+}M_{(p,K)}
\vec \rho _K.}
Thus we can write the $U(N_i)$ instanton factors, in terms of the dynamical
scales $\Lambda _i$ of the original high energy theory, as in \uninst:
\eqn\uninstt{\Lambda _i ^{\sum _j C_{ij}N_j}=\Lambda _i ^{\vec N\cdot
\vec e_i}= \Lambda _i ^{\vec M \cdot \vec e_i}.}

We determine the scales
$\Lambda _{(p,K)}$ of the low energy $U(M_{(p,K)})$ theory in \higgsf
by naive threshold matching relations at the scales of all massive
$U(M_{(p,K)})$ matter and W-boson fields.  The result we thus obtain is
\eqn\matchgen{\Lambda _{(p,K)}^{3M_{(p,K)}}=[W''_K(a_{(p,K)})]^{M_{(p,K)}}
\prod _{(q,L)\neq (p,K)}(a_{(p,K)}-a_{(q,L)})^{-\vec
\rho _K \cdot \vec
\rho _L M_{(q,L)}}\prod _{i=1}^r \left(\Lambda _i ^{\vec M \cdot \vec e_i}
\right) ^{n^i_K},}
where we define $W_K''(a_{(p,K)}) \equiv \sum _{i=1}^r n^i_KW_i''(a_{(p,K)})$.
Note that the RHS of \matchgen\ properly has mass dimension
$M_{(p,K)}(1+\vrho _K \cdot \vrho _K)=3M_{(p,K)}$
(since all positive roots of the
simply laced $ADE$ satisfy $\vrho _K\cdot \vrho _K=\sum _{i,j=1}^rn_K^i
n_K^jC_{ij}=2$).

To see how \matchgen\ is obtained, write the exponent of
$(a_{(p,K)}-a_{(q,L)})$ in \matchgen\ as $M_{(q,L)}\sum
_{i,j=1}^r n^i_K n^i_LC_{ij}$; the terms involving $C_{ii}=2$ are
associated with $W$ boson threshold matching, whereas the $C_{ij}=-1$
terms are associated with threshold matching for matter fields coming
from components of $Q_{ij}$.  The products of $\Lambda _i ^{\vec
M\cdot e_i}$, with exponent $n_K^i$, appearing $\Lambda _i$ in
\matchgen\ results from the fact that $U(M_{(p,K)})$ arises as the diagonal
subgroup, as in \umkd, so the $U(M_{(p,K)})$ gauge coupling is
\eqn\gumd{g_K^{-2}=\sum _i n^i_K g_i^{-2}}
at the scale of the Higgsing \umkd, and using \uninst.

We can now plug \matchgen\ into \Wgcn\ to get the final expression
\eqn\wgca{\eqalign{W_{g.c}&=\sum _{(p,K)}
S_{(p,K)}\left( M_{(p,K)}+M_{(p,K)}\log \left({m_{(p,K)}\prod _i
\Lambda _i ^{n^i_Kn^j_KC_{ij}}
\over S_{(p,K)}}\right) \right)\cr &+\sum _{(p,K)}
\sum _{(q,L)\neq (p,K)}
\sum _{i,j=1}^rS_{(p,K)}M_{(q,L)}n^i_Ln^j_KC_{ij}\log
\left({\Lambda _i \over a_{(p,K)}-a_{(q,L)}}\right).}}

We stress again that this is only an approximation, valid in the regime where
the gauge couplings are weak at the scale determined by the superpotentials.
Nevertheless, the non-trivial monodromies of \wgca\ are expected to be exact,
as the additional quantum corrections are single valued.

\newsec{Geometric Construction}

In this section we will study the geometric realization of the ${\cal
N}=1$ A-D-E quiver theories, and connect with the field theoretic
analysis of these theories presented in the previous sections.  The
geometric description allows the formulation of large N duals via
transitions of the form $S^2\rightarrow S^3$, which we interpret as
the field theory developing gaugino condensates. The dynamics of the
gauge theory can be mapped to geometric language in a beautiful
way. In particular, we show that the running of the gauge couplings is
imposed upon us by the log divergences in the periods of the
holomorphic three form on non-compact 3-cycles. The superpotential
obtained in \wgca, from naive integrating in, is shown to be the
leading order approximation in a weak coupling expansion of the exact
superpotential given in terms of geometric periods. This leading order
approximation can be obtained from the geometry via a monodromy
analysis in the form of the Picard-Lefschetz formula.

\subsec{Review of Geometric Engineering of ${\cal N}=1$ A-D-E Quiver
theories}

The geometric engineering of ${\cal N}=1$ A-D-E quiver theories is
done in two steps. The first is to consider Type IIB on an ALE space
with a blown up A-D-E singularity. Wrapping D5-branes around different
non-trivial 2-cycles will give rise to ${\cal N}=2$ gauge theories on
the world volume. Likewise, adding D3-branes transverse to the ALE
space will give ${\cal N}=2$ affine A-D-E quiver theories on their
worldvolumes. The second step is to realize that these ALE spaces can also
be made nonsingular by adding relevant deformations. These
deformations can then vary over the complex plane transverse to the
ALE space, D3 and D5 branes. This fibration induces a
superpotential in the theories, breaking the supersymmetry down to
${\cal N}=1$.

\bigskip
{\it A-D-E singularities in dimension 2}
\bigskip

Blown down singular
ALE spaces can be viewed as hypersufaces $f(x,y,z)=0$ of ${\bf C}^3$:
$$
\eqalign{
G=A_r:& \qquad f=x^2+y^2+z^{r+1}\cr
G=D_r:& \qquad f=x^2+y^2z+z^{r-1}\cr
G=E_6:& \qquad f=x^2+y^3+z^4\cr
G=E_7:& \qquad f=x^2+y^3+yz^3\cr
G=E_8:& \qquad f=x^2+y^3+z^5}
$$
These spaces can be made smooth by adding relevant deformations of the
form,
$$\sum_{i=1}^r P_{c_i(G)}(t_1,\ldots, t_r)R_{C_2(G)-c_i(G)}(y,z),$$
where the subscripts are the degrees of the polynomials under the
scaling where $t_i$'s have degree one and $f(x,y,z)$ has degree
$C_2(G)$, the dual Coxeter number of G ($c_i(G)$ are the degrees
of the Casimirs of $G$).  Notice that there are $r=$rank$(G)$
deformation parameters $t_i$'s.
For generic $t_i$'s, there are $r$ independent classes of
non-vanishing $S^2$'s and their intersection can be chosen to
correspond to the G Dynkin diagram. The holomorphic volumes of the
$S^2_i$'s are
denoted by,
$$ \alpha_i = \int_{S_i^2}{dy dz \over x}  \qquad {\rm for} \qquad
i=1,\ldots ,r $$
The $\alpha_i$ are in 1-1 correspondence with the simple roots of G,
and are linearly related to the $t_i$.

The deformed ALE space is simple to write in the $A$ and $D$ cases, namely,
\eqn\Adef{A_{r}  \quad : \quad x^2+y^2+\prod_{i=1}^{r+1}(z+t_i)  \qquad
\sum_{i=1}^{r+1}t_i = 0}
\eqn\Ddef{D_{r}  \quad : \quad x^2+y^2z
+{\prod_{i=1}^{r}(z+t_i^2)-\prod_{i=1}^rt_i^2 \over z}+2 \prod_{i=1}^r t_i
y }
and the holomorphic volumes are given by,
\eqn\labA{A_{r}: \quad \alpha_i = t_i - t_{i+1}  \quad  i=1, \ldots ,r}
\eqn\labD{D_{r}: \quad \alpha_i = t_i - t_{i+1}  \quad  i=1, \ldots ,r-1 \quad
{\rm{and}} \quad \alpha_r = t_{r-1}+t_r}
The corresponding equations for $E_6$, $E_7$ and $E_8$
deformations in terms of $t$'s (which are chosen
to be linearly related to $\alpha$'s) are more complicated
and we refer the reader to \km.

\medskip
{\it Fibration}
\medskip

We want to obtain a Calabi-Yau 3-fold by fibering the ALE space described
above over a complex plane whose coordinate we denote by $t$. This
fibration is implemented by allowing the $t_i$'s to be polynomials in
$t$. Therefore, the holomorphic volumes $\alpha_i$ will also be functions
of $t$. Wrapping $N_i$ D5 branes around the $S^2_i$ fiber, but with world
volume transverse to the t-plane, will induce a classical superpotential in
the gauge theory satisfying,
$$
W'_i(t)=\alpha _i(t)
$$
where $t$ corresponds to $\langle \Phi_i \rangle$, the expectation value of
the adjoint of the $U(N_i)$.

Notice that without the superpotentials, i.e., with a trivial fibration,
the normal bundle over each $S^2$ is ${\cal O}(-2)\oplus {\cal O}(0)$. However, with the
introduction of superpotentials, the geometry will have points where a
given cycle can have zero holomorphic volume. These points, which are
singular in the geometry, can be blown up, giving rise to $S^2$'s with normal
bundle ${\cal O}(-1)\oplus {\cal O}(-1)$. If the degree of $W'_i(t)$ is $k$, there will
be $k$ points in the t-plane for each positive root $\v{\rho}_K$ of G where
the holomorphic volume vanishes:
\eqn\solu{\alpha(\v{\rho}_K)=\sum_{i=1}^r n_K^i\alpha_i(t) =
\sum_{i=1}^r n_K^iW'_i(t)
=0. }
These are the supersymmetric vacua corresponding to
the Higgsing \genhiggs\ where we wrap $M_{p,K}$ D5 branes around the cycle
at the $p$-th solution of \solu. Let us rewrite \genhiggs,
$$\prod _i U(N_i)\rightarrow \prod _{K=1}^{R_+}\prod _{p=1}^k
U(M_{(p,K)}).$$
Clearly, the charge conservation condition is \NiK,
$$N_i=\sum _{K=1}^{R_+}\sum _{p=1}^k
M_{(p,K)}n_K^i.$$

\subsec{Large $N$ duality}

In the IR limit of the gauge theory we are left with pure ${\cal N}=1$
SYM with gauge group $\prod _{K=1}^{R_+}\prod _{p=1}^k
U(M_{(p,K)})$. This theory is expected to have gaugino condensation in
each factor of the gauge group, as discussed in section 5. As in
\refs{\stkl,\mn, \vaaug,\civ}, the proposal is that the geometry
realizes this process by geometric transitions of the form
$S^2_{(p,K)}\rightarrow S^3_{(p,K)}$. It is important to notice that
all these are conifold-like transitions since the $S^2_{(p,K)}$'s
being blown down have normal bundle ${\cal O}(-1)\oplus {\cal O}(-1)$.

The number of singular points after blowing down all $S^2_{(p,K)}$'s
is $kR_{+}$, where $k$ is the degree of $W'_i(t)$'s. The large N dual
is therefore achieved by deforming the complex structure of the
singular Calabi-Yau 3-fold,
\eqn\sing{ x^2 + F(y,z,t_1(t), \ldots , t_r(t)) =0}
by normalizable deformations (including the log normalizable).  These
normalizable deformations correspond to dynamical fields, as opposed
to fixed parameters \gvw.  The dynamical fields which they correspond to
are precisely the $SU(M_{(p,K)})$ glueball fields $S_{(p,K)}\sim \Tr
_{SU(M_{(p,K)})}W_\alpha W^\alpha$.  In \CKV, it was shown that the
total number of these normalizable deformations is exactly $kR_+$, the
expected number of $S^3$'s. This matches the natural idea that the
$kR_+$ gaugino condensates are independent, dynamical, and control the
sizes of the $S^3$, parameterizing the deformation of the geometry.

The normalizable deformations can be easily found by noting
that \sing\ has the form
\eqn\land{ f(x,y,z) + a t^{kC_2(G)} + \ldots = 0. }
Charges can be assigned to $x$, $y$, $z$, and $t$ such that the above equation
has charge 1. In particular, $t$ will always have charge
$1/kC_2(G)$. Thinking about \land\ as the superpotential of a
Landau-Ginzburg theory, the central charge is given by,
\eqn\central{ {\hat c}  = (1-2Q(x))+ (1-2Q(y))+ (1-2Q(z))+ (1-2Q(t)) =
{2\over k C_2(G)}\left(k(C_2(G)-1)-1\right).}
The normalizable deformations are those monomials
$t^{\beta}y^{\delta}z^{\gamma}$ with charge $Q(t^\beta y^\delta z^\gamma)<
{{\hat c}\over 2}$; we also include the log normalizable deformations,
with charge  $Q(t^\beta y^\delta z^\gamma)={{\hat c}\over 2}$ \gvw.

\bigskip
{\it Periods and Superpotential}
\medskip

The geometry after the deformation is smooth and contains $kR_+$
non-trivial $S^3$'s. These 3-cycles form a natural basis for $A_{(p,K)}$ cycles
in the Calabi-Yau geometry and we define $kR_+$ non compact cycles $B_{(p,K)}$
dual to the $A_{(p,K)}$'s producing a symplectic pairing. An important role in
the sequel is played by the periods of the holomorphic three form $\Omega$
over $A_{(p,K)}$'s and $B_{(p,K)}$'s. We denote the periods,
\eqn\ABper{ \int_{A_{(p,K)}} \Omega \equiv S_{(p,K)}
\qquad   \int^{\co}_{B_{(p,K)}} \Omega \equiv
\Pi_{(p,K)} =  {\del {\cal F}\over \del S_{(p,K)}},}
where $\co$ is a cutoff needed to regulate the divergent $B_{(p,K)}$
integrals.  The $kR_+$ periods
$S_{(p,K)}$ are determined by \ABper\ in terms of the coefficients of the
$kR_+$ normalizable deformations.  One can then invert these relations, to
write the coefficients of the normalizable deformations in terms of
the $S_{(p,K)}$.

After the transition the D branes have disappeared and have been replaced by
fluxes on the $S^3$'s of a suitable 3-form $H$. This leads to a
superpotential \refs{\tayva, \mayr},
\eqn\wfluxi{W= \int H\wedge \Omega=\sum _{p=1}^k \sum _{K=1}^{R_+}\left(
\int _{A_{(p,K)}}H\int _{B_{(p,K)}}\Omega - \int _{B_{(p,K)}}H\int _{A_{(p,K)}}
\Omega\right).}
This thus gives for the full effective superpotential
\eqn\supgeo{ -{1\over 2\pi i}W =\sum _{p=1}^k \sum_{K=1}^{R_+}
\left( M_{(p,K)}\Pi_{(p,K)} +
{\alpha_{K}\over 2\pi i}S_{(p,K)}\right)}
where $\alpha_K$'s are related to the bare coupling constants of the original
$U(N_i)$'s with $i=1,\ldots, r$ of the quiver theory.
The precise correspondence will be given
below when we show that the logarithmic dependence on $\co$ of $\Pi$'s can
be absorbed in the $\alpha$'s, rendering the superpotential finite (up to
irrelevant constant terms) as we send the cut off $\Lambda _0$ to infinity.

\subsec{Dynamics of the theory}

It was shown in \civ\ for the $G=A_1$ quiver theory case
that \supgeo\ is the exact effective superpotential of
the $X(k,A_1)$ theory, and that the superpotential obtained
from naive integrating in is the leading order approximation of \supgeo\ in
a weak coupling expansion. Here will see that the same is true for the
general class of A-D-E quiver theories we have geometrically engineered in
this section.

\bigskip
{\it Renormalization of gauge couplings}
\bigskip

The superpotential \supgeo\ contains the periods of $\Omega$ over
non-compact cycles. These periods are divergent and need a cut off
$\co$ to be well defined. These are long distance (IR) divergences and
therefore we expect them to be related to short distance (UV)
divergences in the field theory. This was the case for $X(k,A_1)$
\civ, where the renormalization of the gauge coupling constant in
field theory was forced upon us in the geometric set up by the IR
divergence. This is also true for the general A-D-E cases as we now
proceed to show.

The periods of $\Omega$ can be computed using the fact that the
Calabi-Yau under consideration is an ALE fibration over a complex
plane. The three cycles in this geometry project to lines in the
$t$-plane where, over each point, there is an $S^2$. Compact $S^3$
cycles are those for which the projection is a line segment and the
holomorphic volume of the $S^2$ vanishes at each end. Non-compact
cycles on the other hand are semi- infinite lines in the limit when
$\co$ is infinite. The periods can be computed as
integrals of the holomorphic volume of a given $S^2$ over the  path in
the $t$-plane, i.e.,
$$ \int_{B_{(p,K)}}^{\co}\Omega  = \int_{C_{(p,K)}}
{\widetilde\alpha}(\v{\rho}_K)dt,$$
where $C_i$ is an appropriate contour and ${\widetilde\alpha}(\v{\rho}_K)$ is
the volume \solu\ after
the deformations are introduced, so we should have $\widetilde \alpha
(\v{\rho}_K)\rightarrow \alpha(\v{\rho}_K)$ when
the deformations are turned off.

Let us expand ${\widetilde\alpha}$ in a Laurent expansion in $t$,
$${\widetilde\alpha}(\v{\rho}_K) =  \sum_{m=-\infty}^{\infty}\sigma_m t^m $$
The charge of the LHS can be seen to be $kQ(t)$ by setting all deformations
to zero. This implies that
$$Q(\sigma_m)= (k-m)Q(t) = {k-m \over kC_2(G)}.$$

Our aim is to find the possible dependence of $\sigma_m$ on the deformation
parameters. Recall that deformation parameters $d_{\beta\delta\gamma}$ are the coefficients of the
allowed monomials $t^{\beta}y^{\delta}z^{\gamma}$. In the following we will
suppress the subscripts since only the charge will be important. The charge
of deformation parameters is therefore,
$$Q(d) = 1 - Q({\rm monomial}) \ge 1 - {{\hat c}\over 2} =
{k+1 \over kC_2(G)} $$
where equality holds for the log normalizable deformations.

Finally, imposing the condition that ${\widetilde\alpha}_i(t)
\rightarrow \alpha_i(t)$ upon turning off the deformations,
$d\rightarrow 0$,  implies that
$$  \sigma_m = 0 \qquad {\rm for}  \quad m > k $$
$$  \sum_{m=0}^k \sigma_m = \alpha(\v{\rho}_K)$$
$$   \sigma_{-1} = \sum_{i=1}^r g_i d^{\rm log}_i \qquad {\rm where}\quad
d^{\rm log}_i \quad {\rm are \; log \; normalizable,} $$
$g_i$ are classical superpotential parameters and $\sigma_{m}$ for $m\le
-2$ depend on normalizable as well as log normalizable deformations.

The conclusion is then that the $\co$ dependence of the non-compact periods
is
$$\Pi_{(p,K)} = \int^{\co}{\widetilde\alpha}(\v{\rho}_K)dt
=\sum_{i=1}^{r}n_K^iW_i(\co )
+ \sigma_{-1}{\rm Log}(\co ) + {\cal O}({1\over \co}) +\ldots .$$
The first term on the RHS is an irrelevant constant, which is
independent of the deformation parameters. So the only dangerous
divergence is the log one, with coefficient $\sigma _{-1}$.  The only
parameters in the superpotential \supgeo\ which can be renormalized to
absorb these log divergences are the $\alpha_K$'s.  It is non-trivial
for this to be possible, as the $\alpha_K$'s are the coefficients of
very special functions of $S_{(p,K)}$'s; so we need to show that the
$\sigma_{-1}$'s conspire to give this same $S_{(p,K)}$ dependence.

Let us choose the orientation of all the contours for computing the
compact periods to be counter-clockwise and for the non-compact dual
periods to go {}from $\co$ on the lower sheet to $\co$ on the upper
sheet crossing the branch cuts defined to be between the two points
that split when the deformation is tuned on.  The notion of upper
sheet and lower sheet refers to the fact that for each $S^3$ we have a
double point on the $t$ plane and the fibered geometry has naturally
related to a double covering.

Now we only have to remember that at each point on the $t$-plane we
have a fiber with a basis of two cycles intersecting according to the
Cartan matrix of the corresponding A-D-E root system. Let us pick one
of the non-compact periods $\Pi_{(p,K)}$ and keep track of how it
changes as we change $\co\rightarrow e^{i\theta}\co$ with $\theta \in
\{ 0, 2\pi\}$.

Using the Picard-Lefschetz formula \arn, the cycle corresponding to
the positive root $\v{\rho}_K$ will change as the contour crosses the
vanishing cycles
\foot{Vanishing cycles in Picard-Lefschetz formula refer to cycles that can
shrink by changing the complex structure. In our case by setting to zero
the deformations.}  according to their intersection. This can be made very
precise by denoting $\v{\rho}_{L}$ the class of the compact cycle
$\v{\rho}_{K}$ is
crossing. The change in the period is then,
$$ \Delta\Pi_{(p,K)} = (\v{\rho}_{K}\cdot \v{\rho}_{L})S_{(m,L)}$$
where $m=1,\dots ,k$ refers to the particular solution of
$\sum_{i=1}^r n_L^iW'_i(t)=0$ which corresponds to the cycle which we
are crossing.  Now we can write the total change in the non-compact
period as $\co$ goes around as
$$  \Delta\Pi_{(p,K)} = \sum_{L\in \Delta_+}(\v{\rho}_{K}\cdot
\v{\rho}_{L})\sum_{m=1}^kS_{(m,L)}. $$
This implies that $\Pi_{(p,K)}$ has a logarithmic dependence on $\co$ as
expected:
\eqn\pelog{ \Pi_{(p,K)} = {1\over 2\pi i}\left( \sum_{L\in \Delta_+}
(\v{\rho}_{K}\cdot
\v{\rho}_{L})\sum_{m=1}^kS_{(m,L)}\right){\rm Log}(\co ) + \ldots }
where $\ldots$ are the cut-off single valued pieces.

Recall that the second term in \supgeo\ was obtained by the identification
$S_{(p,K)} \leftrightarrow {\Tr (W_{\alpha}^2)_{(p,K)}}$, the $SU(M_{(p,K)})$
glueball field. Therefore,
$\alpha_K$ is also identified with the bare coupling of the
corresponding gauge factor ${8\pi^2 \over \left( g^{\rm YM(K)}_{0}\right)^2}$.
This implies that only $r$ of all $\alpha_K$'s are linearly
independent. Let us choose as basis $\alpha_i$ with $i\in \Delta^0$, the
set of simple roots. The other $\alpha_K$'s corresponding to positive
roots $\v{\rho}_K = \sum_{i=1}^r n_K^i \v{e}_i$ are given by,
\eqn\othr{\alpha_K = \sum_{i=1}^r n_K^i \alpha_i.}
Clearly, each $\alpha_i$ has to have a logarithmic dependence on
$\co$. In order to have a dimensionally sensible expression we need to
include new parameters $\Lambda _i$, which will be identified with the
dynamically generated scales of the high energy $\prod U(N_i)$ theory.
Let us assume the simplest ansatz for the basis,
\eqn\runn{ \alpha_i = - {8\pi^2 \over \left( g^{\rm YM(i)}_{0}\right)^2} =
\beta_i {\rm Log}\left( {\Lambda_i}\over
\co \right)  \qquad {\rm with} \qquad i=1,\ldots , r }
where $\beta_i$ are yet to be determined. This is the same phenomenon as
dimensional transmutation in field theoretic language.

Let us collect the possibly log-divergent pieces of the superpotential
\supgeo,  using the result from \pelog:
$$ -W _{divg}=\sum_{L\in \Delta_+}\sum _{m=1}^kS_{(m,L)}
\left(
\sum_{p=1}^k\sum_{K\in\Delta_+}M_{(p,K)}(\v{\rho}_K\cdot
\v{\rho}_L){\rm Log}(\co ) +
\alpha_{L}\right).$$
The $\alpha _L$ appearing in the above must cancel these divergences
term by term in $L$, requiring that
$$ \alpha_L = - \sum_{p=1}^k\sum_{K\in\Delta_+}M_{(p,K)}(\v{\rho}_K\cdot
\v{\rho}_L){\rm Log}(\co ) + \ldots$$
where $\ldots$ denote cut-off independent pieces. Specializing to $L=i\in
\Delta^0$ and using \runn\ we get that
\eqn\begeo{ \beta_i =
\sum_{K\in\Delta_+}\left( \sum_{p=1}^kM_{(p,K)}\right)
(\v{\rho}_K\cdot
\v{e}_i) =
\sum_{j=1}^rC_{ij}\sum_{K\in\Delta_+}\left( \sum_{p=1}^k
M_{(p,K)}\right)
n_K^i = \sum_{j=1}^rC_{ij}N_j.}
The geometry has thus reproduced the 1-loop holomorphic beta functions
\uninst.

It is simple to see that with \begeo\ and \othr\
the superpotential does not have logarithmic divergences. As
a by-product we have learned that the superpotential also depends
on $r$ scales $\Lambda_i$ in the following form,
\eqn\scaleW{ W =-\sum_{L\in \Delta_{+}}{\hat \alpha}_{L}\left(\sum_{p=1}^k
S_{(p,L)}\right) + \ldots }
where
$${\hat \alpha}_{L} = \sum_{i=1}^r n^i_{L}\beta_i {\rm Log}(\Lambda_i)
\qquad {\rm with}  \qquad  \v{\rho}_L = \sum_{i=1}^r n_L^i \v{e}_i$$
\bigskip
{\it Leading order superpotential}
\bigskip

The exact effective superpotential \supgeo\ can be studied in the weak
coupling limit. This means that the dynamically generated scales
$\Lambda_i$ with $i=1, \ldots ,r$ are small compared to the scales set by
the superpotentials $W_i(t)$'s. In geometrical terms this means that the
compact $S^3$'s are small compared to their separation in the $t$-plane. In
order to be more precise let us introduce some notation. For zero
deformation parameters we get $kR_+$ singular points located at the
solutions of
$$ W'_K(t) \equiv
\sum_{i=1}^r n_K^iW'_i(t) \equiv g_K \prod_{p=1}^k(t-a_{(p,K)}) =0 $$
for $K\in \Delta^+$, the set of positive roots.

After the deformation each singular point $t=a_{(p,K)}$ splits into
two giving rise to $S^3_{(p,K)}$. Let us denote the new two points by
$a_{(p,K)}^{+}$ and $a_{(p,K)}^{-}$.  Now the periods can be written
more explicitly as follows,
$$ S_{(p,K)} = {1\over 2\pi
i}\int_{a_{(p,K)}^{-}}^{a_{(p,K)}^{+}}{\widetilde\alpha}(\v{\rho}_K)dt
\qquad {\rm and} \qquad \Pi_{(p,K)} = {1\over 2\pi i}
\int_{a_{(p,K)}^{+}}^{\co }{\widetilde\alpha}(\v{\rho}_K)dt$$

The weak coupling regime can therefore be defined by the following
conditions $\mid a_{(p,K)}^{+}- a_{(p,K)}^{-}\mid \ll \mid
a_{(m,L)}-a_{(p,K)}\mid$ for all $(p,K)\neq (m,L)$.

Following \civ, using monodromy arguments one can compute the
Log$(S_{(p,K)})$ and Log$(a_{(p,K)}-a_{(m,L)})$ dependence of
$\Pi_{(p,K)}$ and therefore of the superpotential \supgeo.

Consider first the geometry close to $a_{(p,K)}$, this geometry can be
thought of as that of a single conifold in the limit we are considering,
therefore, the $S_{(p,K)}$ period should look like \civ,
$$ S_{(p,K)} = {1\over 2\pi
i}W_{K}''(a_{(p,K)})\int_{a_{(p,K)}^{-}}^{a_{(p,K)}^{+}}
\sqrt{(t-a_{(p,K)})^2-\mu_{\rm
eff}}dt $$
Using Picard-Lefschetz formula for $\mu_{\rm eff}\rightarrow e^{2\pi
i}\mu_{\rm eff}$, we get that the corresponding dual period changes as
$\Delta \Pi_{(p,K)} = S_{(p,K)}$, therefore one can conclude that,
$$\Pi_{p,K} =  {1\over 2\pi i}S_{(p,K)} {\rm Log}{S_{(p,K)} \over
W_{K}''(a_{(p,K)}) }+\ldots $$
Finally, let us consider how $\Pi_{(p,K)}$ changes when we move one
$a_{(q,L)}$ around $a_{(p,K)}$, again using P-L formula gives that,
$$ \Delta\Pi_{(p,K)} = (\v{\rho_K}\cdot \v{\rho_L})S_{(q,L)}$$
Notice that the coefficient in front of $S_{q,L}$ does not depend on $m$ or
$p$, this is because the intersection formula only sees the classes and
for a given $K$ all $p$ have the same class.

Now we can collect all these partial results to write,
\eqn\noncom{2\pi i \Pi_{(p,K)} = S_{(p,K)} {\rm Log}
\ln {S_{(p,K)} \over W_{K}''(a_{(p,K)}) } +
\sum_{L\in \Delta^+}\sum_{m=1}^k  (\v{\rho_K}\cdot \v{\rho_L})S_{(q,L)}
{\rm Log} (a_{(p,K)}-a_{(q,L)}) + \ldots}
in this formula the sum over $L$ and $m$ runs over all $(q,L)\neq (p,K)$.

The leading order superpotential can then be obtained by combining \supgeo,
\noncom, and \scaleW\ to get,
$$
\eqalign{W = & \!\sum_{K}\sum_{p=1}^k M_{(p,K)}\left( S_{(p,K)}{\rm Log}{
W_{K}''(a_{(p,K)})\over S_{(p,K)}} + \! \sum_{L}\sum_{m=1}^k
(\sum_{i,j=1}^rC_{ij}n_K^in_L^j)S_{(q,L)}{\rm Log}{1\over
(a_{(p,K)}-a_{(q,L)})}\right) \cr & \! +
\sum_{K}(\sum_{p=1}^kM_{(p,K)})\sum_{i,j=1}^r(C_{ij}n_K^j{\rm
Log}\Lambda_i)\sum_{L}n_L^i\sum_{m=1}^kS_{(q,L)} + \ldots} $$

In order to compare with the gauge theory answer from naive integrating in,
let us write $W$ collecting all terms with
$S_K$ together,
$$
\eqalign{W &=\sum _{(p,K)}
S_{(p,K)}\left( M_{(p,K)}\ln \left( {W_{K}''(a_{(p,K)})\prod _i
\Lambda _i ^{n^i_Kn^j_KC_{ij}}
\over S_{(q,K)}}\right) \right)\cr &+\sum _{(p,K)}
\sum _{(q,L)\neq (p,K)}
\sum _{i,j=1}^rM_{(p,K)}S_{(q,L)}n^i_Ln^j_KC_{ij}\ln
\left({\Lambda _i \over a_{(p,K)}-a_{(q,L)}}\right)}
$$
We thus find perfect agreement with the gauge theory answer \wgca. Notice
that \wgca\ contains linear terms in $S_{(p,K)}$. These and possibly
an infinite power expansion in $S_{(p,K)}$'s can not be derived using
monodromy arguments.  A more detailed analysis of the geometry result
shows that the superpotential indeed generally contains an infinite
power expansion of terms which are missed by the naive integrating
in analysis, as was computed for $X(k,A_1)$ in \civ.

Finally, one has to check that the weak coupling approximation is
self-consistent. For this it is necessary to identify the expansion
parameters that enter in the infinite power series mentioned before.
Let us assume that all the relevant scales set by the classical
superpotentials are of the same order equal to $\Delta$. This means
that $(a_{(p,K)}-a_{(q,L)})\sim W_J''  \sim \Delta$ for all $K,L,J\in
\Delta_+$. Moreover, let us assume that all the scales of the
individual $U(N_i)$ factors are of the same order $\Lambda_i \sim
\Lambda \ll \Delta$ for $i=1,\ldots ,r$. Let us show that the natural
dimensionless expansion parameter for the computation of periods is
${\Lambda \over \Delta}$.

The leading order superpotential \Wgcn\ implies that
$\ev{S_{(p,K)}}^{M_{(p,K)}}=\Lambda _{(p,K)}^{3M_{(p,K)}}$.  Then, using
\matchgen\ with $W_K''\sim (a_{(p,K)}-a_{(q,L)})\sim \Delta$,
and taking all $\Lambda _i \sim \Lambda$, we find for the
expectation value of the gaugino fields, or in geometric
language, the sizes of the $S^3$ cycles:
$$ \left( {\ev{S_K} \over \Delta^{3}}\right)^{M_K} =
\left( {\Lambda \over \Delta }
\right)^{\sum_{J}M_J\v{\rho}_J\cdot \v{\rho}_K}$$
This implies that the power expansion in $\Lambda /\Delta$, and hence
the superpotential \wgca, are valid approximations when
$\sum_{J}M_J\v{\rho}_J\cdot \v{\rho}_K > 0$. Since
$$ \sum_{J}M_J\v{\rho}_J\cdot \v{\rho}_K = \sum_{j=1}^r n_K^j\left(
\sum_{i=1}^rN_iC_{ij}\right) = \sum_{j=1}^r n_K^i \beta_i $$
with $n_K^i \ge 0$, and $n_j^i =
\delta_j^i$ for $K=j$ a simple root,
the necessary condition is thus that all $U(N_i)$'s have to be
asymptotically free.

This analysis shows that in cases when no weak coupling expansion is
possible in terms of the parameters of a given theory two
possibilities can occur. The first is that the exact superpotential
\supgeo\ might still be computable in a power expansion in terms of
the parameters of a different (dual) theory and the second is that no
simple gauge theoretic interpretation exists even though the geometric
description still yields exact results.

\newsec{Duality Predictions From Geometric Construction of the A-D-E
Quiver Theories}

Consider the geometric engineering of the quiver theory.  Consider
blowing down the cycles (i.e. where the inverse couplings
$1/g_i^2=0$).  If we are just given this geometry together with some
data about which classes the branes wrap (or how much flux is coming
out of each vanishing $S^3$) we cannot uniquely determine the quiver
theory corresponding to it.  The reason for this is that in order to
decipher the gauge theory we have to identify certain parameters in
the geometry with a choice of simple roots of the A-D-E, and this is
unique only up to the choice of a Weyl group action.  This implies that
with this data we cannot quite give a unique description of the quiver
theory, however we can give descriptions in seemingly different
looking gauge theories which {\it have to be equivalent because they
are describing the same underlying string theory}.  Our constructions
apply equally well to A-D-E as well as the affine case.  This is how
geometry predicts gauge theory dualities, in one to one correspondence
with elements of the Weyl group.  As is well known the Weyl group is
generated by Weyl reflections about simple roots, and this we identify
as Seiberg-like dualities in the corresponding quiver theory.

In the original geometric engineering we have blown up $S^2$'s and
which $S^2$'s we blowup picks a particular `preferred' description
for which gauge couplings $1/g_i^2>0$. Of course they can be
viewed as analytic continuation of the other dual descriptions where
some of the gauge couplings squared are negative.  This phenomenon,
taking into account the dimensional transmutation, becomes part of the
data of matching of scales between the dual theories.

Let us consider a given theory with branes $N_i$ wrapping the
corresponding dual cycles, undergoing a transition to Higgs branch
with branch number degeneracies $M_{(p,K)}$ where $K$ labels the
positive roots and $p$ an integer between $1,...,k$.  Now consider a
different choice of positive roots given by Weyl reflection about
$\vec e_{i_0}$.  This affects the roots by
\eqn\weylflip{\vec e_j'=\vec e_j-(\vec e_j\cdot \vec e_{i_0})\vec e_{i_0}.}
The conservation of brane charge determines the rank of the gauge
groups after transitions, as in \refs{\beret ,\ov }\ and we find
\eqn\Nconso{\sum N_i \vec e_i=\sum N_i'\vec e_i'.}
It follows from this that $N'_j=N_j$ for $j\not= i_0$, and
$$N_{i_0}'=N_f-N_{i_0}$$
where $N_f=\sum_{i\not =i_0} (-\vec e_i\cdot \vec e_{i_0})N_i$ denotes
the number of flavors of the $U(N_{i_0})$ theory.  The Weyl
group also acts on the couplings, which correspond to Kahler volumes
of the $e_i'$, as
\eqn\gweyl{{1\over {g'}_i^2}={1\over g_i^2 }-{\vec e_i\cdot \vec
e_{i_0}\over g_{i_0}^2}.}
Similarly it acts on the superpotentials by the integral of the holomorphic
3-form over the relevant cycle which is
\eqn\ssupo{W_{i}\rightarrow W_{i}-(\vec e_i\cdot \vec e_{i_0})W_{i_0}}
In the IR, i.e. at scales below the scale of the superpotential
we also have to choose which branches we are in.  This makes sense
assuming that the coupling of the gauge theory is weak at the
scale of the superpotential, so that the classical analysis is
reliable.  In this case we have branches labeled by the positive
roots $\v{\rho}_K$.  Under the Weyl reflection the positive
roots get permuted except for $\v{\rho}_K=\vec e_{i_0}$ which goes to minus
itself (it is also easy to see, using \ssupo\ that the
choices within a given branch get mapped in a canonical
way). Thus $M_{p,K}=M'_{p,w_{e_{i_0}}(K)}$, for $K\not =
e_{i_0}$ where
$w_{e_{i_0}}$ denotes the Weyl reflection by $e_{i_0}$, and
$M'_{p,K}=-M_{p,K}$ for $K=e_{i_0}$.  Note that this latter action
on the branches would yield negative multiplicities unless
$M_{p,e_{i_0}}=0$.  So only for this case we can formally use
the dual.  We will elaborate on the geometric meaning of this later.
However, we emphasize that even if $M_{p,e_{i_0}}\not=0$ in a formal
sense the dual theory makes sense. What we mean by this is that
when we set up the dual geometry and write the corresponding superpotential,
replacing the flux coming from the branch corresponding to $\vec e_{i_0}$
with a negative number does make sense, and would yield an identical
description of the geometry.  Thus at the level of setting up the dual
geometry description we simply have an ambiguity of reading off the
gauge theory.  Thus the geometry {\it predicts} gauge theoretic
dualities which we will verify in the next section.

\newsec{Dualizing a gauge group factor}

Consider a particular $U(N_{i_0})$ gauge group factor
in our general $\N=1$
quiver labelled by $k$ and $G$ or $\widehat G$.   We write the superpotential
for the fields charged under $U(N_{i_0})$ as
\eqn\welecok{W={s\over k+1}\Tr \phi ^{k+1}+
\Tr \phi \overline Q Q+\Tr m Q \overline Q,}
where $Q$ is a $N_f\times N_c$ matrix, with $\overline Q Q$ in the
adjoint of $U(N_c)$ singlet under $U(N_f)$ and $M=Q\overline Q$ a
$U(N_c)$ singlet and in the adjoint of $U(N_f)$.  The $N_f=\sum _{i\neq
i_0}(-\vec e_i \cdot \vec e_{i_0})N_i$
fundamentals arise from the bi-fundamentals connecting to the
neighboring nodes of the quiver diagram, and the mass $m$ in
\welecok\ is a matrix in the flavor space, which is actually given by
the expectation values of the adjoints of the neighboring nodes' gauge
groups.  We treat the neighboring nodes as weakly gauged flavor
symmetries.

As we briefly review, the above theory can be dualized to a $U(N_f-N_{i_0})$
gauge theory for all $k$.  This is naturally related to the
$U(N_f-N_{i_0})$ which arises in the $\N =2$ theory (setting $s=0$ in
\welecok) at the base where the ``baryon branch'' intersects the
Coulomb branch \ArgyresEH.  Before discussing the details of the
duality, we note a few of the most important features.

As seen in the geometry, the duality corresponds to a $G$ Weyl
reflection, or $\widehat G$ Weyl reflection in the affine case.  The
duality does not act on the $N_i$ of the other nodes, which correspond
to unchanged flavor symmetries, and takes $N_{i_0}\equiv N_c$ to 
$N_f-N_c$, i.e.
\eqn\Nduali{N_i{}'=N_i\quad\hbox{for} \quad i\neq i_0,\qquad N_{i_0}{}'
=N_{i_0}-\sum _j\vec e_{i_0}\cdot \vec e_j N_j.}
As discussed in the previous section, we can write this as
\eqn\vecNis{\vec N\equiv \sum _{i=1}^r N_i \vec e_i=\sum _{i=1}^r N_i{}'
\vec e_i{}',}
with
\eqn\etransf{\vec e_i{}'=\vec e_i-(\vec e_{i_0}\cdot \vec e_i )\vec e_{i_0},}
which is precisely the action of a  Weyl reflection about the simple root
$\vec e_{i_0}$.
Such transformations for all the nodes generate the entire Weyl group (or
affine Weyl group for the case of the affine quiver diagrams).

To see how \gweyl\ occurs in the field theory duality, consider the
holomorphic beta functions of the $\N =1$ quiver diagram theories
(above the scale $\Delta$ where the adjoints get masses); these
coincide with \niibeta, and can
be written as in \betv.
The beta functions
of the theory after the duality transformation are
\eqn\betaitr{\beta _i{}'=\vec e_i \cdot \sum _j \vec e_j N_j{}'
=\vec e_i{}'\cdot \sum _j \vec e_j{}'N_j{}'
=\vec e_i{}'\sum _j \vec e_j N_j =\beta _i -(\vec e_{i_0}\cdot \vec e_i)
\beta _{i_0},}
where we used \etransf,
$\vec e_i{}'\cdot \vec e_j{}'=\vec e_i\cdot \vec e_j=C_{ij}$, and
\vecNis.  So the holomorphic
functions transform precisely as under the Weyl transformation
\etransf.  We can formally integrate these beta functions to get
the similar transformation of the couplings $g_i^{(-2)}$, as in
\gweyl.  A similar transformation as \betaitr\ would hold for the
exact physical beta functions \betae\ if all $\gamma (\phi _i)$ are
equal and $\beta (\lambda _{ij})=0$; likewise for \betael\ if $\widehat
\gamma (Q_{ij})=0$.

Consider matching the running gauge couplings, given by \uninst, before
and after the duality transformation on some particular $U(N_{i_0})$;
the matching occurs at the scale $\mu =\Lambda _{i_0}$ where $U(N_{i_0})$
gets strong:
\eqn\matchik{e^{2\pi i \tau _j(\mu)}=
\left({\Lambda _j\over \mu}\right)^{\beta _j}=\left ({\Lambda _j'}\over \mu
\right)^{\beta _j'}\quad\hbox{at}\quad \mu =\Lambda _{i_0}.}
Using \betaitr\ the matching relation obtained from \matchik\ is
\eqn\lamduali{\Lambda _i'{}^{\beta _i'}=\Lambda _i ^{\beta _i}
\Lambda _{i_0}^{-(\vec e_i \cdot \vec e_{i_0})\beta_{i_0}},}
i.e. (aside from the case $\widehat A_1$ where
$C_{01}=-2$)
\eqn\niill{\Lambda _{i_0}^{\beta _{i_0}}\Lambda '_{i_0}{}^{\beta '_{i_0}}=1,
\qquad \Lambda' _j{}^{\beta '_j}=
\Lambda _{i_0}^{\beta _{i_0}|s_{i_0j}|}\Lambda _j^{\beta _j}\qquad j\neq i_0.}

The first relation \niill, which gives $\Lambda _{i_0}=
\Lambda _{i_0'}$, is similar to the duality relation
\lref\IntriligatorID{
K.~Intriligator and N.~Seiberg,
``Duality, monopoles, dyons, confinement and oblique confinement
in supersymmetric SO(N(c)) gauge theories,''
Nucl.\ Phys.\ B {\bf 444}, 125 (1995)
[hep-th/9503179].
}
\IntriligatorID\ for ${\cal N}=1$ SQCD without the adjoint $\phi$
\eqn\nill{\Lambda _{SQCD}^{3N_c-N_f}\widetilde \Lambda _{SQCD}
^{3\widetilde N_c-N_f}
\sim \mu ^{N_f},}
where $\mu$ is the scale appearing in the dual superpotential as
$W_{mag}=\mu ^{-1}Mq\overline q$.  Indeed, for $k=1$ the adjoint $\phi _{i_0}$
has mass $m=s$ from \welecok\ and can be integrated out from both the electric
and magnetic theories, giving $\Lambda _{SQCD}^{3N_c-N_f}=m^{N_c}\Lambda _{i_0}
^{\beta _{i_0}}$ and $\widetilde \Lambda _{SQCD}^{3\widetilde N_c-N_f}=
m^{\widetilde N_c}\Lambda _{i_0}'{}^{\beta '_{i_0}}$, and then \niill\
agrees with \nill\ for $\mu =m$.

Integrating the beta function equations, in order to have all $g_i^{-2}\geq 0$,
we should have
\eqn\scalineq{\Lambda _{NAF}>\mu >\Lambda _{AF},}
where $\mu$ is the energy scale and $\Lambda _{NAF}$ is the dynamical
scale $\Lambda _i$ of those $i$ which are not asymptotically free,
$\beta _i<0$, and $\Lambda _{AF}$ is that of those $i$ which are.  In
particular, the $\Lambda _{NAF}>\Lambda _{AF}$.  As we lower the scale
$\mu$, eventually we get to $\Lambda _{i_0}$ of the asymptotically
free $U(N_{i_0})$, which we dualize as above.  According to
\betaitr,
$U(N_{i_0}')$ is not asymptotically free and $U(N_i)$ is more
asymptotically free than it was before if nodes $i$ and $i_0$ are
linked.  The relation \lamduali\ ensures that the new scales satisfy
\scalineq, e.g. if $U(N_j)$ is NAF we have $\Lambda _j>\Lambda _{i_0}$
and then we get $\Lambda _j'>\Lambda _{i_0}$ if $U(N_j')$ is NAF or
$\Lambda _j'<\Lambda _{i_0}$ if $U(N_j')$ is AF.

A final relation, which will occupy the rest of this section is the
transformation \ssupo\ of the superpotential:
\eqn\Wtransfi{W_i(\phi _i)\rightarrow W_i(\phi _i)-(\vec e_i \cdot
\vec e_{i_0})W_{i_0}(\phi _i).}  To show that this is indeed the
case, we need to show that the dual of
our theory \welecok\ for the $U(N_{i_0})$ charged fields
is $U(N_f-N_{i_0})$ with the
superpotential
\eqn\wdualdem{\widetilde W=-{s\over k+1}\Tr \widetilde \phi ^{k+1}
+{s\over k+1}\Tr m^{k+1}+\Tr \widetilde \phi q\overline q +\Tr
m \overline q q.}
Here $\widetilde \phi$ is the $U(N_f-N_{i_0})$ adjoint of the dual
theory and $q$ and $\overline q$ are the $N_f$ dual matter fields.
The opposite sign of the first term in \wdualdem, as compared with
\welecok, corresponds to the result of \Wtransfi\ for $i=i_0$: $W_{i_0}
\rightarrow -W_{i_0}$.  The transformation in \Wtransfi\ for the
nodes $i\neq i_0$ corresponds to the second term in \wdualdem.  This
is because the mass $m$ in \welecok\
are actually the adjoints $\phi _i$ of the nodes linked to $i_0$,
so the second term in \wdualdem\ will properly lead to \Wtransfi\
for the nodes $i\neq i_0$ with $\vec e_i \cdot \vec e_{i_0}=-1$.

We will first outline how the predicted superpotential \wdualdem\
indeed arises for the case of $k=1$; after that we'll discuss $k>1$.

\subsec{$k=1$ case}

Consider first the case $k=1$, where $\phi _i$ is massive, with mass
$s$, and can be integrated out for scales $\mu <s$.  The relevant
duality for the low-energy theory is then that of \SeibergPQ.
When $s$ is large,
the low energy theory is $\N =1$ SQCD with $N_f$ flavors and the additional
tree-level superpotential
\eqn\weleci{W_{elec}=-{1\over 2s}\Tr
(\overline QQ)^2+\Tr mQ \overline Q,}
obtained by integrating out $\phi$ from \welecok\ via its equation of motion.
For $s$ large it's a good description to simply add this extra superpotential
to the usual SQCD dynamics.

For $N_f>N_c$, we can dualize the SQCD theory \SeibergPQ\
to $U(N_f-N_c)$, with superpotential
\eqn\wmagi{W_{mag}={1\over \mu}
M \overline q q -{1\over 2s}\Tr M^2+\Tr mM.}
$M$ is massive and can be integrated out by its equation of motion,
$M=s(\mu ^{-1}\overline q q+m)$,
leading to
\eqn\wmagii{W_{mag}={s\over 2} \Tr (m+{1\over \mu} \overline
q q )^2={s\over 2\mu ^2}\Tr (\overline q q)^2 +{s\over \mu}
\Tr (m\overline q q)+\half s \Tr m^2.}
Taking $\mu =s$, this superpotential is precisely what we would obtain
from \wdualdem\ upon integrating out the massive adjoint $\widetilde
\phi$.  In particular, corresponding to the Weyl reflection,
the sign of the quartic term in \wmagii\ is opposite to that of \weleci,
and we have the additional term $W_i(m)$ in \wmagii.

As an aside, we briefly review the vacuum structure of the $U(N_c)$
theory with superpotential \welecok, for $k=1$, thinking of the
linked nodes as a $U(N_f)$ flavor symmetry.  The relevant detailed
analysis has been presented in \refs{\CarlinoUK ,\kentaro}.
A semi-classical analysis of the vacua, for general quark masses
$m$ leads to $\pmatrix{N_f \cr r}$ vacua where the gauge group is Higgsed
as $U(N_c)\rightarrow U(N_c-r)$ for $r=0,\dots ,$min$(N_c,N_f)$; each
unbroken $SU(N_c-r)$ has no massless flavors and thus has $N_c-r$ susy
vacua via gaugino condensation.

Consider the quantum theory in the limit of large $s$, where we simply
add \weleci\ to the usual $U(N_{i_0})$ dynamics.  For example, for
$N_f<N_c$ the theory is described by the mesons $M$ with
superpotential
\eqn\weleckiii{W=-{1\over 2s}\Tr M^2+\Tr mM+
(N_c-N_f)\left({s^{N_c}\Lambda ^{2N_c-N_f}\over \det
M}\right)^{1/(N_c-N_f)}.}  This superpotential has $\half
(2N_c-N_f)\pmatrix{N_f \cr r}$ vacua where $\ev{M}$ expectation values
break $U(N_f)\rightarrow U(N_f-r)\times U(r)$, even in the
$m\rightarrow 0$ limit, for every $r=0,\dots ,N_f$.  These give all the
vacua for $N_f<N_c$ \CarlinoUK.
For $N_f>N_c$ we can analyze the vacua using the $U(N_f-N_c)$
dual.  The result (see \CarlinoUK) are vacua of two types.  One type
is visible semi-classically in the dual theory, with $U(N_f-N_c)$
Higgsed to $U(N_f-N_c-r)$ and $U(N_f)$ is unbroken in the
$m\rightarrow 0$ limit.  The other comes from strong coupling dynamics
in the dual theory: when rank$(M)=N_f$, the dual quarks are all
massive and a dynamical superpotential is generated in the dual,
e.g. via gaugino condensation; as usual, this superpotential is the
continuation of \weleckiii\ to $N_f>N_c$.  These vacua again have
$U(N_f)\rightarrow U(N_f-r)\times U(r)$ for $r=0,\dots ,N_f$.

One can also analyze the problem in the limit
where the adjoint mass $s\ll \Lambda$, with $\Lambda$ the scale of the
theory with $\phi$ included.  The theory can then be usefully analyzed
in terms of the curve of the $\N =2$ SQCD theory, breaking to $\N =1$
by the small adjoint mass $s$.  This analysis again shows two sorts of
vacua \refs{
\ArgyresEH, \CarlinoUK}.  One set, existing for all $N_f$, are vacua
with the entire gauge
group confined, and the flavor symmetry broken as $U(N_f)\rightarrow
U(N_f-r)\times U(r)$ for all $r\leq [N_f/2]$ via monopole
condensation.  The other set exists for $N_f>N_c$ and have unbroken
$U(N_f)$; they are visible semi-classically in the dual $U(N_f-N_c)$
theory of \ArgyresEH.

\subsec{$k>1$}
Now consider \welecok\ with $k>1$.
As discussed in \AharonyNE\ for $k=2$ and more generally in
\ElitzurHC, these theories,
without the term $m\overline QQ$ in \welecok, are dual to a
$U(N_f-N_i)$ theory with superpotential
\eqn\wdualoi{\overline{W}=-{s\over k+1}\Tr \widetilde \phi _i ^{k+1}
+\Tr \widetilde \phi  q \overline q ,}
(Comparing with \ElitzurHC, we have normalized $\overline q$ and $q$ so
that the coefficient of the Yukawa term in \wdualoi\ is the same as in the
electric theory.)  As discussed in \ElitzurHC, this duality can be
obtained from that of \refs{\KutasovVE,\KutasovNP,\KutasovSS} by
deforming by the $\overline Q \phi _i Q$ term in \welecok.  The
dual theory is of the same form as the original theory, and does not
contain the gauge singlet mesons found in the original ${\cal N}=1$
dualities of \refs{\SeibergPQ ,
\KutasovVE,\KutasovNP,\KutasovSS}; all of the mesons usually required
in the dual are massive for $0\neq s_i<\infty $ \ElitzurHC.

Following the $\Tr mQ\overline Q$ in \welecok\ through the duality is
a little more involved.  As was discussed in \AharonyNE\ for the case
$k=2$, one finds various vacua.  Our interest is in showing that one
of these vacua has $U(N_f-N_c)$ gauge group, with the terms involving
$m$ in the superpotential, as in \wdualdem.

As in \ElitzurHC, we obtain the duality by flowing from that of
\refs{\KutasovVE,\KutasovNP,\KutasovSS}, which relates the theory
\weleci\ to a magnetic $U(kN_f-N_c)$ theory
with superpotential
\eqn\wiispd{\overline W=-{s\over k+1}\Tr Y^{k+1}+{s\over
\mu ^2}\sum _{j=1}^{k}M_j\widetilde q Y^{k-j}q+\lambda M_2+mM_1.}
The $\Tr \phi \overline Q Q$ perturbation in \wiispd, with coefficient
$\lambda$ which we'll take to equal $1_{N_f}$, leads to a
Higgsing of the magnetic theory to $U(N_f -N_c)$ \ElitzurHC.
We now consider the effect of the added $m$ perturbation
in \wiispd. The $F$-term conditions required for a vacuum of the
theory \wiispd\ are
\eqn\meom{\eqalign{{s_i\over \mu ^2}\widetilde q Y^pq&=-m\delta _{p,k-1}-
\lambda \delta _{p, k-2},\cr
Y^k&=\mu ^{-2}\sum _{j=1}^{k-1} (k-j)M_j q\widetilde q Y^{k-j-1},\cr
\sum _{j=1}^k M_j \widetilde q Y^{k-j}&=0.}}
The vacuum solution of \ElitzurHC\ for $m=0$ Higgses $U(kN_f-N_c)$
to $U(N_f-N_c)$.   This solution can now be modified to account for
$m\neq 0$.  For simplicity, we just discuss the case $k=2$.  Considering
first the first flavor, the vacuum of \ElitzurHC\
has $q_1^\alpha =b\delta ^{\alpha ,1}$ and $\widetilde q^1_\alpha =b\delta
_{\alpha 1}$, with $b^2=-\lambda _1 \mu ^2/s$, satisfying \meom\ for
$p=0$.  We can satisfy \meom\ for $p=1$ by taking $Y_1^1=-m_1/\lambda _1$.
In order to satisfy the other equations in \meom\ we also need $(M_1)_1^1$
and $(M_2)_1^1$ to be non-zero; these non-zero values will not contribute
to the low-energy superpotential, since the linearity of \wiispd\ in the $M_j$
ensures that the coefficients of the $M_j$ have zero expectation value.

We now expand \wiispd\ around this vacuum, where $U(2N_f-N_c)$ is
Higgsed to $U(2N_f-N_c-1)$.  Though the $q_1$ and $\widetilde q^1$
flavor is eaten, we get back a flavor from $F^\alpha \sim Y_1^\alpha$ and
$\overline F_\alpha \sim Y_\alpha ^1$.
Expanding out the $-{s\over 3}\Tr Y^3$ term of the
$U(2N_f-N_c)$ theory gives
\eqn\higgspart{-{s\over 3}\Tr Y^3\rightarrow -{s\over 3}
\left( (-{m_1\over \lambda _1})^3+
\Tr \widehat Y^3 \right)+\lambda _1 \Tr \overline F \widehat Y F + m_1
\overline F F,}
where $\widehat Y$ is the part of $Y$ in the unHiggsed $U(2N_f-N_c-1)$
adjoint, and $F$ has been normalized so that the Yukawa
coupling in \higgspart\ ccoefficient is $\lambda _1$.  Continuing this
process for all flavors, and taking the $\lambda =1_{N_f}$, we
eventually get a $U(N_f-N_c)$ theory with superpotential precisely as
in \wdualdem, just as we wanted to verify.

We can also see the above $U(N_f-N_c)$ dual in the limit where we
treat the coefficient $s$ of the $\N =2 \rightarrow \N =1$
superpotential term in \welecok\ as being small, via an analysis
similar to that of \ArgyresEH.  In the undeformed $\N =2$ theory, at
the root of the baryon branch, there is a free-magnetic
$U(N_f-N_c)\times U(1)^{2N_c-N_f}$ theory.  Deforming by the term
$W_i={s\over k+1}\Tr \phi ^{k+1}$ leads to various vacua, the one of
interest for us being that where the $U(N_f-N_c)$ remains unbroken and
the $U(1)^{2N_c-N_f}$ is Higgsed entirely by monopoles, which condense
due to the $W_i={s\over k+1}\Tr \phi ^{k+1}$ deformation.  Carrying
out this analysis along the lines of \ArgyresEH, it can be seen how
all the terms in the expected superpotential \wdualdem\ can indeed
arise.

\newsec{$A_2$ example}

In this section we study the $A_2$ quiver theory with $k=1$ as an
example of how the dualities enter the description of the theory
both in the field theory analysis and in the geometric analysis.
We first present the field theory analysis and then discuss how
it is realized geometrically.  The other A-D-E cases
work in a similar fashion.

\subsec{QFT analysis of $A_2$ with $k=1$}
Consider the $A_2$ quiver theory $U(N_1)\times U(N_2)$ for $k=1$, i.e.
\eqn\wiex{W_i(\phi _i)=m_i(\half \Tr \phi _i^2-a_i\Tr \phi _i)}
for $i=1,2$.  As in \higgsf, vacua have
\eqn\aiiv{U(N_1)\times U(N_2)\rightarrow U(M_1)\times U(M_2)\times U(M_3),}
with $N_1=M_1+M_3$ and $N_2=M_2+M_3$ and $U(M_3)$ is diagonally
embedded in $U(N_1)\times U(N_2)$.   In these vacua $\phi _1=diag(a_11_{M_1},
a_3 1_{M_3})$ and $\phi _2=diag(a_21_{M_2}, a_31_{M_3})$ with
$a_3=(m_1a_1+ m_2a_2)/(m_1+m_2)$ (using \aKis).  Using \matchgen\ the low
energy scales are
\eqn\scalele{\eqalign{\Lambda _{M_1}^{3M_1}&=
m_1^{M_1}(a_1-a_3)^{- M_3}(a_1-a_2)^{M_2}\Lambda _{N_1}^{2M_1-M_2+M_3},\cr
\Lambda _{M_2}^{3M_2}&=
m_2^{M_2}(a_2-a_3)^{-M_3}(a_2-a_1)^{M_1}\Lambda _{N_2}^{2M_2-M_1
+M_3},\cr
\Lambda _{M_3}^{3M_3}&=m_3^{M_3}
(a_3-a_1)^{-M_1}(a_3-a_2)^{-M_2}\Lambda _{N_1}^{M_3-M_2+2M_1}
\Lambda _{N_2}^{M_3-M_1+2M_2}.}}

The description of this theory in terms of the Higgs
branches presented above, which uses a classical
analysis, make most sense when the couplings
of the gauge theories are weak at the scale relevant for the superpotential.
Let us call this scale $\Delta $ (simplifying the description
by assuming that there is only one physical scale associated
with the $W_i$.  So the analysis above is valid if $\Lambda_{N_i}<<\Delta$.
We then expect to get a gaugino condensation for the three
remaining gauge groups $U(M_i)$ with scales given by
$\Lambda_{M_i}<<\Delta$.  The running of the coupling of various groups
is depicted in Fig. 1. For scales  $\mu >\Delta$ we have the
$N=2$ running of the $U(N_1)\times U(N_2)$ and for scales $\mu <\Delta$
we have to take into account the superpotential and the Higgsing
to the three  $U(M_i)$ branches.  Note that the couplings of the
$U(M_i)$ groups at the scale $\Delta $ are given by ${1\over g_1^2},
{1\over g_2^2},{1\over g_1^2}+{1\over g_2^2}$ which explains
the values of the three coupling constant at $\mu =\Delta$
shown in Fig. 1.

\bigskip
\centerline{\epsfxsize=0.65\hsize\epsfbox{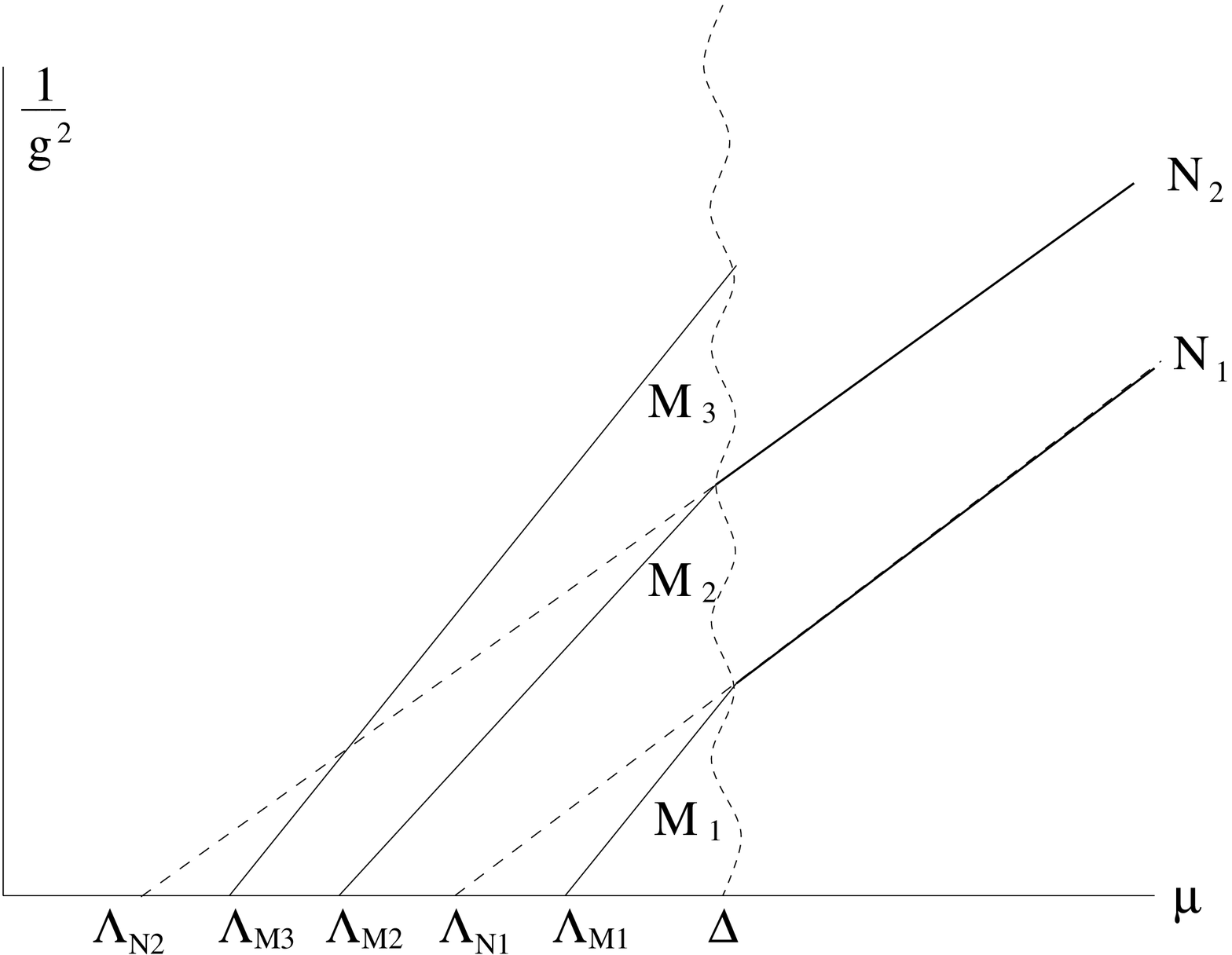}}
\noindent{\ninepoint\sl \baselineskip=8pt {\bf Figure 1}:{\sl Running of
gauge couplings for $\Lambda_{N_1}\sim \Lambda_{N_2}\ll \Delta$. Notice that ${1\over g^2_3(\Delta )}={1\over
g^2_1(\Delta)}+{1\over g^2_2(\Delta)}$. The energy axis is plotted in
logarithmic scale.}}
\bigskip

\bigskip
\centerline{\epsfxsize=0.65\hsize\epsfbox{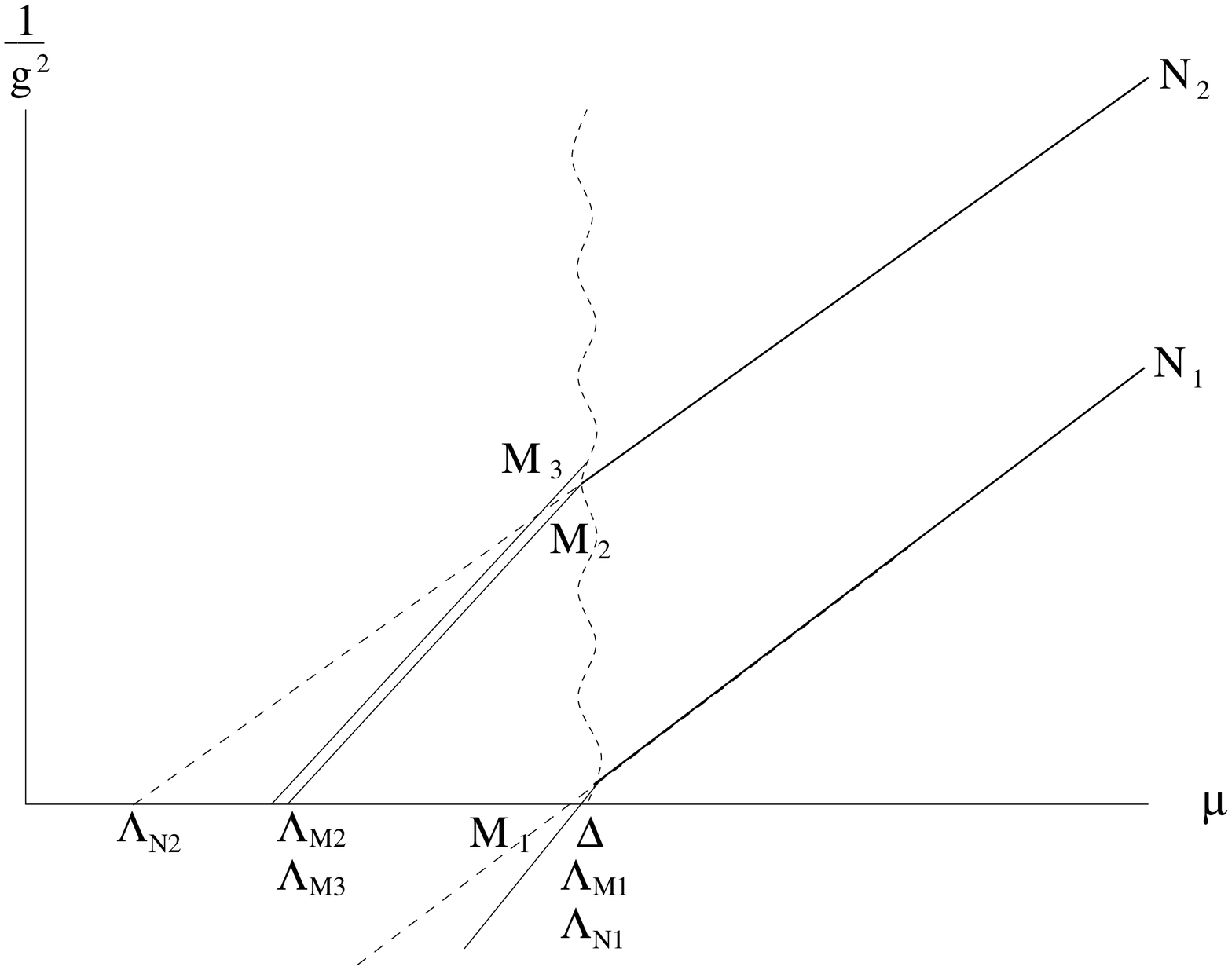}}
\noindent{\ninepoint\sl \baselineskip=8pt {\bf Figure 2}:{\sl Running of
gauge couplings for $\Lambda_{N_1}\sim \Delta$, $\Lambda_{N_2}\ll
\Delta$. Notice that since ${1\over
g^2_1(\Delta)}\rightarrow 0$ one gets ${1\over g^2_3(\Delta )}
= {1\over g^2_2(\Delta)}$.}}
\bigskip

When is it appropriate to dualize, say, $U(N_1)\rightarrow U(N_2-N_1)$
as in the previous section?
  Suppose e.g. that $N_2>N_1$.  We start changing
  the scales of the theories such that $\Lambda_{N_1}$
becomes bigger, with $\Lambda _{N_2}$ held fixed.  As we change $\Lambda_{N_1}$
we reach a point where $\Lambda_{N_1} \sim \Delta$.  In this case we
have the coupling constants given by  Fig. 2.

Now we ask what happens if we move $\Lambda_{N_1}$ to a region where
$\Lambda_{N_1}>>\Delta$?  The coupling $1/g_1^2$ of the $U(N_1)$ group
then formally goes to zero and becomes negative.  It is then
natural to use Seiberg duality to obtain a description in terms of a
more weakly coupled theory with positive $1/{\widetilde g_1}^2$.  Note
that now we are in a situation where $\Lambda _{N_1}\gg
\Lambda _{N_2}$ so that, for energy scales $\Delta <\mu \leq  \Lambda _1$, the
$U(N_1)$ dynamics are important (with ``negative $1/g_1^2$'') and the
$U(N_2)$ can effectively be treated as a spectator flavor symmetry,
which is weakly gauged.  Thus we replace $U(N_1)\times U(N_2)$ with
its dual gauge group $U({\widetilde N}_1)\times U({\widetilde N}_2)$
where ${\widetilde N_1}=N_2-N_1$ and ${\widetilde N}_2=N_2$, with
$W_1\rightarrow -W_1$ and $W_2\rightarrow W_1+W_2$ and ${1\over
{\widetilde g_1}^2}=-{1\over g_1^2}$ and ${1\over {\widetilde
g_2}^2}={1\over g_2^2 }+{1\over g_1^2}$.  However this cannot be a
good description if $M_1\not=0$; in this case (as an
approximate relation) we have $\Lambda_{M_1}\sim \Lambda_{N_1}$ so
the good description of the theory for scales $\mu < \Lambda_{N_1}$
should include the gaugino condensation (and various corrections
associated with $W$) in an $\N=1$ factor. Thus, for $M_1\neq 0$, once we go
to scales $\mu < \Delta $ we will encounter strong coupling quantum
corrections (involving gaugino condensation etc.) and thus we do not
have a weak coupling classical description.  We now consider
the branch where $M_1=0$.  Then $\Lambda_{M_1}=0$ and there is
no gaugino condensate associated with that branch.  Thus we will have
the running of the couplings depicted in the figure below:

\bigskip
\centerline{\epsfxsize=0.65\hsize\epsfbox{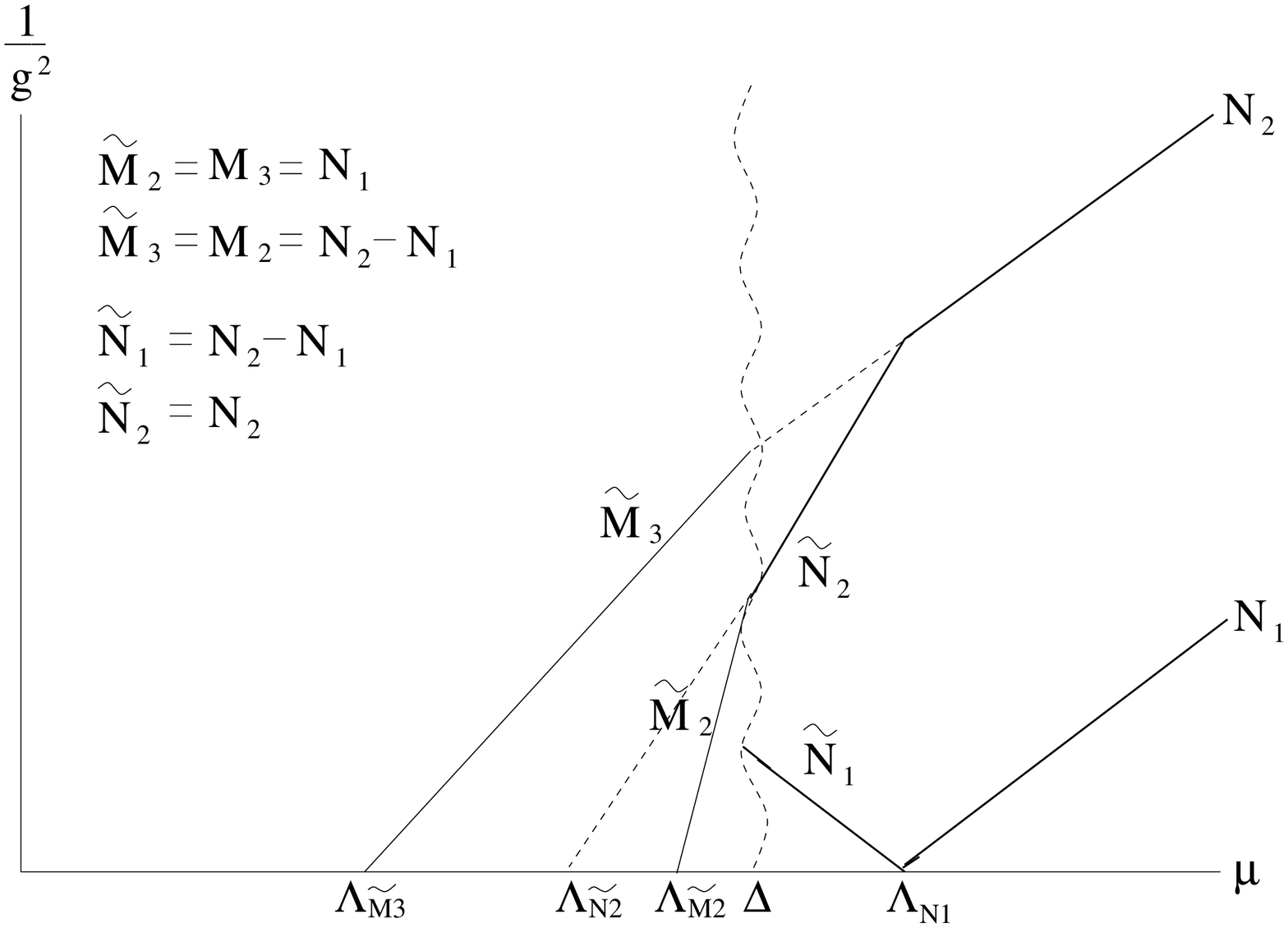}}
\noindent{\ninepoint\sl \baselineskip=8pt {\bf Figure 3}:{\sl Running of
gauge couplings. Notice that at $\mu =\Lambda_{N_1}$ a Seiberg duality is
necessary to keep ${1\over g^2_1(\mu)}>0$ for $\mu<\Lambda_{N_1}$.}}
\bigskip

Note that in this case after we hit the scale $\mu =\Delta$ we will
Higgs the theory to $U({\widetilde M}_2)\times U({\widetilde M}_3)$
(with ${\widetilde M}_3={\widetilde N}_1=N_2-N_1$ and ${\widetilde
M}_2={\widetilde N_2}-{\widetilde N}_1=N_1$, after which the theory
undergoes gaugino condensate in each factor.  Note however in terms of
the original assignments of branches what we had called $M_2,M_3$ have
switched roles: ${\widetilde M_2}=M_3$ and ${\widetilde
M_3}=M_2$. This is exactly the action of the Weyl group generated by
$e_1$ on the corresponding roots.  Note that this is consistent with
the field theory analysis for scales as well.  According to \niill,
after the duality $\Lambda _{\widetilde N_1}=\Lambda _{N_1}$ and
$\Lambda _{\widetilde N_2}^{2N_2-{\widetilde N_1}}=\Lambda
_{N_2}^{2N_2-N_1}\Lambda _{N_1} ^{2N_1-N_2}$.  We see from \scalele\
that for ${\widetilde M_1}=M_1=0$ the duality maps $\Lambda
_{{\widetilde M_2}}=\Lambda_{M_3}$ and $\Lambda _{{\widetilde M_3}
}=\Lambda _{M_2}$.  The $M_1=\widetilde M_1=0$ branch corresponds, in
terms of the discussion at the end of sect. 8.1, to the vacua which
are semi-classically visible in the dual $U(N_f-N_c)$ theory, with
unbroken $U(N_f)$.

\subsec{Geometry}

Let us describe the same case from the geometry side.  We will
first present qualitatively how the geometry analysis works.  However
we will be able to do more, namely we will provide also a basis
for an exact quantitative analysis of the vacuum structure of the
theory.

Geometrically over the $t$ plane we have three double points
where the three $S^3$'s will emerge with sizes given by $\Lambda_{M_i}$,
separated by distance of the order of $\Delta$.  In the limit
 $\Lambda_{M_i}<<\Delta$ we have three small $S^3$'s.  Now we analyze
 the theory with respect to scales:  For very high energies $
 \mu >>\Delta$ , this translates to probing the geometry at $t\sim \mu>>0$
(using the holographic picture) which basically means that the sizes of $S^3$'s
and the scale of $W$ is negligible and we have the $\N=2$ description.
This means far away from the ``tips'' of the cone we have a description
of the geometry in terms of the $S^2$'s with normal bundle which
is essentially ${\cal O}(-2)\oplus {\cal O}(0)$ with
varying $B$-field which gives the variation of
$1/g_i^2$.  This in turn can be captured by $
\Delta (1/g^2(t))=\int_t^{\Lambda_0} H$ which gives the expected
running.  As we approach the scales $t=\mu \sim \Delta$ this description
breaks down and we can distinguish the three distinct points which
correspond to blow downs of the three $S^2$'s corresponding to three
branches.   Now we see the corresponding $S^2$'s and their running
in accordance with $\N=1$ running of the couplings.
As we approach each double point we begin to see that
each double point has led to an $S^3$ which is identified with
the gaugino condensation in the gauge theory.
This is assuming all the $M_i\not=0$.  If any of the $M_i=0$
the corresponding $S^3$ does not emerge and has vanishing
size as there is no flux to support the $S^3$ (which is dual
to the statement that there
is no gaugino condensate if there is no gauge group!)

If we now consider increasing the scale of ${\Lambda_{N_1}}$, if
none of the $M_i$ are zero, this makes the size of $S^3$ corresponding
to $M_1$ to become larger.  Now suppose ${\Lambda_{N_1}}>\Delta$.  In this
case when we consider the description of the $B$ field as we come
towards the tip of the cone we see that the corresponding $S^2\rightarrow 0$.
However this happens at a scale where the $S^3$ corresponding
to branch one is large. Thus as we approach closer to the tip
of the geometry we do not have any gauge theoretic description, as the
gaugino condensation in the first factor would have already taken place.
However, suppose $M_1=0$.  In this case the size of the $S^3$ corresponding
to the first branch remains vanishing as we change $\Lambda_{N_1}$
even after we pass to ${\Lambda_{N_1}}>\Delta$.  In this case when
we describe the geometry as we approach the tips of the cone
as we get to the scale $t\sim \Lambda_{N_1}$ we have the $S^2
\rightarrow S^2$ transition giving us a dual geometry.  This is still
a good description because the $S^3$'s are much smaller and we can ignore
them, even as we approach $t\sim \Delta$.  As we go towards
the tips of the cone we find two $S^3$'s with finite size, which is
the scale the gaugino condensates take place.  In terms of the new
branches, the role of the two $S^3$'s have switched in accordance
with the Weyl reflection.

This was a qualitative description
of how the geometry sees the gauge theory description
given above.  However the geometry
also yields precise quantitative information which
we now discuss.
Before turning to the $X(1,A_2)$ example let us start by illustrating the
deformation procedure of section 6 corresponding to the large N dual for
$X(k,A_r)$ and then we set $r=2$ and $k=1$.

Recall the equation for the $A_r$ ALE space fibered over the t-plane given
by \Adef, where the fibration data and tree level superpotentials are related
explicitly by \labA,
$$ W'_i(t) = t_i(t) - t_{i+1}(t) \qquad {\rm for} \qquad i=1,\ldots ,r
\qquad {\rm with}  \qquad \sum_{i=1}^{r+1} t_{i}(t) = 0 $$
The charges of $x$, $y$, $z$ and $t$ are
$$Q(x)={1\over 2} \qquad Q(y)={1\over 2} \qquad Q(z)={1\over r+1} \qquad
Q(t)={1\over k(r+1)}$$
giving for the central charge (in agreement with \central\ for $C_2(A_r) =
r+1$)
$$ {\hat c} = (1-2Q(x))+(1-2Q(y))+(1-2Q(z))+(1-2Q(t)) =
2{(rk-1)\over k(r+1)} .$$

The (log) normalizable deformations are easily computed, by looking for
monomials with charges (equal to) less than ${{\hat c}\over 2}$, to be,
$$ P_{k-1}(t)z^{r-1}+P_{2k-1}(t)z^{r-2} + \ldots +
P_{(r-1)k-1}(t)z+P_{rk-1}(t)$$
where $P_j(t)$ are polynomials of degree $j$ in $t$. The number of
deformation parameters is then given by $k\sum_{j=1}^{r} j = k {r(r+1) \over
2}$. This is the same as $kR_+$ as expected. Moreover, the log normalizable
ones correspond to the leading coefficient of each $P_j$ and there are
exactly $r$ of them.

Let us now specialize to $X(1,A_2)$. The deformed geometry corresponds to,
\eqn\Atdef{ x^2+y^2 + (z+t_1(t))(z+t_2(t))(z+t_3(t)) + a z + b t + c = 0 }
where $a z$ and $b t$ are log normalizable
deformations and $c$ is normalizable.
{}From the tree level superpotentials \wiex, the fibration data is given
as,
$$ t_1(t) = {1\over 3}(2W'_1(t)+W'_2(t))  \qquad t_2(t) = {1\over
3}(-W'_1(t)+W'_2(t)) \qquad  t_3(t) = -{1\over 3}(W'_1(t)+2W'_2(t))$$
For zero deformation parameters, i.e., $a=b=c=0$ there are three singular
points in the geometry, given by the location of double roots of the
discriminant of the cubic equation in $z$ for $x=y=0$. The discriminant is
given by,
$$\Delta = (t_1(t)-t_2(t))^2(t_2(t)-t_3(t))^2(t_3(t)-t_1(t))^2$$
The solutions of $\Delta=0$ are then given by solving,
$$ t_1(t)-t_2(t) = W'_1(t) = 0 \quad t_2(t)-t_3(t) = W'_2(t) = 0 \quad
{\rm and} \quad t_1(t)-t_3(t) = W'_1(t)+W'_2(t) = 0$$
This makes explicit the 1-1 correspondence of $S^3$'s
with the positive roots of
$A_2$: after tuning on $a,b$ and $c$, the geometry is smooth, since
each double root of $\Delta$ splits into two, giving rise to an $S^3$ of
non-zero volume. Let us write $\Delta$ as,
$$\Delta = (t-a_1^+)(t-a_1^-)(t-a_2^+)(t-a_2^-)(t-a_3^+)(t-a_3^-) $$

If we write \Atdef\ as,
$$ x^2+y^2 + (z-z_1(t))(z-z_2(t))(z-z_3(t)) = 0 $$
then the compact periods of the holomorphic three form $\Omega$ are given by,
$$ S_K = {1\over 2\pi i}\int_{a_K^-}^{a_K^-} (z_I(t) - z_J(t)) dt  \quad
{\rm for} \quad  (K,I,J) \quad {\rm cyclic \; permutations \; of}
\quad (1,2,3)$$
and non-compact periods by,
$$ \Pi_K =  {1\over 2\pi i}\int_{a_K^+}^{\co } (z_I(t) - z_J(t)) dt$$
with $(K,I,J)$ as before.

The superpotential \supgeo\ is in this case,
$$ {1\over 2\pi
i} W_{\rm eff} = M_1 \Pi_1 + M_2 \Pi_2 + M_3 \Pi_3 + {\alpha_1\over 2\pi
i}S_1+ {\alpha_2\over 2\pi
i}S_2+{\alpha_3\over 2\pi
i}S_3$$
with $\alpha_3 = \alpha_1+\alpha_2$,
$$\alpha_1 = {1\over g^2_1(\co )} = (2N_1 - N_2) {\rm Log}\left(
{\Lambda_{N_1}\over \co} \right) \quad {\rm and} \quad \alpha_2 =
{1\over g^2_2(\co )} = (2N_2 - N_1) {\rm Log}\left(
{\Lambda_{N_2}\over \co} \right)$$

Let us describe the two interesting cases given in Figure 1 and Figure
3. The first corresponds to the case where $\Lambda_{N_1}$ and
$\Lambda_{N_2}$ are much smaller than $m_1$, $m_2$, $a_1-a_2$, $a_2-a_3$ and
$a_1-a_3$.

As discussed for general $X(k,G)$ in section 6, if the two theories
are asymptotically free, i.e, $2N_2-N_1>0$ and $2N_1-N_2>0$, then the exact
superpotential admits an expansion of the form,
$$ W_{\rm eff} = \sum_{K=1}^3 M_K \left( S_K{\rm Log}\left({\Lambda_K^{3}\over
S_K} \right) + \sum_{n,m,p=1}^{\infty }h^K_{nmp}S_1^nS_2^mS_3^p \right)$$
where $h^K_{nmp}$ only depends on the classical superpotential parameters, but
not on the particular Higgsing, and $\Lambda_K$'s
 are the scales from the threshold
matching conditions.  This also yields the running of the coupling
constants as depicted in the figure 1.

The second interesting case is when $\Lambda_{N_1}$ is taken to be
larger than the scales set by the superpotentials keeping $\Lambda_{N_2}$
fixed as before.  For $M_1 > 0$, this limit implies that $S_1$ grows
indefinitely and no weak coupling description is available
anymore. However, for $M_1 = 0$, as we will now argue there is a
vanishing $S^3$ corresponding to that branch.  At a leading order this
is obvious, because the gaugino condensate analysis suggests $S_1\sim
e^{-1/g^2M_1}$ which as $M_1\rightarrow 0^{+}$ gives $S_1\rightarrow
0$. However, in order to argue that this also continues to be the case
even when ${\Lambda_{N_1}}$ gets bigger we need to show that in an
analytic expansion $S$ is still zero.

The weak coupling approximation leads to an expression for
$S_1$ in terms of a power expansion with no order zero term in,
$$T = \left( {\Lambda_{N_1}\over \Delta } \right)^{{2N_1-N_2\over M_1}} $$
In the limit $M_1\rightarrow 0^+$,  $T \rightarrow  0$ since
${\Lambda_{N_1}\over \Delta }<1$ and $2N_1-N_2>0$.  Thus we have
$S_1=0$ in an analytic neighborhood, which by analytic
continuation implies its vanishing for all values of ${\Lambda_{N_1}}$.
In this situation the weak coupling description for the
$S_2,S_3$ will still be valid.  Here
as we increase $\Lambda_{N_1}$ we encounter the situation shown
in Fig. 3 where the couplings run until we get to a situation
where we have to describe it in terms of a dual description.
Here since the $S_2,S_3$ are small and $S_1$ is vanishing
we can still continue pretending to be in the phase where
the $S^2$'s are blown up with a dual description.
This is just the same geometry, of course, but interpreted
differently.

In order to see explicitly this description notice that the same geometry
with the same units of fluxes can be interpreted in a different way by the
following simple identification,
$$  z_1(t) = {\widetilde z}_2(t) \qquad  z_2(t) ={\widetilde z}_1(t)
\qquad z_3(t)
= {\widetilde z}_3(t)$$
In terms of the new identification we are led to write the superpotential
corresponding to this geometry and fluxes as follows,
$$ {1\over 2\pi i}W_{\rm eff} =
M_3 {\widetilde \Pi}_2 + M_2 {\widetilde \Pi}_3 -
{\alpha_1\over 2\pi i}{\widetilde S}_1 +
{\alpha_3\over 2\pi i}{\widetilde S}_2 +
{\alpha_2\over 2\pi i}{\widetilde S}_3$$
with the following identification:
$$ {\widetilde M}_1 = 0 \quad {\widetilde M}_2=M_3 \quad {\widetilde
M}_3=M_2 \quad {\widetilde \alpha}_1 = - \alpha_1 \quad {\widetilde
\alpha}_2 = \alpha_3 \quad {\widetilde \alpha}_3 = \alpha_2$$
This dual description arises as the IR of an $U({\widetilde N_1})\times
U({\widetilde N_2})$ $\N =2$ gauge theory with superpotentials,
$$ {\widetilde W}'_1(t) = -W'_1(t) \quad {\widetilde W}'_2(t) =
W'_1(t)+W'_2(t)$$
where,
$$ {\widetilde N_1} = {\widetilde M_1} + {\widetilde M_3} = N_2-N_1
\qquad {\widetilde N_2} = {\widetilde M_2} + {\widetilde M_3} = N_2$$

As discussed in section 6, the original weak gauge theory description
is invalid because $U(N_1)$ is asymptotically free and $\Lambda_{N_1}>
\Delta$. On the other hand, the dual theory also has ${\widetilde
\Lambda}_{{\widetilde N}_1} =
\Lambda_{N_1} > \Delta$, however, $U({\widetilde N}_1)$ is now IR free, and the
proof of validity of the weak coupling description is valid.

As we approach scales given by $S_2,S_3$ the weak coupling
gauge theory description ceases to make sense and we have the geometry
involving the two blown up $S^3$'s.

\newsec{Affine A-D-E quivers, RG cascades and affine Weyl reflections}

Up to now, as far as the quantum analysis, we have mainly concentrated
on the non-affine case.  In this section we will discuss aspects
of affine quiver theories.  We will also indicate the relation between
affine case and the non-affine case.  In fact we will motivate
the discussion of the affine case from this viewpoint.

Consider a non-affine case we have studied.  Suppose we are in a
situation where the ranks $N_i$ are such that all the $\N=2$ quiver
gauge groups are asymptotically free.  Let us discuss the theory at
scales $\mu >> \Delta$ where $\Delta $ denotes the scales relevant for
the superpotentials.  In this situation the couplings get weaker and
weaker as we go backwards towards the UV.  However, it is crucial to
remember that these field theories were obtained from a decoupling
limit of some string theory and even though ${1\over g_s}$ has been
taken to be much larger than any of the inverse square gauge couplings,
eventually those will become comparable to ${1\over g_s}$ in the UV.
This implies that stringy modes are not negligible and the field
theory description in terms of the non-affine quiver theory is
inadequate.  However we can view this as a special case of the affine
quiver theory, and we know that a weighted sum of the inverse coupling
constants squared is $1/g_s$. This means that at some
point we can get a situation where $1/g_0^2 <0$, i.e. the node
associated with the affine extension would have a negative coupling.
Formally we may not be bothered by this because there is no brane
wrapping it before.  However, if we wish that the string coupling be
weaker than all the other couplings we will have to do an inverse
duality of the form corresponding to the case $N_f=N_c$, corresponding
to a Weyl reflection on the affine node.  Continuing this towards the
UV we will end up, as we will discuss below with the inverse RG
cascade.  Of course the infrared physics does not get modified from
the non-affine case and so our discussion of the superpotential
etc. for the non-affine case would still be valid.  Note however that
there are other affine cases which do not end up as a UV completion of
non-affine case.  In particular, the projection of branches for the
positive affine roots to non-affine roots, leads to both positive and
negative roots.  This implies that in the affine case we
will end up with the superpotential where $M_i$ are the {\it net}
number of branches for a given positive root, which thus can be a
positive or negative integer.  This is the only new ingredient in the
context of affine case as far as the superpotential analysis is
concerned.

In the remainder of this section we will explain how the RG cascade
arises in the affine case and its relation to affine Weyl group.

\subsec{RG cascade and affine quivers}

Consider the general ${\cal N}=1$ affine quiver theory, with
$W_i={s_i\over k+1}\Tr \phi _i ^{k+1}$, and gauge group $\prod
_{i=0}^r U({\hat N}_i)$, with ${\hat N}_i=N_0d_i+N_i$, for general
$N_i$ with ${\hat N}_0 = N_0$.  In the AdS/CFT correspondence, the
$N_i\neq 0$ correspond to introducing wrapped branes.  If all $N_i=0$,
the theory could be scale invariant.  For some $N_i\neq 0$, some of
the groups are asymptotically free and others are IR free.  The gauge
couplings of the asymptotically free groups increase in the IR until,
eventually, it becomes appropriate to dualize the group.  This leads
to a RG cascade, via the various affine Weyl reflections, which
generalizes the case $\widehat A_1$, with $k=1$, found in \stkl.

The cascading effectively reduces $N_0$.  But nothing holomorphic
depends on $N_0$.  For example, the holomorphic beta functions
\uninst\ for ${\hat N}_i=N_0d_i+N_i$ are independent of $N_0$
since $d_i$ is a null vector of $C_{ij}$.  Also, though the theory is
not conformal when the $N_i\neq 0$, the holomorphic beta function
for the diagonally
embedded $U(N_0)_D$, whose gauge coupling \taudis\ is given by the string
coupling, always vanishes
\eqn\betalco{\beta _D\equiv \sum _{i=0}^r d_i \beta _i =0.}
One might wonder whether or not similar statements apply for the
physical beta functions, e.g. the physical beta functions
\betae\ will be independent of $N_0$ as long as all $\gamma (\phi _i)$ are
equal.  Likewise the physical beta functions will satisfy \betalco\ if
$\beta (\lambda _{ij})=0$ and all $\gamma (\phi _i)$ are equal.  These
statements should indeed be true in the limit where $N_0$ is large
compared to the $N_i$, i.e. away from the IR limit where $N_0$ has
cascaded away.  But in the far IR this description is no longer even
useful, as the theory breaks up into the vacuum branches, with gaugino
condensation, discussed in the earlier sections.  We'll discuss the RG
flow as the theory cascades, above the scale where it eventually
confines.

As already discussed, we
parameterize the gauge couplings as
\eqn\giparm{g_i^{-2}(\vec x)={1\over g_s}\delta _{i,0}+\vec e_i \cdot \vec x,}
with $\vec x $ in the $\widehat G$ Coxeter box, defined by the
condition that all $g_i^{-2}(\vec x)\geq 0$.  The vanishing of the
beta function \betalco\ corresponds to the fact that
\giparm\ gives
\eqn\gdiag{g_D^{-2}\equiv \sum _{i=0}^r d_i g_i^{-2}(\vec x)={1\over g_s},}
independent of $\vec x$.  In the AdS/CFT correspondence, for large
$N_0$, we can also see this equality \gdiag, and the statement
\betalco\ corresponds to the fact that, even with the wrapped branes,
the solution has constant dilaton.

Now we write the beta functions for the couplings \giparm\ as $\beta _i=
-8\pi ^2\dot{g}_i^{-2}$, with $\dot{}\equiv -d/d(\ln\mu)$.  Using \giparm,
we get
\eqn\betavel{\beta _i =-8\pi ^2 \vec e_i \cdot \dot{\vec x}
\quad\hbox{and thus}\quad -8\pi^2\dot{\vec{x}}=\v{N}}
where $\dot{}$ represents the flow towards IR, and we have used
\betv .
Think of $\vec x$ as the position of a particle living in the Coxeter
box; according to \giparm, this position gives the gauge couplings.
The particle is moving with a velocity $\dot{\vec x}$, which
corresponds to the beta functions as in \betavel.

In fact, we can make a more complete mechanical analogy. The walls of
the Coxeter box are in one-to-one correspondence with the simple roots
$\vec e_i$, since $\vec e_i$ is the normal vector to the wall where
$g_i^{-2}(\vec x)=0$.  Label the wall associated with $e_i$ by $i$.
Consider attaching the particle to the $i$-wall of the Coxeter box by
$N_i$ strings of equal tension.  For a fixed position of the particle
in the Coxeter box the lowest energy state for any such configuration
arises when the strings are perpendicular to the walls.  The $U(N_i)$
coupling constant $g_i^{-2}(\vec x)$ is identified with the length of
the corresponding string (i.e. the distance of the particle to the
$i$-th wall).  See Fig. 4B.  It is an easy exercise to see that if
$N_i=N_0d_i$ there is no net force on the particle due to the strings.
This is the case when the beta function vanishes and $\dot{\vec x}=0$.
If $N_i\not= N_0d_i$ there would be a net force.  Noting that the
direction of the $N_i$ string is in the $-\vec e_i$ direction (as that
is a vector perpendicular to the corresponding wall and in the correct
orientation) we see that the net force is given by (choosing the
tension to be $1/8\pi^2$)
$$\v{F}=-{1\over 8 \pi^2 }N_i\v{e}_i=-{1\over 8\pi^2} \v{N}$$
Now, using \betavel\ we see that
$${\dot{\vec x}}={\vec F}$$
and this captures the RG flow; this is not quite the Newton's
equation, as the force is giving the velocity.  The similarity of this
to brane constructions of gauge theory is striking.  In fact at least
in one case (which we discuss in section 12) it is identical to one.

\bigskip
\centerline{\epsfxsize=0.75\hsize\epsfbox{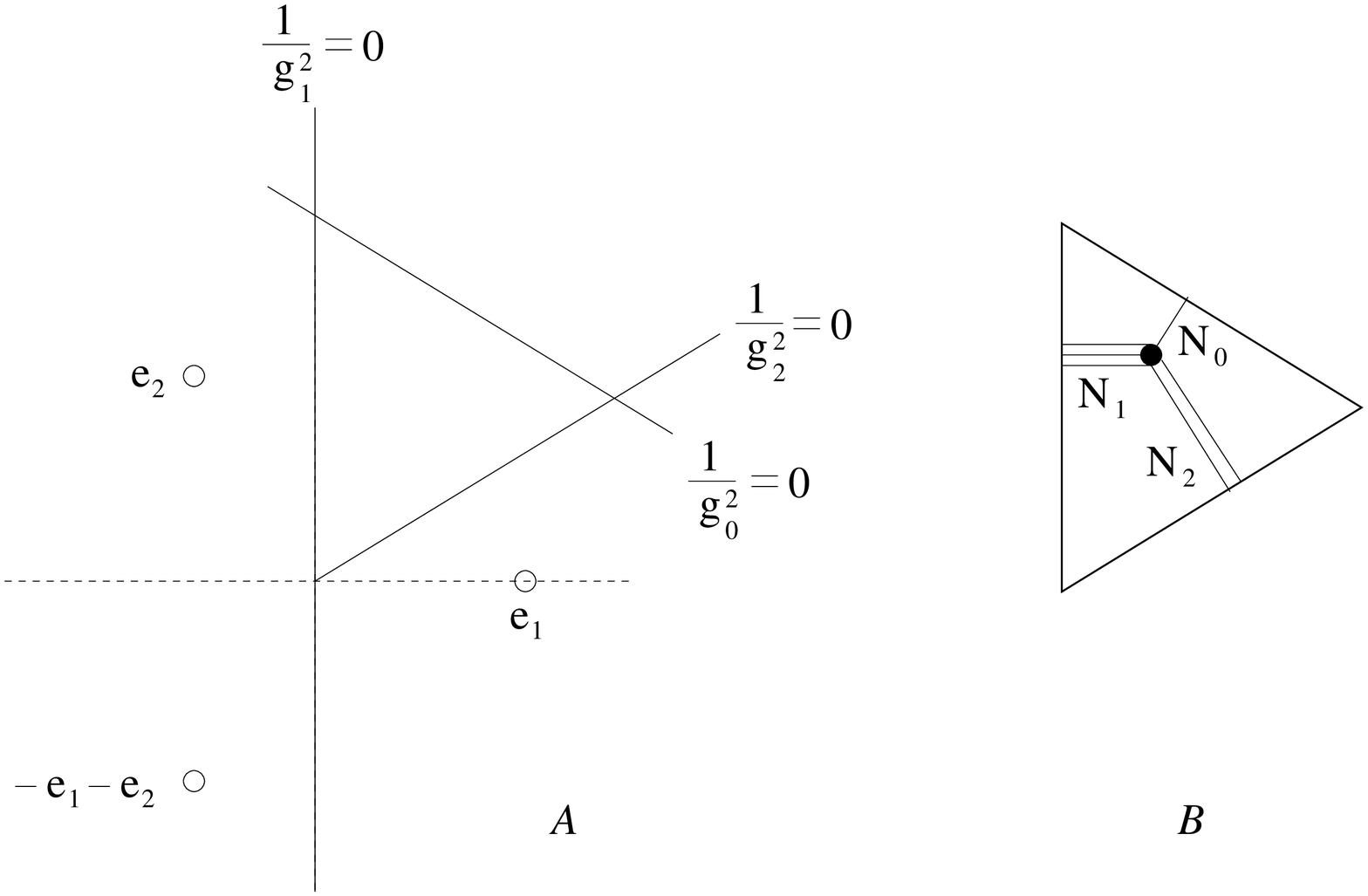}}
\noindent{\ninepoint\sl \baselineskip=8pt {\bf Figure 4}:{\sl {\it A})
Construction of the Coxeter Box for ${\hat A}_2$. $\v{e}_1$ and $\v{e}_2$
are the simple roots of $A_2$. {\it B}) Particle living in the Coxeter box
attached to the walls where ${1\over g^2_i}=0$ by $N_i$ strings.}}
\bigskip

If there are wrapped branes and the theory is not conformal, the
particle moves in the direction of ${\vec F}$ until eventually it hits
one of the walls of the Coxeter box, which is defined where one of the
$g_{i_0}^{-2}$ given by \betavel\ vanishes.  At this wall the gauge
group $U({\hat N}_{i_0})$ gets strong and should be dualized.  As
discussed earlier, this duality corresponds to a Weyl reflection.  It
is precisely a Weyl reflection about the corresponding wall of the
Coxeter box.

In other words, the particle bounces off the wall, with angle of
incidence equal to angle of reflection.  This can be seen from the
beta function transformation \betaitr, which gives
$$\vec e_i\cdot \dot{\vec x}~{}'=\vec e_i \cdot \dot{\vec x}-(\vec e_i
\cdot \vec e_{i_0})(\dot{\vec x}\cdot \vec e_{i_0}),$$
with $\dot{\vec x}~{}'$ the particle's velocity after bouncing off the wall
of the Coxeter box.  We thus obtain
\eqn\refl{\dot{\vec x}~{}'=
\dot{\vec x}-\vec e_{i_0}(\dot{\vec x}\cdot \vec e_{i_0}),}
which is precisely a Weyl reflection of the velocity $\dot{\vec x}$,
reversing the component normal to the wall and preserving the component
parallel. So the particle bounces off the walls with angle of incidence
equal to angle of reflection.

Of course, which wall the particle hits depends on both its velocity and
its initial position.  The initial position of the particle is given by
the scales $\Lambda _i$ of the $U({\hat N}_i)$ groups, as these are the
integration constants in integrating the beta function equations.  Suppose
that we integrate the 1-loop beta functions as
\eqn\giinteg{g_i^{-2}(\mu )={\beta _i\over
8\pi^2}\ln \left({\mu \over \Lambda _i}\right)} where the
$g_s^{-1}\delta _{i,0}$ term in \giparm\ can be thought of as scaled
into $\Lambda _0$.  In order to have all $g_i^{-2}\geq 0$, the scales
should satisfy the condition \scalineq\ which, as discussed in sect. 8,
is properly preserved upon dualizing.

If we flow to the IR, decreasing $\mu$, the first group $i_0$
to be dualized is that whose scale $\Lambda _{i_0}$ which $\mu$ hits
first.  After the duality transformation, the scales are modified as
in \niill.  The condition which led to \niill\ is precisely
equivalent to the condition that the particle's position $\vec x$ is
continuous across the bounce, even though the velocity is reflected.
The picture of the bouncing particle captures the details of
the RG flow in a simple fashion.

For $\widehat A_1$, the Coxeter box is the interval $0\leq x\leq
g_s^{-1}$ and the cascade of \stkl\ (which was the $k=1$ case, but the
present picture applies for all $k$), corresponds to the particle
bouncing back and forth in this interval.  For the $\widehat A_2$ case
the Coxeter box is the inside of an equilateral triangle (see Fig. 4A)
and for the $\widehat A_3$ case it's the inside of a tetrahedron.  For
generic velocity, the particle will bounce off of each of the $r+1$
walls in succession.

In all cases, the cascading reduces the number $N_0$ of units of $F_5$
flux.  As discussed before \vecNis, after bouncing off of the wall
where $g_{i_0}^{-2}=0$, the ${\hat N}_i$ change according to \Nduali;
note that ${\hat N}_{i_0}'={\hat N}_{i_0}-\sum _j C_{i_0j} {\hat
N}_j=N_0d_{i_0} -\sum _j C_{i_0j}N_j$.  The number of units of $N_0$
charge is also reduced in correspondence with the root lattice
translations $T$ in the affine Weyl group $\widehat W\cong
W\semidirect T$.  More precisely suppose we start in the UV with a
quiver theory with $U(N_i)$ as $i=1,...,r+1$.  As we go towards the IR
we undergo Weyl reflections to stay in the Coxeter box.  However
suppose we continued the line not worrying about staying in the Coxter
box (i.e. use the variables of the original gauge theory even passed
the negative coupling squared).  After a while we can bring the
particle back to the fundamental domain given by the Coxeter box by a
translation vector $\vec{R}$ in the root lattice. This would be of
course in the direction of $-\v{N}$ and so let us write it as
$\v{R}=-a\v{N}$ with $a >0$.  The change in the 3-brane charge is
given by
$$\Delta N_0 =\sum_i{N_i {\vec e_i}}\cdot {\vec R}=-a\v{N}
\cdot \v{N}$$
This follows, just as in the case \stkl\polsh\ from the fact that
$$\Delta N_0={ 8\pi^2}\int H_{NS}\wedge H_R$$
and using the fact that $H_R$ is characterized by the vector $\sum _i N_i
{\vec e_i}$ and $\int H_{NS}$ which is the vector $\Delta {1\over
g_i^2}$ is given by the vector ${\vec R}$.  Note that in this way it
is clear the $\Delta N_0 <0$ as we proceed towards IR.  Thus cascade
phenomenon is directly related to the translation part of the affine
Weyl group.  Also note that the above formula for the change of the
threebrane charge is also valid continuously as we continuously change
$a$ because $a=log(\mu_1/\mu_2)$ as we change the scale from
$\mu_1$ to $\mu_2<\mu_1$.

\newsec{Affine $\N=1$ theories: the conformal case}

Consider the theory of 3-branes in the presence of a point-like
singularity of the Calabi-Yau.  One expects to get a conformal
theory.  For example considering  $N_0$ 3-branes at the conifold singularity
\klebwi\ it was argued that one gets a conformal theory,
with a natural AdS dual description.  In general this is expected
from the geometry if the singular limit admits a smooth deformation with
no 2-cycles left. In this case there are no allowed instantons and so there
are no quantum corrections to the geometry.  Thus also in the limit
that the deformation disappears one expects to have no quantum
deformations for the singular classical geometry.  Typically for probes
in the presence of singularities
protected by quantum corrections one expects to obtain conformal theories.

Let us now consider a class of examples which yield
singular 3-folds, where singularity is at a point. Of course
all the geometries we considered admit deformations which have
no 2-cycles, and so the singular limit is expected to be
a good description also quantum mechanically.
Consider taking all
\eqn\wihom{W_i(\phi _i)={s_i\over k+1}
\Tr \phi _i ^{k+1}} homogeneous.
Then it is possible to see using our geometric description that
we have a local description of Calabi-Yau which has an isolated
singularity at the origin given by a quasi-homogeneous hypersurface
in ${\bf C}^4$.  Thus we expect in all such cases that the
3-brane probe to yield $\N=1$ conformal theories.
  Note that not all 3-fold point-like singularities with 3-brane
 probes yield $\N=1$ superconformal {\it gauge theories}.  For example
as discussed in \CKV\ the threebranes in the presence
of $x^2+y^2+z^2+w^{l}=0$ even though it is expected
to be conformal, has no conventional gauge theoretic description
for $l$ odd (for $l=2k$  it corresponds to affine $A_1$ theory
with superpotential $\Phi^{k+1}$).

We now provide evidence for the existence of the conformal fixed point
for this class of quiver theories from the gauge theory analysis.
 The exact beta function for the
$s_i$ is
\eqn\betasi{\beta (s_i)=-3+(k+1)\Delta (\phi _i)\equiv -3+(k+1)(1+\half
\gamma (\phi _i)).}
The superpotential respects
a classical $U(1)_R$ symmetry under which
\eqn\Risi{R(\phi)={2\over k+1}\qquad R(Q_{ij})={k\over k+1}}
and the gauginos have charge $+1$.
Generally this $U(1)_R$ symmetry is anomalous, which is the case
if any $U(N_i)$ instanton 't Hooft vertex has non-zero $U(1)_R$
charge.  The total $U(1)_R$ charge of the $U(N_i)$ instanton 't Hooft vertex
is
\eqn\Rlami{R(\Lambda _i^{\vec e_i \cdot \vec N})={2\over k+1}
(\vec e_i \cdot \vec N),} with $\vec N=\sum _{i=0}^r \vec e_iN_i$.
These are only all zero if the diagram is the affine $\widehat G$
diagram and $N_i=N_0d_i$, with $N_0$ arbitrary.  More generally,
though some of the \Rlami\ are non-zero, the 't Hooft vertices could
be invariant under a discrete subgroup of this $U(1)_R$, corresponding
to an anomaly free discrete R-symmetry.

The $\widehat G$ affine quiver theories with $N_i=N_0d_i$ thus satisfy
the necessary conditions for a ${\cal N}=1$ superconformal field
theory: there is an anomaly free $U(1)_R$ symmetry, which is needed
for the superconformal current multiplet.  Indeed, in the $\widehat G$
affine quiver theory with $N_i=N_0d_i$, giving $\phi$ and $Q_{ij}$
dimensions via the superconformal chiral primary relation $\Delta
=3R/2$ with the conserved anomaly free R-charges \Rlami,
\eqn\Dis{\Delta(\phi _i)={3\over k+1}, \quad \Delta (Q_{ij})=
{3k\over 2(k+1)},} implies that all exact beta functions vanish.  The
exact beta functions which vanish are \betae\ for the gauge couplings,
\betaij\ for the superpotential couplings $\lambda _{ij}$, and \betasi\
for the couplings $s_i$ in $W_i(\phi _i)={s_i\over k+1}\Tr \phi _i^{k+1}$.

The conditions \Dis\ are necessary and sufficient for having an $\N =1$
superconformal theory.  For the moment we will assume that, for general
$k$ in \wihom, the
equations \Dis\ have solutions for some values of the couplings,
$s_i^*$, $g_i^*$, and $\lambda _{ij}^*$, and discuss the resulting
$\N =1$ superconformal theories.

Note that, as in \LeighEP, the conditions \Dis\ for a superconformal
theory are fewer in number than the number of adjustable coupling
constants $g_i$, $s_i$, and $\lambda _{ij}$.  This is because the
$\beta (g_i)$ given by
\betae\ are proportional to the $\beta (\lambda _{ij})$ for the affine
quiver theories with $N_i=N_0d_i$, for any $N_0$.  Since the $\beta
(g_i)$ vanishing equations are redundant, for all $i=0\dots r$ running
over all nodes of the affine quiver diagram, there are $r+1$ fewer
conditions on the fixed point couplings than there are unknowns.
Thus, if there is any solution $s_i^*$, $g_i^*$, and $\lambda
_{ij}^*$, there will be a $r+1$ dimensional moduli space of such
solutions, with $r+1$ exactly marginal operators.  Including the theta
angles, this is a $r+1$ complex dimensional moduli space, with $r+1$
complex moduli.

Thus, for all $k$, the superconformal field theories are expected to
have a $r+1$ complex dimensional moduli space of couplings, similar to
the $r+1$ dimensional moduli space of couplings of the ${\cal N}=2$
affine quiver theories. The modulus corresponding to the diagonal
$U(N)$ is related to the IIB string coupling, with its $SL(2,Z)$
S-duality group. The others are expected to have an S-duality group
given by the $G$ Weyl group.

Assuming that these fixed points exist, we can determine their $a$
and $c$ central charges, as in \AnselmiYS, in terms of the 't Hooft
anomalies of the $U(1)_R$ symmetry in the supercurrent multiplet:
\eqn\acis{a-c={1\over 16}\Tr U(1)_R, \qquad 5a-3c={9\over 16}\Tr U(1)_R^3.}
The 't Hooft anomalies $\Tr U(1)_R$ and $\Tr U(1)_R^3$ get
contributions only {}from the massless chiral fermions.  Consider
first the $\N =2$ $\widehat G$ affine quiver theories with $N_i=N_0d_i$.
The appropriate $U(1)_R$ symmetry to use in \acis\ is that under
which $R(\phi)=R(Q)=2/3$, and \acis\ then gives
\eqn\aandcii{a_{\N =2}=c_{\N =2}=a_{free}={9N^2|\Gamma _G|\over 32}
\left(1+(-{1\over 3})^3+2(-{1\over 3})^3\right)
={N_0^2 |\Gamma _G|\over 4},} where the terms are from the gauginos,
and fermionic components of $\phi$ and $Q_{ij}$ respectively and we
used $\sum _i d_i^2=\half \sum _{ij}|s_{ij}|d_id_j=|\Gamma _G |$.
The central charge is independent of the moduli $\tau _i$, which is
why it had to agree with the free field values, as in \aandcii.

Now consider the ${\cal N}=1$ theories with deforming superpotential
as in \wihom, for general $k$. The appropriate $U(1)_R$ symmetry assignments
are as in \Risi\ and then \acis\ gives for the central charges
\eqn\aandci{\eqalign{a_{\N =1}(k)=c_{\N =1}(k)&={9N^2|\Gamma _G|\over 32}
\left(1+({2\over k+1}-1)^3+2({k\over k+1}-1)^3\right)\cr
&={27k^2N_0^2|\Gamma _G|\over 16(k+1)^3}.}}
This expression has a maximum value at $k=2$, where $a_{\N =1}(k=2)=
a_{\N =2}=a_{free}$, coinciding with \aandci.  For all other $k\neq 2$,
we have $a_{\N =1}(k)<a_{free}$.  According to the conjectured $a$
theorem, $a$ decreases along RG flows to the IR.  Since $a_{N=1}\leq a_{N=2}$,
the conjectured $a$ theorem is compatible with a RG flow from the
${\cal N}=2$
superconformal fixed points in the UV to our claimed ${\cal N}=1$ RG fixed
points in the IR.  The fact that $a_{\N =1}<a_{free}$ for $k\neq 2$
gives evidence that these theories indeed flow to an interacting RG
fixed point rather than a free field theory.

Note that $a_{\N =1}(k=2)=a_{free}$, so the existence of a $\N=1$
RG fixed point for $k=2$ could potentially violate a strong form
of the $a$ theorem, where $a$ must decrease along flows unless
the theories are related by an exactly marginal operator.  The $\N =1$
theory with $k=2$ looks like a deformation of the $\N =2$ theory by
the operators $\sim \Tr \phi _i^3$, which are not exactly marginal
(they're marginally irrelevant near the $\N=2$ line of fixed points).
It could be, though, that the $\N=1$ theories exist, but just can't be
obtained by starting from the $\N=2$ fixed points.   Or it could be
that the strong form of the $a$ theorem does not hold here (even in
2d, it's known that a non-compact target space can invalidate the
$c$-theorem \PolchinskiDY\ ).

The AdS/CFT correspondence supports the existence of the $\N =1$
superconformal theories for general $k$ in \wihom\ and general
$\widehat G$ quiver diagram with $N_i=N_0d_i$.  The AdS/CFT correspondence
for the $k=1$ theories was discussed in \GubserIA, and this can be
generalized for all $k$.  Write the metric near the singularity
of $X_6(k, \widehat G)$ as $ds_6^2=dr^2+r^2 ds_5^2$, with $ds_5^2$
the metric of the base $M_5(k,\widehat G)$, with $M_5$ a 5d Einstein
manifold.   The  AdS/CFT correspondence here would be
between IIB string theory on $M_{k, \widehat G}$ and our general
$(k, \widehat G)$ superconformal theories.

\lref\GubserVD{
S.~S.~Gubser,
Phys.\ Rev.\ D {\bf 59}, 025006 (1999)
[hep-th/9807164].
}
\lref\BergmanQI{
A.~Bergman and C.~P.~Herzog,
hep-th/0108020.
} We can also check the above field theory exact results \aandci\ for $a$
and $c$ with our associated geometry by using the AdS/CFT relation of
\GubserVD : regarding the 3-fold $X$ associated with the ${\cal N}=1$
affine quiver theories with $W_i\sim \Tr \phi _i^{k+1}$
as a cone over base $M_5$, the prediction is that
\eqn\aadsc{a_{{\cal N}=1}=c_{{\cal N}=1}={\pi ^3 N_0^2\over 4{\rm Vol}(M_5)}}
normalized so that the $\N =4$ case is $M_5=S^5$ of unit radius.

Though the Calabi-Yau metric is not known, the relevant volume
appearing in \aadsc\ can still be found.
The case $k=1$ was recently analyzed in \BergmanQI, and the discussion
there can be immediately generalized to our general $k$ case.
Consider a general Calabi-Yau $n$-fold $X$ with a conical singularity
which can be written as $F(z_0, \dots z_n)=0$, up to resolving terms,
with $F(\lambda ^{w_0}z_0, \dots , \lambda ^{w_n}z_n)=\lambda ^d
F(z_0, \dots z_n)$ homogeneous.  Then the $2n-1$ real dimensional
base $B$ of the cone has volume
\BergmanQI\
\eqn\volformi{{\rm Vol}(B)={2d\over (n-1)!\prod w_i}\left({\pi (\sum w_i -d)
\over n}\right)^n.}

We can apply this formula to our deformed geometry, where $n=3$,
with $z_0\rightarrow t$ and $z_i\rightarrow X_i$ .  So
$d=kC_2(G)$ and the weights are
\eqn\weightl{\matrix{\Gamma _G & w_0&w_1&w_2&w_3&d\cr
A_{r} &1&\half (r+1) k &\half (r+1) k &k&(r+1)k\cr
D_r &1&(r-1)k&(r-2)k&2k&2(r-1)k\cr
E_6&1&6k&4k&3k&12k\cr
E_7&1&9k&6k&4k&18k\cr
E_8&1&15k&10k&6k&30k\cr}}
with  $w_1=\half C_2(G)k$ and $d=kC_2(G)$.
Plugging these into \volformi, note that in all cases $\sum w_i -d=k+1$
and $\prod _iw_i=k^3C_2(G)|\Gamma _G|/4$, so we get
\eqn\volmv{{\rm Vol}(M_5)={4\pi ^3(k+1)^3\over 27k^2|\Gamma _G|}.}
Using \volmv\ in \aadsc\ gives a result which agrees perfectly with
the field theory result \aandci\ for all $G$ and $k$.

Though we have given some arguments supporting the existence of the
general $\N=1$ RG fixed points discussed above, we should also mention
the possibility that there might be no solution of the conditions \Risi\
needed for a superconformal theory when $k\geq 2$.  One can see a
possible problem for $k=2$ by considering the
perturbative expressions for the anomalous dimensions\foot{We thank
I. Klebanov and
J. Polchinski for pointing this out to us, and for related
discussions.}.  Considering a single $U(N_c)$ factor in the quiver
theory, which has $N_f=2N_c$, the exact beta functions are (up to
positive proportionality factors)
\eqn\betagen{\beta (s)\sim \gamma (\phi), \qquad -
\beta (g)\sim \beta (\lambda)
\sim \gamma (\phi)+2\gamma (Q).}
Thus a RG fixed point requires $\gamma (\Phi)=\gamma (Q)=0$.  For
$s=0$ there is an IR attractive fixed line of arbitrary $\lambda =g$,
which is the ${\cal N}=2$ superconformal fixed line.  Deforming by
small $s$ one finds $\gamma (Q)\sim \lambda ^2-g^2$ and $\gamma
(\phi)\sim \lambda ^2-g^2+cs^2$ with $c$ a positive constant,
suggesting that there is no solution of $\gamma (Q)=\gamma (\phi )=0$
for $s\neq 0$.  The signs of the beta functions are such that $s$ and
$\lambda$ are driven to smaller values, whereas $g$ is driven to
larger values.  Eventually the theory could perhaps hit a RG fixed
point outside the region of validity for the above approximations for
$\gamma (Q)$ and $\gamma (\Phi)$.  The $\N =1$ fixed points could also
exist outside the basin of attraction of the deformed $\N =2$
theories.  Also, we should anticipate the need to include the dynamics
of the coupled gauge group factors in order to see the possible $\N
=1$ RG fixed points -- turning off the dynamics of the neighboring
gauge groups can lead to a free theory.

One can also consider this issue by using
the duality of \refs{\AharonyNE, \ElitzurHC}\ between
${\cal N}=1$ supersymmetric $SU(N_c)$ gauge theory, with $N_f$
fundamentals and adjoint $\phi$ and superpotential
\eqn\Weleco{W_{elec}={s\over k+1}\Tr \phi ^{k+1}+\lambda \overline Q\phi Q,}
and the $SU(N_f-N_c)$ magnetic dual, with similar superpotential.  For
$N_f=2N_c$, rather than flowing to a non-trivial fixed point, both the
electric and magnetic theories could be IR free (which is possible
here since both have the same spectrum).  But the non-trivial duality
map suggests that, especially upon gauging the flavor symmetries,
there could in fact be a non-trivial fixed point for general $k$.

\newsec{Quiver Theories from Local CY 3-folds}

So far we have discussed type IIB probe theories involving
3-brane and 5-branes wrapping cycles in a local 3-fold which has
compact 2-cycles, but no compact 4-cycles.  Furthermore we have talked
about transitions in geometry which is interpreted as the large $N$ dual
of the corresponding gauge theory.  On the other hand CY 3-folds are
known also to have transitions where some 4-cycles
and 2-cycles shrink to zero size and where there is a transition
involving emergence of non-vanishing 3-cycles.  A notable
example along this line is the del Pezzo transitions,
where a del Pezzo surface (complex dimension two)
shrinks inside a threefold and some three cycles emerge
on the other side.  More generally, the 4-cycles can have several
components, and each of the components can shrink.  It is
natural
to ask what the corresponding quiver gauge theories are
in such cases where 4-cycles shrink with
wrapped D3,D5 and D7 branes, and
what is the large $N$ dual interpretation of the corresponding
geometric transition involving the emergence of 3-cycles.

There are two well known classes of this type.  One is obtained
by considering $C^3/\Gamma$ where $\Gamma$ is a subgroup of
$SU(3)$, and the other is when we consider the local threefold
to be realized torically.  These two sets have overlap when $\Gamma$
is abelian.  One could also consider more general examples which
are neither toric, nor orbifolds.  Our construction of quiver
theories applies to all these cases.  However  we will mostly
specialize to the case of
local threefolds which can be realized torically.  The advantage of
restricting to the toric case is that the mirror symmetry action
is simpler to study (though it can also be carried out in the other
cases).  These correspond to geometries that are realized in the linear
sigma model approach \witli\
by considering the Higgs
branch of a $G=U(1)^k$ gauge system with $k+3$  matter fields $x_i$
with charges $Q_i^j$, with $j=1,...,k$ satisfying
$$\sum_i Q_i^j=0$$
One also has (complexified) FI terms given by $t^j=r^j+2\pi
i\theta^j$.
The geometry is given by considering
$$M=\{\sum_i Q_i^j |x_i|^2=r^j\}//G$$
which is a local Calabi-Yau 3-fold.

Before we consider branes in such geometries we have to find a good
description of these backgrounds.  The description just given is
adequate when $r^j>>0$.  But our regime of interest is not in large
sizes for the 2- and 4-cycles, but rather when the sizes vanish.  In
this limit quantum corrections, due to worldsheet instantons will be
quite important.  To bypass having to compute these corrections, we
will consider the mirror type IIA description, which already has the
IIB quantum corrections summed up into a classical description.
Following \horv\ we introduce dual periodic variables $Y_i$ satisfying
$Re Y_i= |x_i|^2$.  Define $y_i=e^{-Y_i}$, so that $y_i$ are single
valued complex numbers.  Then the dual theory is given by a LG theory
with superpotential $\sum_i y_i$ with the constraints
$$\prod_i y_i^{Q_i^j}=e^{-t^j}$$
Note that using these relations we can get rid of all the $y_i$'s
except for three.  Let us call them $y_1,y_2,y_3$.
By the homogeneity condition we can write
$$\sum y_i=y_3 f(y_1/y_3,y_2/y_3, t^j)$$
As has been shown in \hiv\ this is equivalent to
(see also the earlier work \klmvw\ )
considering the 3-fold given by (redefining $y_1/y_3\rightarrow y_1$
and $y_2/y_3\rightarrow y_2$)
$$uv=f(y_1,y_2, t^j)$$
where $u,v \in \bf{C}$ and $y_1,y_2\in {\bf C}^*$.  In order
to study Lagrangian D6 branes (wrapping 3-cycles and filling the
spacetime), which is the mirror of D3,D5 and D7 branes,
 it is convenient to define
$$W=f(y_1,y_2, t^j)$$
The 3-fold geometry is thus given by
$$W=f(y_1,y_2, t^j)\qquad uv =W$$
The advantage of writing it in this way is that, as shown in \hiv ,
one can split the study of the Lagrangian branes as a fibered geometry
involving 2-cycles in the $y_1,y_2$ space, times the circle in the
$u,v$ plane.  The  space of compact 2-cycles over $W=f(y_1,y_2)$
with boundary over a point $W=W_0$ is the
subject of singularity theory \arn\ and is in 1-1 correspondence with the
critical
points of $W$,
i.e. with solutions to
$$\partial_{y_1}W=\partial_{y_2}W=0$$
Let us label the critical points by $p_\alpha$ with $\alpha=1,...,r$,
and define $W_{\alpha}=W(p_\alpha)$.
The geometry of these cycles with boundary over $W=W_0$
 can be viewed as a disk where the boundary of the disk
is identified with a circle over $W=W_0$ and vanishing at
the critical points, along a path over the $W$ plane connecting
$W_\alpha$ to $W_0$. Let us call the corresponding cycle $\Delta_\alpha$.
  To obtain 3-cycles we take the product of this disc with the circle
corresponding to circle action on the $uv$ plane.  However this does
not yield a closed 3-cycle.  To remedy this we note that if we take
$W=W_0$ then for a generic point on the path over the $W$-plane we have
an $S^1\times S^1$ fiber, which as we approach $W_\alpha$ one $S^1$
shrinks and as we approach $W=0$ the other $S^1$ shrinks.  This is
therefore an $S^3$.  Let us continue calling the corresponding
class of $S^3$ by $\Delta_\alpha$.
 We thus have $r$ distinct classes of three
cycles.
This construction was already made
in this context in \hiv .   See figure 5.

\bigskip
\centerline{\epsfxsize=0.70\hsize\epsfbox{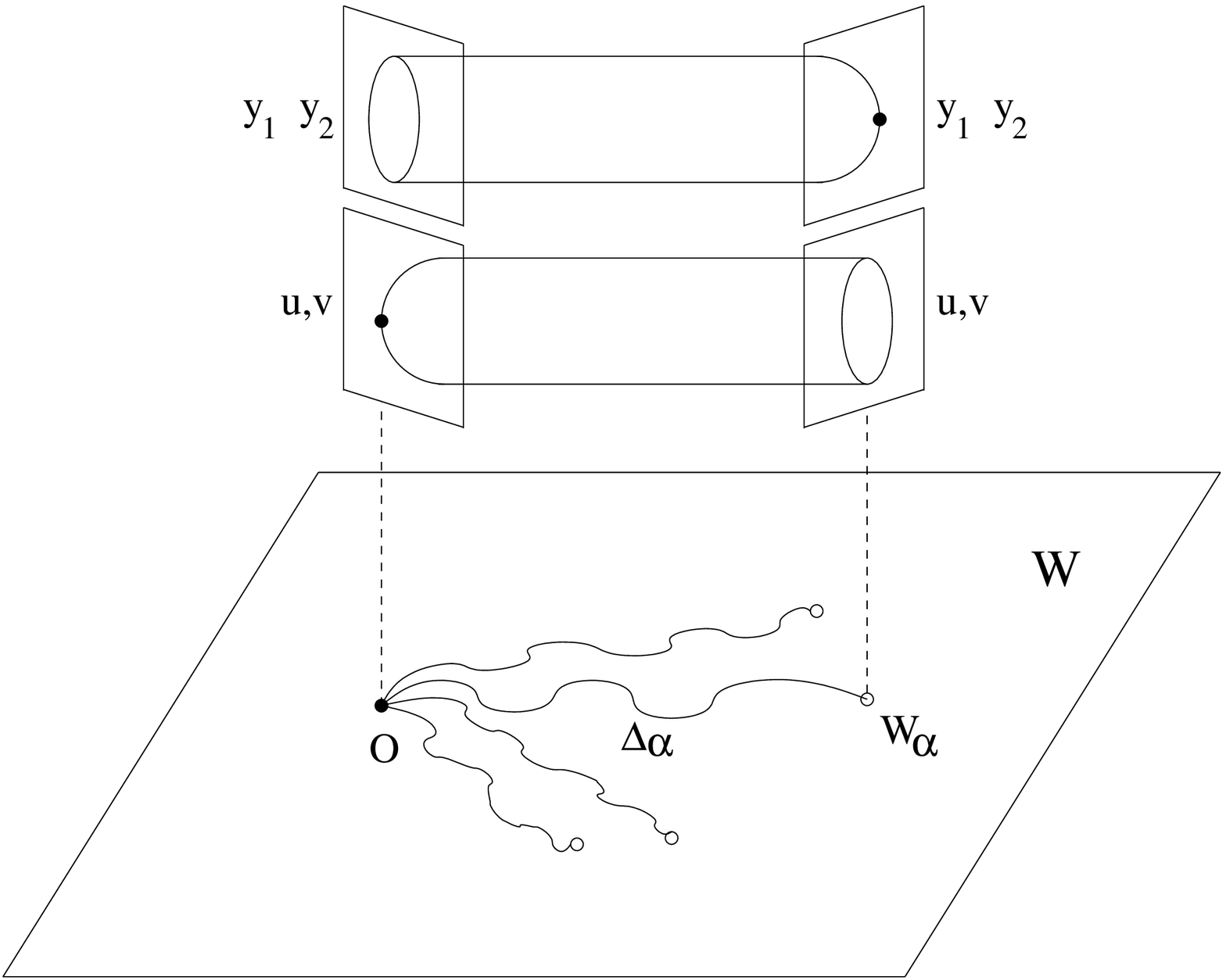}}
\noindent{\ninepoint\sl \baselineskip=8pt {\bf Figure 5}:{\sl
 Identification of 3-cycles with
$S^3$ topology for each path connecting one of the critical points of $W$
with $W=0$.}}
\bigskip

In order to map these cycles to the D3,D5 and D7 brane charges we need
to know how the $r$ classes $\Delta_\alpha$ map to the allowed charges
in the IIB description.  The basic idea of how to determine this was
shown in \hiv\ for the case of toric del Pezzos ($P^2$ blown up with
up to three points, and $P^1\times P^1$).  Namely, each
$\Delta_{\alpha}$, in the large volume limit, can be identified as the
mirror of a brane configuration on the compact 4-cycles in the CY
geometry.
 In fact, for the case of del Pezzos, these theories have been recently
studied
in \ih\ using the ideas in \hiv , and our discussion here for
the construction of the quiver theory is similar to it.
  The case studied in \hiv\ was mainly for the Fano case, but
the same ideas apply to compact cycles inside CY 3-folds.

Let us denote the corresponding brane with $V_\alpha$
which can also be viewed as a bundle (or more precisely a sheaf)
on the type IIB side supported on the compact cycles of the CY.  Using
ideas from mirror symmetry the Chern character for $V_\alpha$ leads
to the brane charges for $\Delta_{\alpha}$.
 The class of D3 brane
is generally simple to determine:  Note that, as in \SYZ\ this should
map to a $T^3$ in the mirror which consists
of the phases of $u,y_1,y_2$.  Note that a representative
of this class can be
chosen to have large $|y_1|,|y_2|$ and so its intersection with all
the $\Delta_\alpha$ which are at finite values of $y_i$ is zero.
Let
$$[T^3]=\sum_{\alpha =1}^r d_\alpha \Delta_\alpha$$
Thus $N_0$ D3 branes get mapped to $N_0 d_\alpha$ D6 branes wrapped
over $\Delta_\alpha$ as $\alpha=1,...,r$.
We will choose $d_\alpha\geq 0$, if necessary by reversing
the orientation of the corresponding $\Delta_{\alpha}$.
Thus the theory of $N_0$
 $D3$ branes gives rise to a gauge theory with gauge group
$$\prod_{\alpha} U(N_0 d_\alpha)$$
This is an ${\cal N}=1$ supersymmetric gauge theory.  There is also
chiral matter which is in one to one correspondence with the
intersection of the cycles.  In particular there are
$$n_{\alpha \beta}=\Delta_\alpha \circ \Delta_\beta$$
net chiral fields transforming in the bifundamental of
$U(N_0 d_\alpha)\times U(N_0 d_\beta)$ (where here we are including
the extra circle coming from $uv=const$ in the definition of
$\Delta_\alpha$).  There is another
interpretation of these intersection
numbers, in terms of the associated
$\N =2$ LG theory in 2d, defined by
the superpotential $W(y_1,y_2)$, namely
$$n_{\alpha \beta}=Tr_{\alpha \beta}(-1)^F F$$
where the right hand side is the index
defined in \cfiv\ for the associated LG
theory which counts the net number of BPS kinks
between the vacua $\alpha$ and $\beta$.
Note that the intersection
number is topological and the actual number of matter fields may
be more, but there is nothing preventing non-chiral pairs to pair up
by mass deformations and we assume to be in this generic situation.
  Note that $n_{\alpha \beta}=
-n_{\beta \alpha}$ and the sign of $n_{\alpha \beta}$ correlates with
the chirality of the matter.  In terms of the type IIB description
the classes $\Delta_{\alpha}$ get mapped to bundles
$V_\alpha$ (or sheaves).  These should
correspond to rigid bundles\foot{These sheaves are not bundles on the
Calabi-Yau, but rather are bundles on the local
4-cycles.  We will refer to
them as bundles for ease of discussion.} (as the topology of the mirror
is $S^3$ and has no $b_1$ which counts the deformation
parameter of the brane \SYZ ).  These rigid bundles are sometimes
referred to as spherical bundles.  More precisely, a spherical bundle (or
sheaf)
${\cal F}$ on a Calabi-Yau threefold is one that satisfies
${\rm Ext}^i({\cal F},{\cal F})=({\bf C},0,0,{\bf C})$ for $i=(0,1,2,3)$,
corresponding to the cohomology of $S^3$, where ${\rm Ext}^i$
is the analog of $H^i$ in the sheaf theoretic context.
An example is ${\cal O}_{{\bf P}^1}$ where the Calabi-Yau is
${\cal O}(-1)\oplus{\cal O}(-1)$.  Another class of examples is any line
bundle on a component of a shrinking 4-cycle in a Calabi-Yau, as we will
check below.

If $V_\alpha$ and $V_\beta$ are bundles on the same component of a shrinking
4-cycle $S$, then the intersection numbers $n_{\alpha \beta}$ are given in this
type IIB setup,
as will be discussed below, by
\eqn\intnum{n_{\alpha \beta}= \chi (V_\alpha , V_{\beta})-\chi (V_\beta, V_{\alpha})}
where $\chi(V_\alpha ,V_\beta)$ denotes the Euler class of the $\overline
\partial$ complex mapping sections of $V_\alpha \rightarrow V_\beta$,
where here we are ignoring the embedding of the 4-cycle inside the
Calabi-Yau.  This relation has also been noted in
\refs{\dfrbrane, \ih}.
This is the IIB dual of the intersection number of the cycles.
By Riemann-Roch, this simplifies to
$$n_{\alpha \beta}=c_1(S)\cdot(c_1(V_\beta)-c_1(V_\alpha)),$$
where $S$ denotes the union of complex surfaces inside the CY
which can shrink and on which the bundles $V_\alpha,V_\beta$ are supported.
Generically there is a choice of bundles, called an exceptional collection,
for which for each pair $\alpha, \beta$
 at most either $\chi ({V_\alpha, V_\beta})$ or
$\chi({V_\beta ,V_\alpha})$
 is non-zero, and moreover all the $H^i$ are zero except
for $H^0(V_{\alpha}^*\otimes V_{\beta})$, corresponding to holomorphic
sections of maps from $V_{\alpha}\rightarrow V_{\beta}$
This implies
that for each pair of branes $\alpha ,\beta$ we choose an
ordering so that $n_{\alpha \beta}\geq 0$.  Then the $n_{\alpha
\beta}$ matter fields $Q^i_{\alpha \beta}$, with
$i=1,...,n_{\alpha \beta}$ matter fields are in 1-1 correspondence
with $n_{\alpha \beta}$ holomorphic maps
$$f^i_{\alpha \beta}: \qquad  V_\alpha \rightarrow V_\beta.$$

Let's now specialize this discussion to our local case.  The actual
cohomologies whose index yields the intersection numbers
$n_{\alpha\beta}$ needs to be worked out.  In this calculation we need
to clearly distinguish between bundles on complex surfaces and sheaves
on the Calabi-Yau, so we introduce temporary notation in the next two
paragraphs for clarification.  If $V_\alpha$ is a line bundle on a
component of the shrinking 4-cycle, we now denote by
$\widetilde{V}_\alpha$ the corresponding sheaf on the local
Calabi-Yau.  It will be helpful first to recall the result of a
computation in
\seith, which says that
$${\rm Ext}^i(\widetilde{V}_\alpha,\widetilde{V}_\alpha)=
{\rm Ext}^i(V_\alpha,V_\alpha)\oplus{\rm Ext}^{3-i}(V_\alpha,V_\alpha)^*.$$
Since $V_\alpha$ is a line bundle, we get ${\rm Ext}^i(V_\alpha,V_\alpha)=
H^i({\cal O})$ which is ${\bf C}$ for $i=0$ and is 0 otherwise.  This shows
that $\widetilde{V}_\alpha$ is spherical, as claimed.

A straightforward extension of this calculation based on Lemma 3.16 of
\seith\ shows that
$${\rm Ext}^i(\widetilde{V}_\alpha,\widetilde{V}_\beta)=
H^i(V_\alpha^*\otimes V_\beta)\oplus H^{3-i}(V_\beta^*\otimes V_\alpha)^*.$$
This immediately implies \intnum\ if we use $n_{\alpha\beta}=\chi
(\widetilde{V}_{\alpha},\widetilde{V}_{\beta})= \sum_{i=0}^3 (-1)^i{\rm dim}{\rm
Ext}^i(\widetilde{V}_{\alpha},\widetilde{V}_{\beta})$.  But we can see more.  A collection of
bundles $\{V_\alpha\}$ defines an exceptional collection if for any pair
of distinct bundles $V_\alpha,V_\beta$, then only one of $V_\alpha^*\otimes
V_\beta$ or $V_\beta^*\otimes V_\alpha$ has cohomology, and that one has
cohomology in only one degree.

Note that this theory
is anomaly free, because the net number of chiral matter for the
gauge group $U(N_0 d_\alpha)$ is given by
$$\sum_\beta N_0 d_\beta n_{\alpha \beta}=
N_0 \sum_{\beta} d_\beta (\Delta_\alpha \circ \Delta_\beta)=
N_0  (\Delta_\alpha \circ [T^3])=0.$$
In addition the theory will have some superpotential involving
the chiral matter fields.\foot{These are in 1-1 correspondence
with the potential `1/4 BPS instantons' in an ${\cal N}=2$ 2d
LG theory with
superpotential $W(y_1,y_2)$ (for a study
of these see \quar\ ).  To see this note
that disk
instanton
configurations can be organized in terms of a cyclic ordering of vacua
at infinity as we go along a big circle, $\alpha_1,...,\alpha_k$
 and a choice of a 1/2 BPS kink between any such pair, which is
given by a chiral field  $Q^i_{\alpha_l \alpha_{l+1}}$,
where $i$ can be a number from $1$ to $n_{\alpha_l \alpha_{l+1}}$
(assuming it is positive).  Thus each such instanton configuration
is in one to one correspondence with an allowed superpotential term.}
Consider a term in the superpotential, which by gauge invariance is
of the form
$$W=a_{i_1...i_k}Tr Q^{i_1}_{\alpha_1 \alpha_2}...Q^{i_l}_{\alpha_l
\alpha_{l+1}}...
Q^{i_k}_{\alpha_k \alpha_1}$$
with
$i_l=1,...,n_{\alpha_l\alpha_{l+1}}$
with $k+1\equiv 1$.
It is natural to try to relate $a_{i_1...i_k}$ to properties
of the ring structure of the corresponding holomorphic
maps in the type IIB description.  Note that this is natural
because there is a natural product structure on the holomorphic maps
between bundles, namely the composition of maps.  Consider
$$f^{i_1}_{\alpha_1 \alpha_2}...f^{i_k}_{\alpha_k \alpha_1}$$
which will be a holomorphic map from $V_{\alpha_1}$ to itself
and by assumption the only non-trivial class here is $H^0$ which is
the identity map.  So this composition will be a multiple
of identity and so we have
$$f^{i_1}_{\alpha_1 \alpha_2}...f^{i_k}_{\alpha_k \alpha_1}=
a_{i_1...i_k}{\bf 1}_{\alpha_1}$$
which we propose as defining
 the couplings of the superpotential.  For
$k=3$ this is obvious as this can be viewed as the overlap of
the corresponding wavefunctions, and this is exact in the type IIB
side.  We believe this result also holds for arbitrary $k$, and
we have verified this in a number of examples.  In some of the examples
we consider we find that the $Ext$ group appears at degree 1.  In such
a case, the same formula applies where we use the $Ext^1$ to deform
the relevant bundles and compute the product of sections for the
deformed bundles. This will be discussed in the context of ${\bf P}^1\times
{\bf P}^1$ below.

In addition the above theories have FI terms.  As we change
the $t^j$, not only do we change the coupling
constants, as the volume of 3-cycles change,
 but we also generate FI terms, as is well known
\modgr.   Note that even though there are
$r$ gauge groups, the number of $t^j$'s is given by the number of
2-cycles in the original type IIB geometry, which is less than $r$ by
the number of 4-cycles $n_4$ plus zero cycle (which equals 1).  Taking
into account the possible variation of the string coupling constant
$\tau$ which is the coupling constant that the D3 brane sees, we see
that the coupling constants and the FI terms of the $r$ gauge groups
satisfy $n_4$ constraints.

So far we have only discussed the quiver theory
for pure D3 branes.  We can also have more general
brane configurations involving D5 and D7 branes, which
leads to a more general quiver theory where everything is as
for the D3 branes, except that the rank of the gauge groups
are now arbitrary (instead of being a multiple of $d_\alpha$).
This is exactly as in the A-D-E quiver theories we studied before.
However there is one difference between this case and the
${\cal N}=1$ A-D-E quiver theories studied earlier:  Whereas the
previous theories were non-chiral and anomaly free, the present theories
are chiral and in some cases are not anomaly free.
This can be easily understood from the type IIB string theory setup as well.
Consider for example a D3 branes, which corresponds to a point
in the transverse 6-dimensional space.  If the 6-cycle were compact
then the theory would have problem supporting the D3 brane as the flux
would have nowhere to go. Namely we consider the 6-cycle and we delete
where the 3-brane is located then the topological fact that the
$S^5$ surrounding the 3-brane is topologically trivial in the
compact 6-cycle
implies that it cannot support any 5-form flux.  However the same argument
applies also to 5-branes and 7-branes, which wrap 2 and 4 dimensional
cycles in the transverse space.  Namely if there is a compact dual
cycle for any of them then we can use that cycle to undue any
$S^3$ or $S^1$ surrounding the corresponding brane.  Thus the only
allowed total brane charges consistent with flux
conservation
are in 1-1 correspondence with cycles in $H_*$ which
intersect no compact cycle.  In particular all the 4-cycles
are ruled out, because they intersect the 2-cycles inside them.  Of
the 2-cycles we can choose classes which do not intersect the 4-cycles
(i.e. are not electric-magnetic duals).
In particular there are $n_2-n_4$ such choices where $n_2$ denotes the
number of inequivalent 2-cycles and $n_4$ the number of 4-cycles.

One has to be careful
that here we are discussing the total charge of the branes.  There
is no problem with having D3,D5 and D7 branes wrap
arbitrary cycles as long as the total class does not intersect the
compact cycles.

We can also understand this result from the condition
of having an anomaly free quiver theory. Suppose we wrap $N_\alpha$
branes over the $\Delta_{\alpha}$ cycle.  Then the condition for having
no anomalies is that
$$\sum_{\beta} N_\beta n_{\alpha \beta}=0$$
for each $\alpha$, which means that the vector $N$ is in the null
subspace of the matrix $n$.  Recall that the matrix $n_{\alpha \beta}$
is, by a change of basis, an anti-symmetrized intersection form
of the 0-,2- and 4-cycles.  Thus indeed this is exactly the condition that the
cycle represented by $N$ has no intersection with any other compact cycle,
as was anticipated.  Thus in general there are
$$r'=1+n_2-n_4=r-2n_4$$
inequivalent integers controlling the rank of
the $r$ gauge groups where $r=1+n_2+n_4$.

Note that the ideas presented above for construction of the quiver
theory could also have been carried out for the case of the general
4-cycle
shrinking inside the CY 3-folds, and not just the toric case.
This can be done directly in the type IIB setup, without appealing
to mirror symmetry.
Namely
we would have to choose a collection of exceptional spherical bundles,
as many as necessary to get all the allowed brane charges, namely $r$.
Of course there would be $2n_4$ constraint for the multiplicities
of these bundles.
Then the quiver diagram can be constructed by computing the ${\rm Ext}$
group between pairs of such bundles, and Yukawa
couplings can be computed as discussed above, by the composition
of the holomorphic maps between bundles.  However the reason
we have discussed the type IIA mirror in the toric
case is that, just as in \ov , the relevant
worldsheet
quantum corrections
become classical and leads to a simple understanding of Seiberg-like
dualities.

\subsec{Two Examples: $\bf P^2$  and ${\bf P}^1\times {\bf P}^1$}

Let us consider two simple examples to illustrate these ideas:
${\bf P}^2$ and ${\bf P}^1\times {\bf P}^1$ inside a CY 3-fold.
These cases were studied in detail in \hiv , and here we use the
discussion above to write down the corresponding quiver theory.

Consider first the $\bf P^2$ case.  Then the threefold is given by
$$uv=W \qquad W=y_1+y_2+{e^{-t}\over y_1y_2}+1$$
This arises from a linear sigma model with a single $U(1)$ gauge
group with 4 fields with charges $(-3,1,1,1)$.
Here $t$ denotes the complexified Kahler class of
${\bf P}^2$. The three critical points of  $W(y_1,y_2)$
are given by $y_i=\omega e^{-t/3}$ where $\omega $ is a third
root of unity, which yields the critical values in the $W$ plane
on three points on a tiny circle of radius $|e^{-t/3}|$ near $W=1$.
See Figure 6.  The three $\Delta$'s are depicted in Figure 6
 and connect each of these three critical values to $W=0$.
As discussed in \hiv\ these are in 1-1 correspondence in the type IIB
setup with  branes wrapping ${\bf P}^2$ and representing
bundles ${\cal O}(-1),{\cal O}(0),{\cal O}(1)$.  If we label these $\alpha$'s by
$1,2,3$ respectively, we have
$$n_{12}=n_{23}=3 \qquad n_{13}=6$$
%

\bigskip
\centerline{\epsfxsize=0.70\hsize\epsfbox{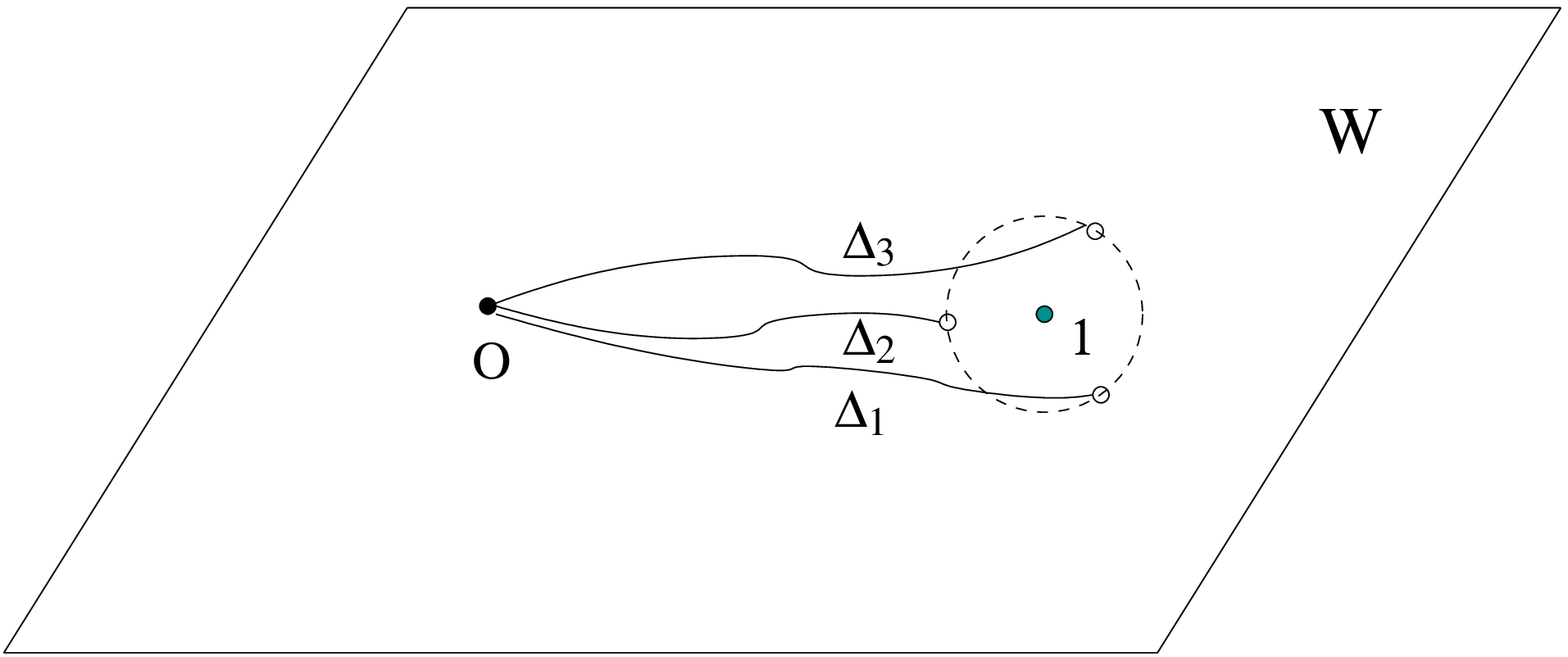}}
\noindent{\ninepoint\sl \baselineskip=8pt {\bf Figure 6}:{\sl Projection of
3-cycles on the $W$ plane for the mirror of a three-fold with a ${\bf P}^2$.}}
\bigskip

Note that these are in one to one correspondence
with the holomorphic maps from bundles, namely
$$A_i:  (12) \qquad {\cal O}(-1)\rightarrow {\cal O}(0) \leftrightarrow
H^0({\cal O}(1))=3$$
$$B_i:  (23) \qquad {\cal O}(0)\rightarrow {\cal O}(1) \leftrightarrow
H^0({\cal O}(1))=3$$
$$C_{ij}: (13) \qquad {\cal O}(-1)\rightarrow {\cal O}(1)\leftrightarrow
H^0({\cal O}(2))=6$$
where $i,j$ take values $1,2,3$ and $C_{ij}$ is a symmetric tensor.
There are no other maps or cohomologies between any pairs of these bundles.
Note that the sections $A_i$ and $B_j$ can be identified with degree
one homogeneous function of $(z_1,z_2,z_3)$ so we can choose  a basis
where
$$A_i\leftrightarrow z_i \qquad B_j \leftrightarrow z_j$$
$$C_{ij}\leftrightarrow z_iz_j$$
The class corresponding to pure D3 brane charge is given,
as follows from
\hiv , by
$$[{\cal O}(-1)]-2[{\cal O}(0)]+[{\cal O}(1)]=[\Delta_1]-2[2\Delta_2]+[\Delta_3]$$
which means that to get positive $d_\alpha$ we need to reorient
$$\Delta_2\rightarrow
\Delta'_2= -\Delta_2,$$
 in which case we obtain $d_1=1,d_2=2,d_3=1$, and moreover
$$n_{21}=n_{32}=3 \qquad n_{13}=6$$
Note that the signs of $n_{12}, n_{23}$ have changed due
to a change in sign in the orientation of $\Delta_2$.
We can now write the corresponding superpotential using
the identification with sections of the bundle and the corresponding
multiplication, which gives
$$W=\sum_{i,j=1}^3A_iB_j C_{ij}$$
Note that in this case $r'=r-2=1$ so there is only one inequivalent
choice for rank, which simply corresponds to putting $N_0$ D3 branes,
giving the gauge group $U(N_0)\times U(N_0)\times U(2 N_0)$.  This
reflects the statement that for $P^2$ the $P^1$ cycle in $P^2$ intersects
$P^2$ and no $D5$ or $D7$ branes are allowed.  It can also be readily
checked that the matrix $n_{\alpha \beta}$ has only one null eigenvector.

\bigskip
\centerline{\epsfxsize=0.40\hsize\epsfbox{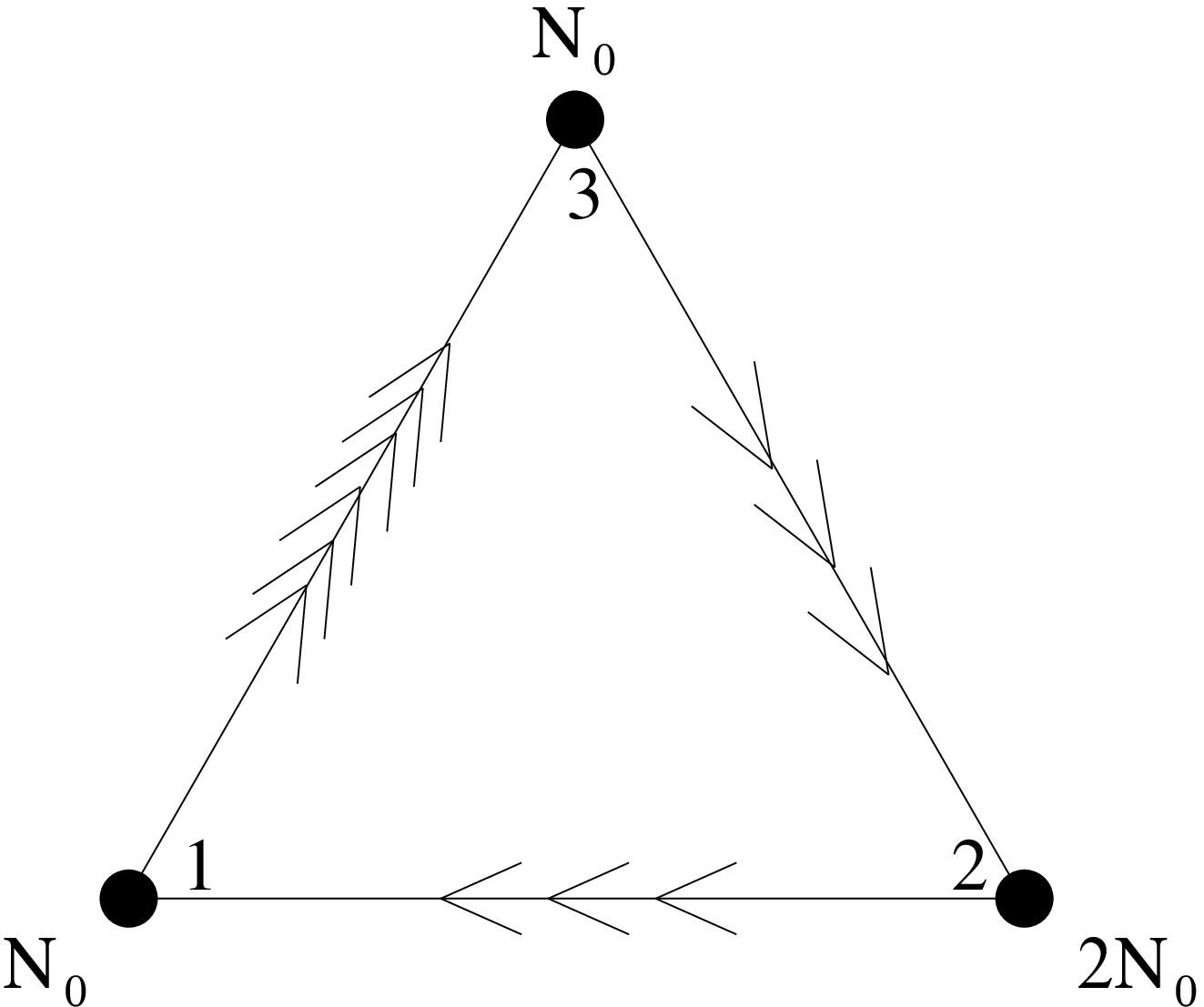}}
\noindent{\ninepoint\sl \baselineskip=8pt {\bf Figure 7}:{\sl Quiver
diagram corresponding to the field theory realized on the world volume of
$D3$ branes.}}
\bigskip

For the ${\bf P}^1\times {\bf P}^1$ we get the mirror description
$$uv=W \qquad W=y_1+{e^{-t_1}\over y_1} +y_2+{e^{-t_2}\over y_2}+1$$
where $t_1,t_2$ denote the complexified volumes of the two ${\bf P}^1$'s.
The linear sigma model corresponding to this has two $U(1)$ charges
and 5 fields with charges $(1,1,0,0,-2)$ and $(0,0,1,1,-2)$.
The ${\bf P}^1$'s are realized by the fields 1,2 and fields 3,4.
Note that the difference of these two $U(1)$'s shows
that in this geometry there is a ${\bf P}^1$ with area
$t_1-t_2$ in the class being
the difference of the two ${\bf P}^1$'s whose normal bundle
is ${\cal O}(-1)+{\cal O}(-1)\rightarrow {\bf P}^1$ since this yields the charges
$(1,1,-1,-1,0)$.  We will use this fact later.

\bigskip
\centerline{\epsfxsize=0.70\hsize\epsfbox{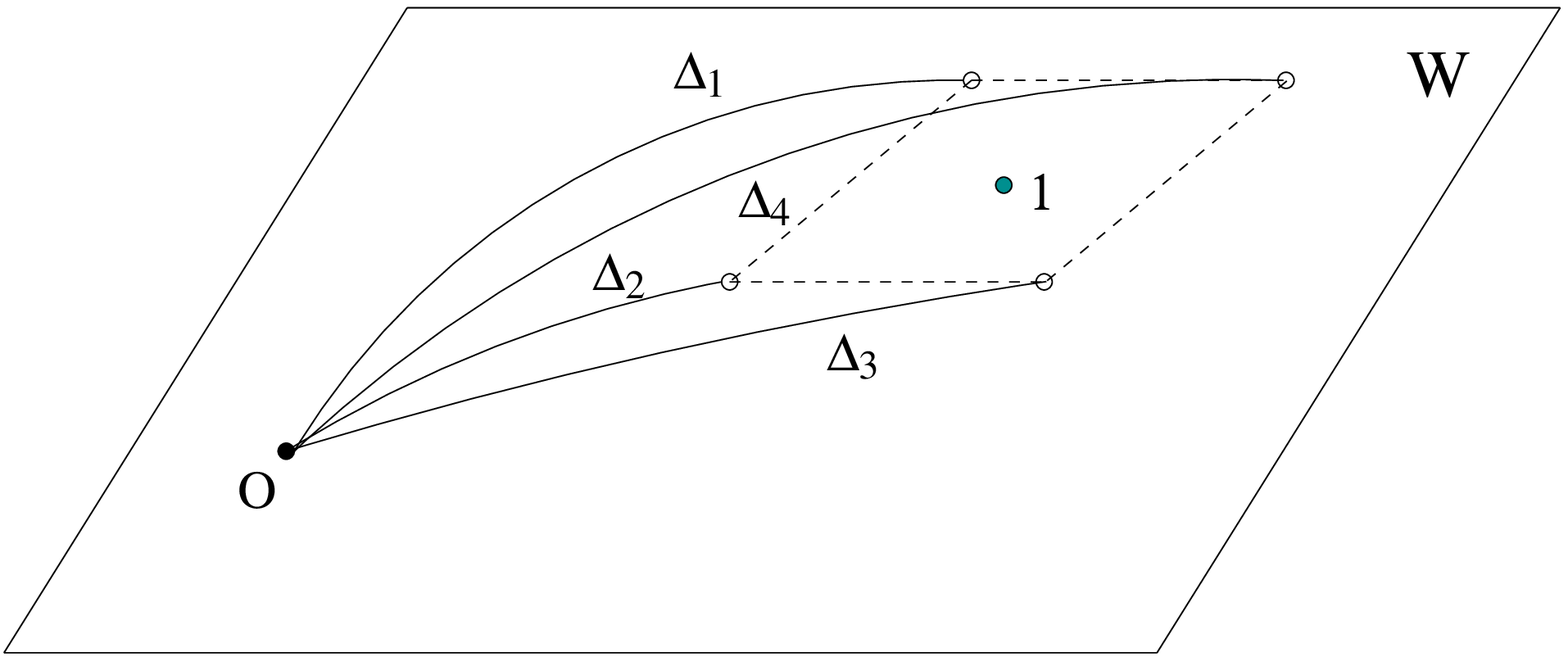}}
\noindent{\ninepoint\sl \baselineskip=8pt {\bf Figure 8}:{\sl  Projection of
3-cycles on the $W$ plane for the mirror of a three-fold with a ${\bf
P}^1\times {\bf
P}^1$.}}
\bigskip

There are four critical points of $W$ and thus there are four $\Delta$'s
as
shown in Fig. 8.  As follows from \hiv\ the $\Delta$'s are in
one to one correspondence with ${\cal O}(-1)_1\otimes {\cal O}(-1)_2 ,{\cal
O}(0)_1\otimes
{\cal O}(-1)_2,{\cal O}(0)_1\otimes {\cal O}(0)_2 ,{\cal O}(-1)_1\otimes
{\cal O}(0)_2$, which gives
$$n_{12}=n_{23}=2 \qquad n_{14}=n_{43}=2 \qquad n_{13}=4$$
moreover they are in 1-1 correspondence with monomials
$$n_{12}, n_{43}: z_1^i \qquad n_{23},n_{14}: z_2^i \qquad n_{13}:
z_1^i z_2^j$$
where $i,j$ are $1,2$ and the subscript of $z$ denotes the choice
of the ${\bf P}^1$, and there are no other maps or cohomologies.
The class corresponding to pure $D3$ brane is given by
$$[\Delta_1]-[\Delta_2]+[\Delta_3]-[\Delta_4]$$
which means that if we wish to get positive $d_{\alpha}$
we switch the orientation of $[\Delta_2]\rightarrow -[\Delta_2],[\Delta_4]
\rightarrow -[\Delta_4]$. Note that this yields
$$n_{21}=n_{32}=2 \qquad n_{41}=n_{34}=2 \qquad n_{13}=4$$
The superpotential can be easily deduced from the multiplication
of sections and we obtain
$$W=\sum_{i,j=1}^2 A_iB_jD_{ij}+A'_iB'_i D_{ji}$$
where $D_{ij}$ corresponds to the 4 diagonal chiral fields
(no symmetry condition on the indices imposed) and $A_i, B_j$
are matter fields on the edges $32,21$ and $A_i',B_j'$ are the
matter fields on the edges $34,41$.

\bigskip
\centerline{\epsfxsize=0.45\hsize\epsfbox{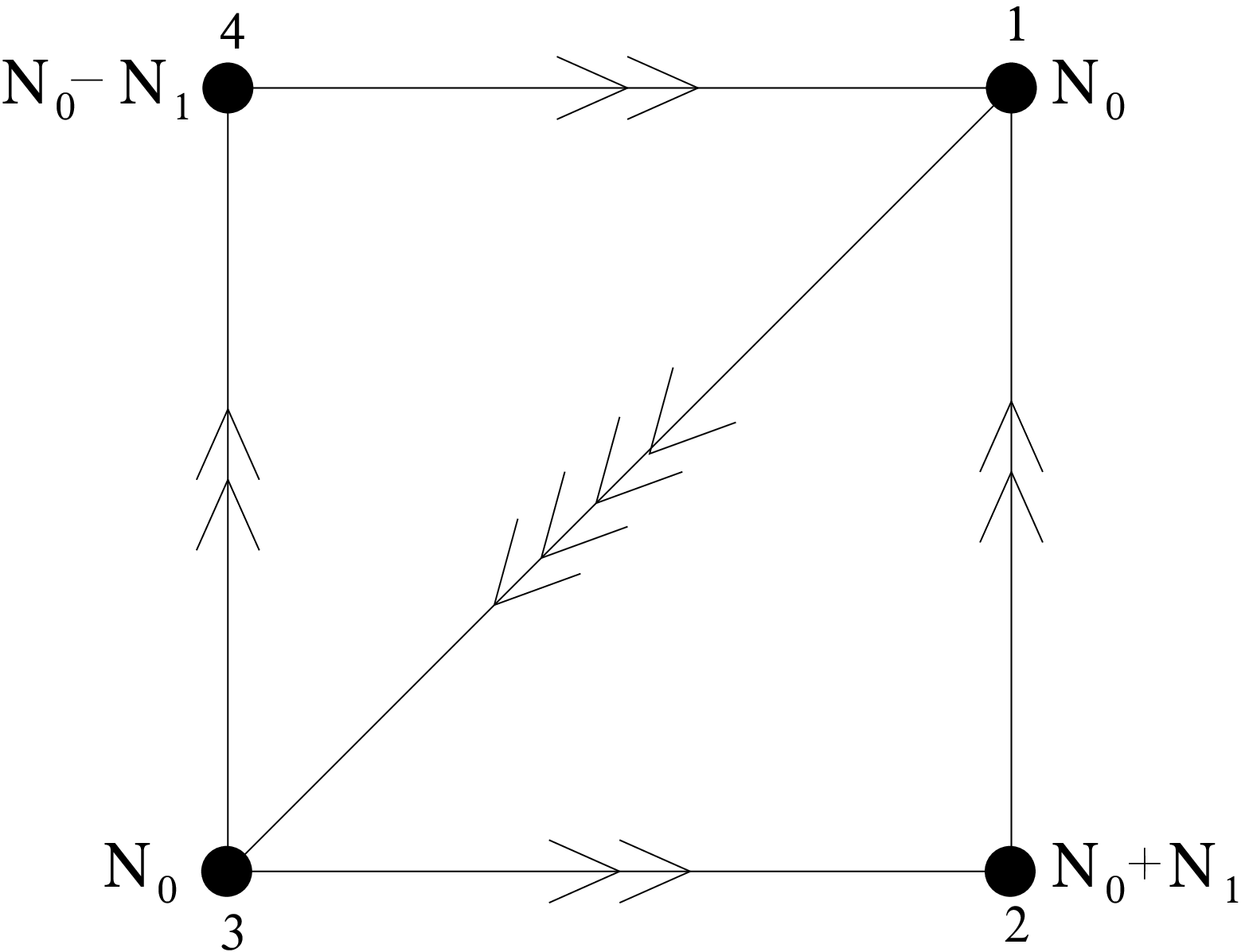}}
\noindent{\ninepoint\sl \baselineskip=8pt {\bf Figure 9}:{\sl  Quiver
diagram for the field theory corresponding to ${\bf P}^1\times {\bf P}^1$.}}
\bigskip

In this case we expect to have the possibility of adding one
more brane charge because $r'=r-2=4-2=2$.  This corresponds
to the fivebrane wrapping the 2-cycle class given by the
difference of the two ${\bf P}^1$, as that will not have any
intersection with the 4-cycle given by ${\bf P}^1\times {\bf P}^1$.
This class is given by the class represented by $[\Delta_4]-[\Delta_2]$.
Suppose we put $N_1$ branes in that class,
and $N_0$ 3-brane.  In this case the gauge theory
will be $U(N_0)\times U(N_0+N_1)\times U(N_0)\times U(N_0-N_1)$
(ordered according to node number).  One can check that this is the only
anomaly free choice of ranks allowed in this example, as expected.

\subsec{Geometric Transitions and Seiberg-like Dualities
for Chiral ${\cal N}=1$ Quiver Theories}

In the type IIB setup the dualities arise by flops in
Kahler cones, as first studied in \beret .  However in cases
we are considering the type IIA mirror is more convenient as
the quantum worldsheet corrections are absent, as noted above.
In this context the dualities in field theories arise by considering
varying the parameters of the bulk and using conservation of brane
charge to deduce a dual description as in \ov , which we follow here:
We consider a quiver theory arising from the toric theories discussed
above which in the mirror type IIA description is given by $N_\alpha$
D6 branes wrapping $\Delta_\alpha$ and filling the spacetime. This
leads to the gauge group $\prod_\alpha U(N_\alpha)$ with
$\Delta_{\alpha} \circ \Delta_{\beta}= n_{\alpha \beta}$ chiral fields
transforming in $(N_\alpha, {\overline N}_{\beta})$ or
$({\overline N}_{\alpha},N_\beta)$ depending on the sign of $n_{\alpha
\beta}$. The theory would also have some superpotential and some FI terms.
The coupling constant $g^2$ of the $U(N_\alpha)$ group is inversely
proportional
to the volume of the $\Delta_\alpha$ cycle.  The dualities typically arise
when the coupling goes to infinity, which occurs when the cycle
$\Delta_\alpha$
goes to zero size, and then it becomes negative, which means that another
cycle has emerged and we wish to find the new gauge description
which would be a dual quiver theory.  Now there are many
inequivalent topological configurations the theory may emerge in:
There are in fact $r-2$ possibilities depending on which wedge, in the
$W$ plane, the new $\Delta'_\alpha$ cycle emerges, as shown in Fig. 10.

\bigskip
\centerline{\epsfxsize=0.95\hsize\epsfbox{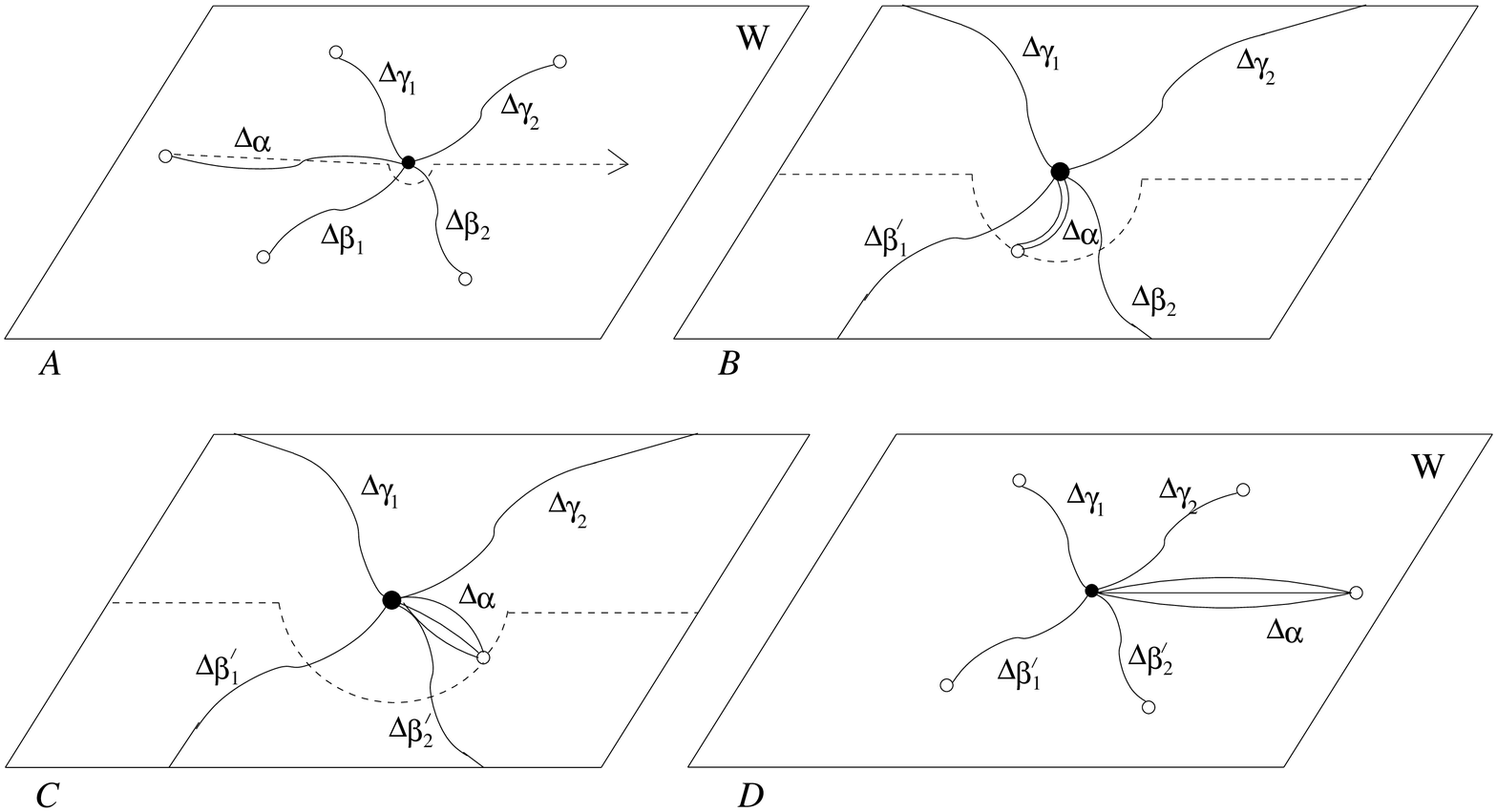}}
\noindent{\ninepoint\sl \baselineskip=8pt {\bf Figure 10}:{\sl (A)
Trajectory of the critical
point of $W$ corresponding to cycle $\Delta_\alpha$. The point actually
crosses the origin, but a small detour is taken in order to compute the
changes in the charges.  (B) Close-up of the first crossing. The class of
$\Delta_{\beta_1}$ changes and brane creation takes place. (C) Close-up of
the second crossing, $\Delta_{\beta_2}$ changes and brane creation takes
place again. (D) Final configuration.}}
\bigskip

The brane charge conservation gives the rank of the dual theory.
In fact, as shown in \hiv\ in this case we create new branes.
Strictly speaking the case studied in \hiv\ did not have the extra
circle coming from $uv=W$, in the D-brane worldvolume, but this does
not affect the arguments of \hiv .

  The easiest
way to find the brane charge is to avoid passing through the singular
point, by going around it.  As will be evident from
the result, the resulting
brane charge
conservation yields the same result independent of which way we go around
the $W=0$ point.  The transition line of the quiver divides the $W$
plane and thus the set of
gauge groups to two parts.  Let us denote the ones on one side by
$\beta_i$ and the other by $\gamma_j$.  As we move the $\Delta_\alpha$
passing through the $\Delta_\beta$'s, it creates new $\Delta_\alpha$
branes given by
$$-N_{\beta_i} n_{\beta_i \alpha}$$
Also after passing through the singularity it is natural
to reorient the brane (for preserving the same
supercharge this is necessary).
Thus the total number of $\Delta_\alpha$ branes are given by
$$N'_\alpha= \sum_i N_{\beta_i} n_{\beta_i\alpha}-N_\alpha$$
Note that the rank of the other gauge groups do not change.
However, the quiver diagram changes because the intersection
of cycles has changed.  This can be found using Picard Lefschetz
theory and one finds the following:  The intersections
of the cycles with $\alpha$ has not changed except for the overall
sign having to do with the reorientation of the $\alpha$ cycle.
This means that the arrows ending or
beginning on $\alpha$ will have the same degeneracy but opposite
orientation.  The intersection
between the cycles among $\Delta_{\beta_i}$ or $\Delta_{\gamma_j}$
do not change,  but the intersections between the pairs from
these groups changes according to
$$(\Delta_{\beta_i}\circ \Delta_{\gamma_j})'=n'_{\beta_i\gamma_j}=
n_{\beta_i\gamma_j}\pm n_{\beta_i\alpha}n_{\alpha \gamma_j}.$$
This is the same as how the soliton numbers change in the corresponding
2d LG theory \cecov, and the sign choice is correlated with the
orientation of the cycles \cecov.   For our present case, we can write
a simple expression which gives this sign choice, and which applies
for any cycles $\sigma$ and $\rho$, whether they're
$\beta_i$'s or $\gamma_j$'s:
\eqn\chdel{n'_{\sigma\rho} = n_{\sigma\rho }+{1 \over
2}(n_{\sigma\alpha}|n_{\alpha\rho}|+|n_{\sigma\alpha}|
n_{\alpha\rho}).}
This expression is properly antisymmetric and captures all of the
changes of the cycle intersections described above.
In the original type IIB description where these cycles are mapped
to a collection of exceptional bundles we get a new collection
of exceptional bundles under the above operation.  This is
well known mathematically and is known as the mutation of exceptional
bundle (for a review of this phenomenon
in the physics literature see \refs{\zas,\hiv} ).

Let us consider a special case of the above transmutation
which turns out to have a simple gauge theoretic interpretation.
Let all the arrows from $\beta_i$ to $\alpha$ be incoming and
all the ones from $\alpha$ to $\gamma_j$ be outgoing.
In this case the
change in the quiver diagram, corresponding to a different
number of holomorphic maps between the mutated bundles as well
as the change in the ring structure
can be described as follows:
Let $A^l_{\beta_i \alpha}$ denote the sections from $V_{\beta_i}\rightarrow
V_{\alpha}$ with $l=1,...,n_{\beta_i\alpha}$.  Let $B^k_{\alpha \gamma_j}$
denote the sections from $V_{\alpha}\rightarrow V_{\gamma_j}$
with $k=1,..., n_{\alpha \gamma_j}$ and
$C^p_{\gamma_j \beta_i}$ denote the sections of $V_{\gamma_j}\rightarrow
V_{\beta_i}$ with $p=1,..., n_{\gamma_j \beta_i}$.  Consider the ring
for the sections which can be captured by the superpotential
$$W=a_{lkp}A^l_{\beta_i \alpha}B^k_{\alpha \gamma_j}C^p_{\gamma_j \beta_i}+
... $$
where $...$ all the other elements given by the ring structure.
In the new quiver theory the $A$'s and $B$'s have disappeared
and in the above superpotential they only appear in $A^lB^k$ combinations
and will be replaced by a single new object $M^{lk}$ (the ``meson field'')
in the $W$.  Moreover instead of $A$'s and $B$'s we have
 new sections $A'^l_{\alpha \beta_i}$, $B'^k_{
\gamma_j \alpha}$ with the same degenerancy
but dual gauge quantum numbers and which appear in the
above superpotential as $W\rightarrow W+ B'^kA'^l M^{lk}$.  This is of course
nothing but the Seiberg duality on the factor represented by the node
$\alpha$.  Note that $W$ now has a quadratic piece given by
$$a_{lkp}M^{lk}_{\beta_i \gamma_j}C^p_{\gamma_j \beta_i}$$
which can be integrated out to lead to a net
$n_{\beta_i \alpha}n_{\alpha \gamma_j}-n_{\gamma_j \beta_i}$
chiral fields in the bifundamental representation of $ (\beta_i \gamma_j)$,
in agreement with \chdel .
The description of $W$ and the objects given above encode
the holomorphic maps between transmuted bundles as well as the ring
structure between them.
 Moreover in the field theory
setup what we have done is Seiberg duality,
as was proposed in the context of $\N=1$ chiral quiver
theories in \hanse.   Here
we see that Seiberg duality is simply a special case of what happens
when a cycle shrinks and another evolves which is realized as mutation
of bundles on the type IIB side, and the brane creation and the Picard
Lefschetz monodromy in the type IIA side.  For a special class of theories
(branes in the del Pezzo background) and
when $N_f=2 N_c$ for the node to be dualized, the connection between
variation of Kahler moduli
and Seiberg duality was noted in \plb\hanse. Also for the same special cases
the connection to Picard-Lefschetz monodromy was pointed out in \hanse .
\foot{An important ingredient here, compared to those studied in \hanse\
is the brane creation effect discussed in \hiv . Taking this effect
into account eliminates the claimed discrepancies between Seiberg
duality and Picard-Lefschetz monodromy.}

Here we have found a more general possibility of quiver duality corresponding
to the brane movements where not all the branes corresponding
to incoming and outgoing
fields are on the same side on the $W$ plane.  This is not necessarily
a field theory duality as the dynamics may or may not relate these
theories to undergo mutation in an arbitrary way. It would be interesting
to see which of these classes of quiver dualities are realized
in field theory.  Of course, this could in principle also be
settled by going to the large $N$ dual holographic description
as we have seen in the context of A-D-E quiver theories in this paper.
It would be
interesting to see if one can also understand the case of quiver
mutation
when not all the $\Delta_\alpha$'s and $\Delta _\beta$'s
correspond to chiral and anti-chiral matter respectively,
from the field theory
viewpoint.  One idea along this line is that effectively one
has given mass to some of the flavors and then one does a Seiberg-like
duality on less flavors and then after duality one takes the mass to zero.
This idea is worth further study.

\subsec{Examples of Seiberg Dualities as Mutations: ${\bf P^2}, {\bf
P}^1\times {\bf P}^1$}

Let us see how this works in the context of the examples we discussed before.
Consider the ${\bf P}^2$ case.  Suppose we change $t$ so that $t\rightarrow 0$
and then becomes negative.  Note that the orbifold limit where the
theory is equivalent to ${\bf C}^3/Z_{3}$ corresponds to $t\rightarrow -\infty$.

\bigskip
\centerline{\epsfxsize=0.40\hsize\epsfbox{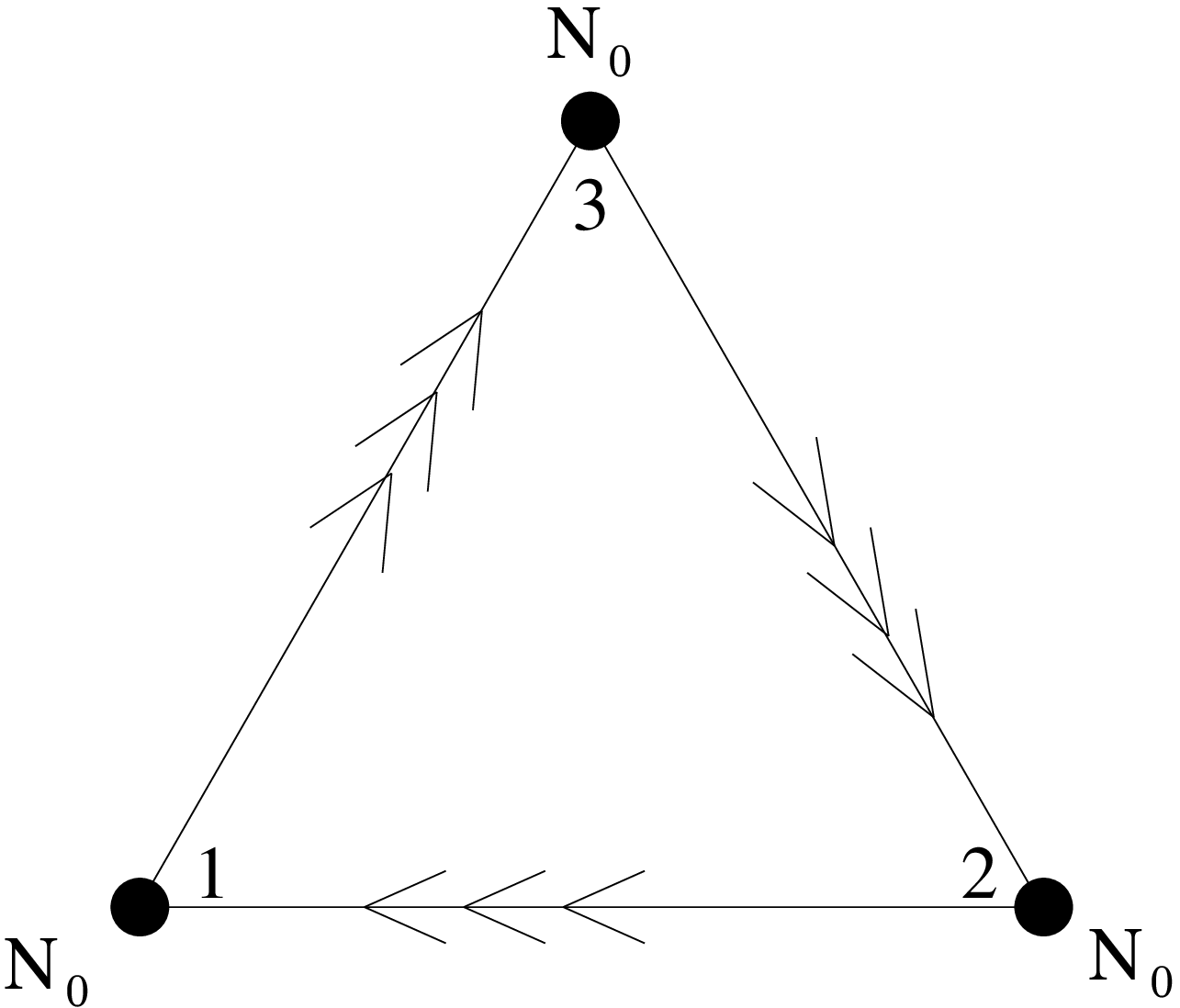}}
\noindent{\ninepoint\sl \baselineskip=8pt {\bf Figure 11a}:{\sl Quiver
diagram after a mutation in the ${\bf P}_2$ example.}}
\bigskip

Since $\Delta_2$ has shrunk and emerged
on the other side where it would have been with negative coupling,
then we will end up dualizing the node $2$, see Fig. 11b.
According to our discussion above we obtain a mutated bundle in the type IIB
configuration with quiver diagram given in Fig. 11a.  Following
Seiberg duality, which is equivalent to the quiver theory of the mutation,
and replacing $A_i B_j\rightarrow M_{ij}$
and introducing the dual matter fields $A_i',B_j'$ we obtain
$$W=\sum_{i,j=1}^3A_iB_j C_{ij}\rightarrow W=M_{ij}C_{ij}+M_{ij}B_j'A_i'$$
Noting that $C_{ij}$ is symmetric and $M_{ij}$ has no particular
symmetry property, we see that the symmetric part of $M_{ij}$ pairs
up with $C_{ij}$ and becomes massive, and we are left with the antisymmetric
part of $M_{ij}$.  Let us define $C_k'=\epsilon_{ijk}M_{ij}$.  Then
the superpotential is
\eqn\psuper{W=\epsilon_{ijk}C_k'B_j' A_i'}
which is the superpotential expected at the orbifold point, obtained
using the methods of \dm\ in \ks.

\bigskip
\centerline{\epsfxsize=0.80\hsize\epsfbox{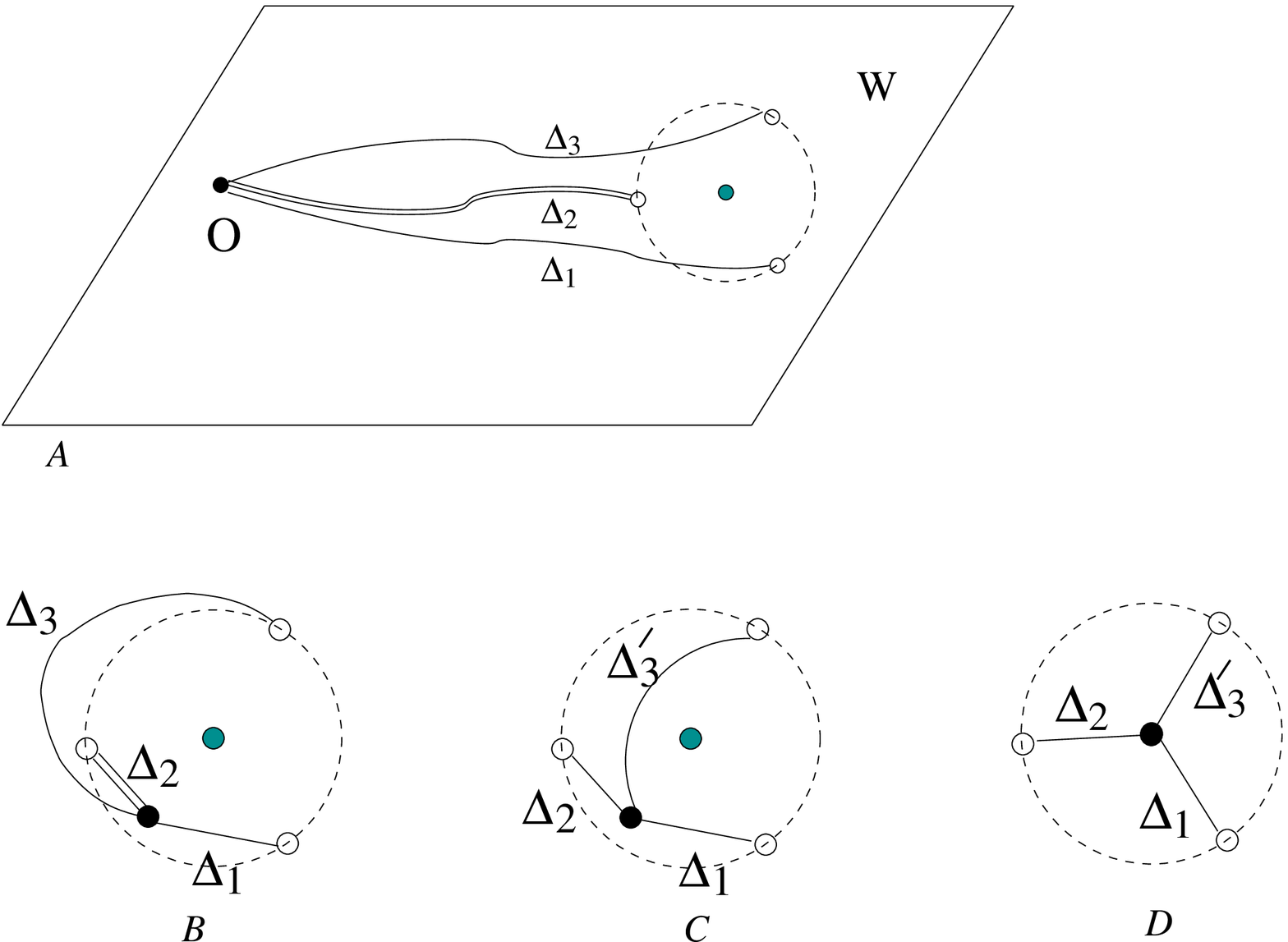}}
\noindent{\ninepoint\sl \baselineskip=8pt {\bf Figure 11b}:{\sl
 Mutation in
the ${\bf P}^2$ example. Notice that the projection of the $\Delta_2$ cycle
actually becomes zero, but a small detour has been taken in order to compute
the new charges.}}
\bigskip

Let's now obtain the quiver of Figure 11a and the superpotential of
\psuper\ by geometry.  Mutating ${\cal O}(1)$ to the left of ${\cal O}$,
we get the rank 2 kernel of the multiplication map
$${\cal O}\otimes {\rm Hom}({\cal O},{\cal O}(1))\to {\cal O}(1),$$
which is identified with the bundle $\Omega^1(1)$, whose cohomologies
are well known.  The resulting
bundles are ${\cal O}(-1),\ \Omega^1(1),\ {\cal O}$.

The intersection numbers are now $n_{ij}=\pm 3$.  The maps are
$$ \eqalign{ &{\rm Hom}({\cal O}(-1),{\cal O})=H^0({\cal O}(1))={\bf C}^3  \cr
             &{\rm Hom}({\cal O}(-1),\ \Omega^1(1))=H^0(\Omega^1(2))={\bf C}^3 \cr
             &{\rm Hom}(\Omega^1(1),{\cal O})=H^0(T(-1))={\bf C}^3.}
$$
Note that $\Omega^1(2)$ is the kernel of the multiplication map
$${\cal O}(1)\otimes H^0({\cal O}(1))\to {\cal O}(2),$$
so its sections are identified with the set of triples
$\{\ell_1,\ell_2,\ell_3\}$ of linear forms such that $x_1\ell_1+x_2\ell_2+x_3
\ell_3=0$, where $x_i$ are the homogeneous coordinates on ${\bf P}^2$.
The combination $x_1\ell_1+x_2\ell_2+x_3
\ell_3$ corresponds to the symmetric part of $M_{ij}=A_iB_j$, so the kernel
is naturally identified with the antisymmetric part as expected.  In other
words, we have basis of sections identified with the triples
$$\{(0,x_3,-x_2),\ (-x_3,0,x_1),\ (x_2,-x_1,0)\}$$
which are obtained antisymmetrically.  The identification of
$(\ell_1,\ell_2,\ell_3)$ with a section of $\Omega^1(2)$ is by
$$(\ell_1,\ell_2,\ell_3)\mapsto \ell_1dx_1+\ell_2dx_2+\ell_3dx_3.$$
The identification of the $A_i'$ with sections of $T(-1)$ is via
$$(A_1',A_2',A_3')\mapsto A_1'{\partial\over \partial_{x_1}}+
A_2'{\partial\over \partial_{x_2}}+A_3'{\partial\over \partial_{x_3}}.$$
So the composition of the maps ${\cal O}(-1)\to\Omega^1(1)\to{\cal O}$ is given
by
$$(A_1'{\partial\over \partial_{x_1}}+
A_2'{\partial\over \partial_{x_2}}+A_3'{\partial\over \partial_{x_3}})\cdot
(\ell_1dx_1+\ell_2dx_2+\ell_3dx_3)=A_1'\ell_1+A_2'\ell_2+A_3'\ell_3$$
under the identification of maps ${\cal O}(-1)\to {\cal O}$ with sections
of ${\cal O}(1)$.  Direct calculation shows that $A_i'C_k'=
\epsilon_{ijk}x_j$.  Identifying $B_j'$ with the dual of $x_j$ completes
the calculation of the superpotential.
Note that the result we have found for what the brane charges are
and the bundles they correspond to in
the orbifold limit is consistent with the result \diagom\ where
it was shown that
the wrapped branes for the orbifold point just discussed can be
identified with the exceptional collection ${\cal O}(-1)$, $\Omega (1)$,
${\cal O}(0)$,
where $\Omega $ is the cotangent bundle of ${\bf P}^2$.

To this exceptional
collection we can apply mutations, generating an infinite
number of exceptional collections. In field theory, this corresponds to
successively performing Seiberg duality at different nodes of the quiver
gauge theory.
By this procedure, we generate an infinite number of gauge theories
that are Seiberg dual to each other. To interpret this infinite number
of exceptional collections geometrically, we have
to consider the Teichm\"uller space of this noncompact Calabi-Yau, rather
than just its moduli space \dfr. The stringy moduli space of ${\bf C}^3/Z_3$
is given by a sphere with three punctures, the large volume
limit, a conifold point and the orbifold point. Since the conifold
point has a monodromy of infinite order, the Teichm\"uller space
would be an infinite cover of the moduli space. In particular, there
are infinite copies of the orbifold point. At each copy of the orbifold
point we have a different set of wrapped branes, which are just the
different exceptional collections, related by mutations.

It is possible to show that the number of chiral fields
of the quiver theory, $x_i$, satisfy a Diophantine equation
\cecov\ (which was studied there in the context of
solitons of ${\bf P^2}$ sigma model):
$$x_1^2+x_2^2+x_3^2=x_1x_2x_3$$
Moreover the rank of the corresponding gauge groups $N_i$ on
each of the three nodes of the quiver are given by
$$N_i=N {x_i\over 3}$$
where $N_i$ denotes the node facing the side with $x_i$
flavors.
Furthermore, we would like to argue that the phenomenon of duality
cascade is present also in this case. This is easiest to see if
we start with the quiver theory of figure 11a, with gauge group $U(N)^3$.
If we apply Seiberg duality to one of the nodes, we obtain the quiver
theory of figure 7, with gauge group $U(N)\times U(N)\times U(2N)$. For
each node of the original theory, we obtain a different copy of this theory.
We can now perform Seiberg duality at each node of the new theories, and
iterate the process. The structure that appears is presented in figure 12

\bigskip
\centerline{\epsfxsize=0.65\hsize\epsfbox{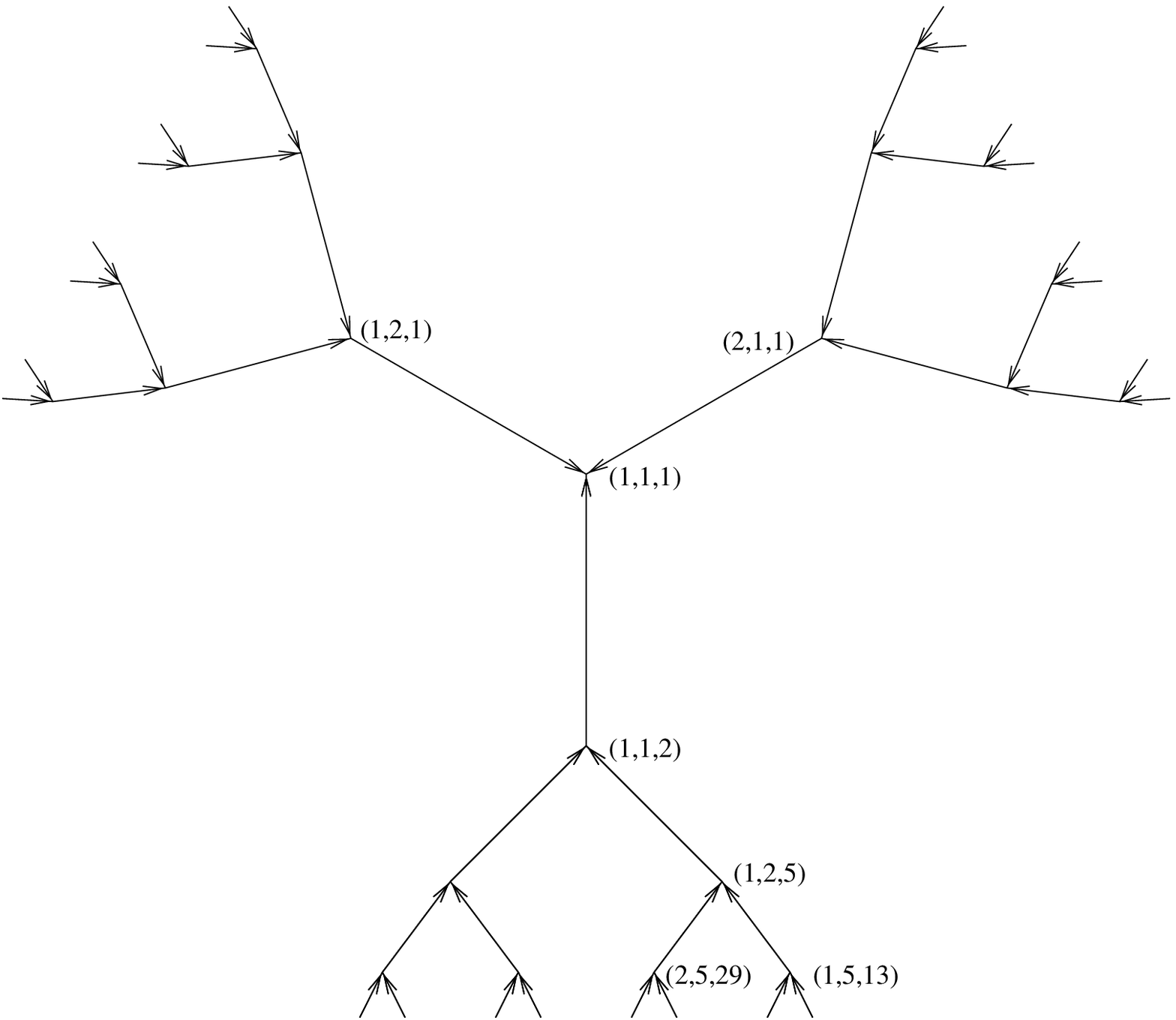}}
\noindent{\ninepoint\sl \baselineskip=8pt {\bf Figure 12}:{\sl
Structure
of the cascade for the ${\bf P}^2$ case. Each node in the figure represents
a gauge theory. The ranks $N_i$ for some of them are displayed
(which can be rescaled by an overall factor of $N$). The arrows
give the direction of the flow.}}
\bigskip

To prove that there is a duality cascade, we proceed as follows. Let's
denote by $N_{1,2,3}$ the ranks of the three groups. First, it is easy
to see that the number of flavors of the $U(N_1)$ gauge theory is given
by $N_2x_3=N x_2x_3/3$.  The condition that $U(N_1)$ be asymptotically
free is that $3N_1>N x_2x_3/3$ which is equivalent, using $N_1=Nx_1/3$ to
$x_1>x_2x_3/3$.  It is not too difficult to show that for the largest
rank, say $N_1$, this condition is satisfied.
{}From this it follows that the group with largest
rank is AF, whereas the other two are not. This implies that given any of
these theories, as we flow to
the IR, the coupling of the group with largest $N_i$ goes to infinity, and
a Seiberg duality for that group is called for. After the duality, the
situation repeats, the new group with largest $N_i'$ is now the only AF theory,
 and
repeatedly applying the same argument we see that as we flow to the IR, we
encounter a cascade of dualities where the ranks of the gauge groups strictly
decrease, until we reach the IR endpoint of the flow, the $U(N)^3$ theory.
However, we should note that this duality cascade is different from the
other ones we have talked about, in the sense that the number of 3-branes is
the same throughout.  Moreover, as we will discuss in the next section
there is no geometric transition corresponding to blowing up 3-cycles
in this case.

\bigskip

For another example consider ${\bf P}^1\times {\bf P}^1$ discussed before.
Let us consider the limit $t_1,t_2\rightarrow -\infty$ with $t_1-t_2=t$
being finite.   This can also be viewed as the orbifold of ${\cal
O}(-1)\oplus {\cal O}(-1)
\rightarrow {\bf P}^1$ by the $Z_2$ acting on the fiber and fixing the ${\bf
P}^1$.  Then the $W$ plane geometry is given in Fig. 13 below,
corresponding to $\Delta_2$ passing through zero.  Doing Seiberg
duality/mutation we obtain the quiver diagram given in Fig. 14.
To obtain the superpotential we introduce the meson field $M_{ij}=A_iB_j$
and the dual matter fields ${\widetilde A}_i, {\widetilde B}_j$ with the superpotential
$$W=\sum_{ij}M_{ij}D_{ij}+A'_iB'_jD_{ji}+M_{ij}{\widetilde B}_j{\widetilde A}_i.$$

\bigskip
\centerline{\epsfxsize=0.75\hsize\epsfbox{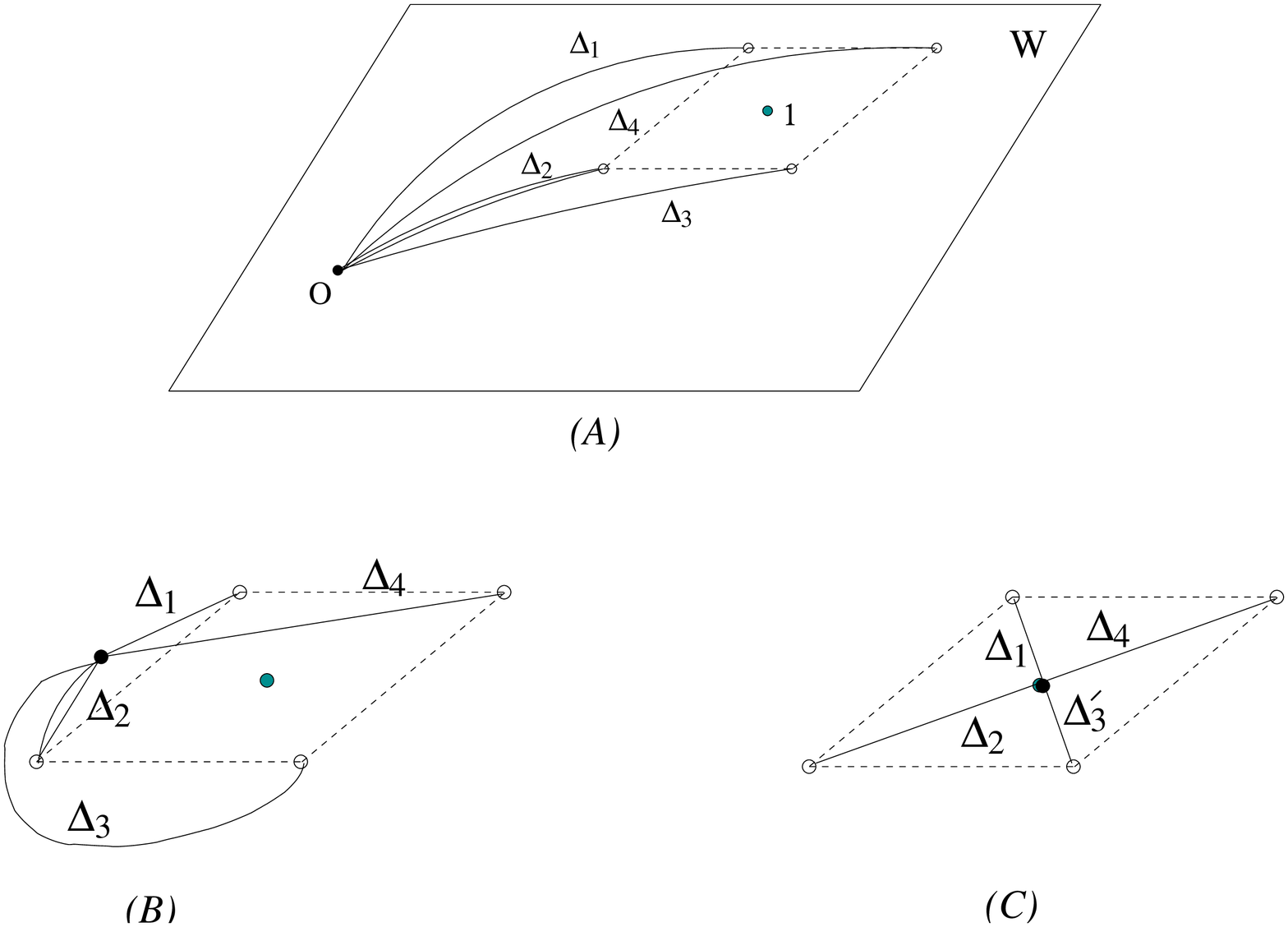}}
\noindent{\ninepoint\sl \baselineskip=8pt {\bf Figure 13}:{\sl Mutation in
the ${\bf P}^1\times {\bf P}^1$ example.}}
\bigskip

\bigskip
\centerline{\epsfxsize=0.45\hsize\epsfbox{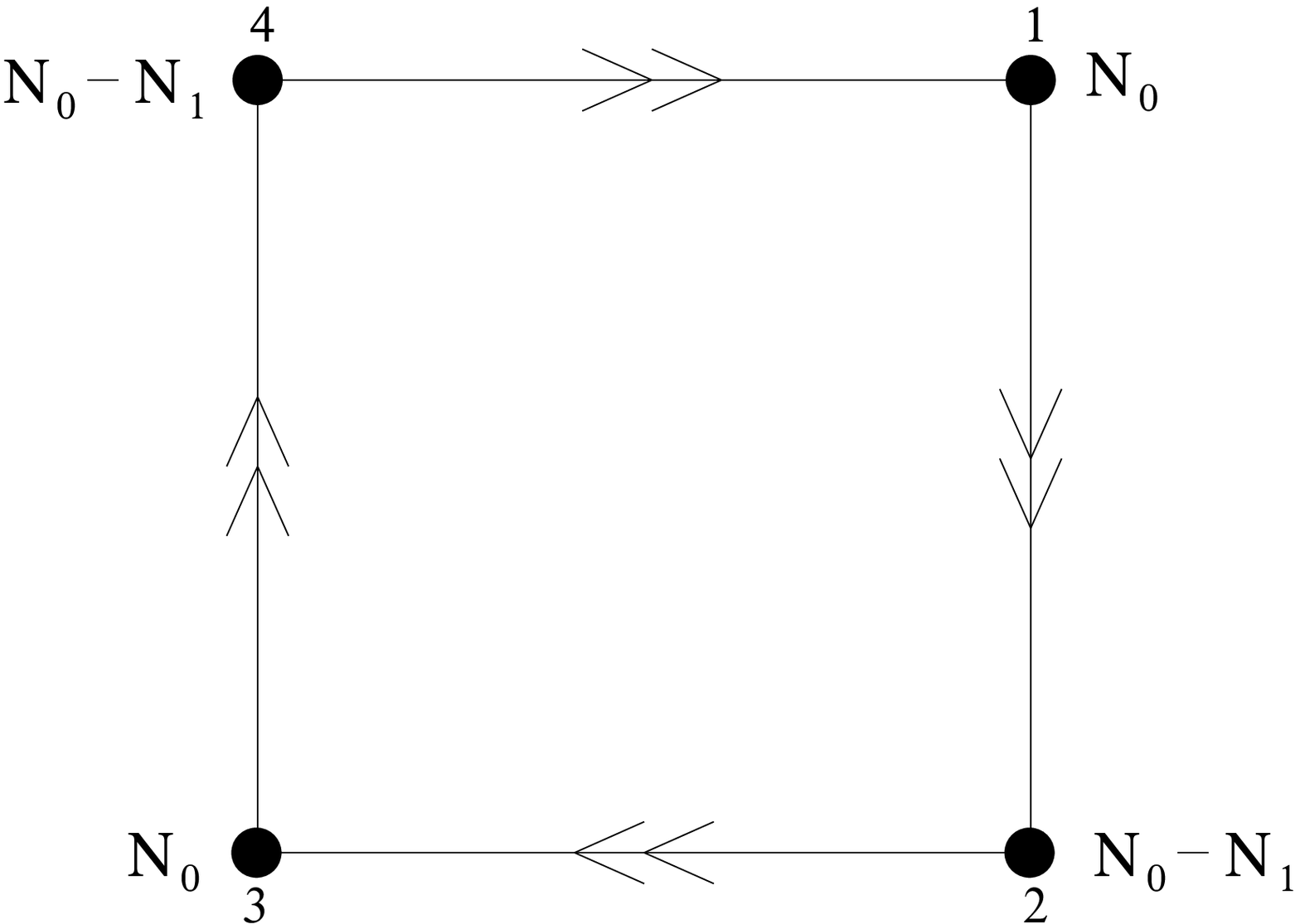}}
\noindent{\ninepoint\sl \baselineskip=8pt {\bf Figure 14}:{\sl Quiver
diagram for the ${\bf P}^1\times {\bf P}^1$ case after the mutation.}}
\bigskip

Integrating out $M_{ij}$ and $D_{ij}$
we obtain
\eqn\suupp{W=-A'_iB'_j{\widetilde B}_i{\widetilde A}_j}
which agrees with the results of \morpl\
(which can be seen by redefining $A'_i=\epsilon_{ik}F_k$
and $B'_j=\epsilon_{jr} G_r$).  Note that here the theory has
$N_0$ three branes and $N_1$ 5-branes wrapping the ${\bf P}^1/Z_2$.
The superpotential for the 3-brane and 5-branes for ${\bf P}^1$
 is the ${\widehat A}_1$ theory and the above theory is its $Z_2$
 orbifold which is how \morpl\ obtained $W$.
This duality was already noted
in \hananpa\  in the case with $N_1=0$, as the explanation
of the two inequivalent quiver theories.  Here we see in which
regime of parameters each is naturally defined.

\bigskip
We can also see from the viewpoint of transmuted bundles how
the quiver diagram and the Yukawa couplings arise.
Performing the mutation, we replace ${\cal O}(0,0)$ with the kernel
of ${\cal O}(0,-1)\otimes {\rm Hom}({\cal O}(0,-1)\to {\cal O}(0,0))$.
Since the Hom is two dimensional, the kernel is a line bundle.  Since its
Chern class is immediately calculated as $(0,-2)$, we see that the
mutated bundle is ${\cal O}(0,-2)$.  The collection of bundles is
given by ${\cal O}(-1,-1),\ {\cal O}(0,-1),\ {\cal O}(0,-2),\
{\cal O}(-1,0)$.  The nonzero Hom's and Ext's are
$$
\eqalign{ &{\rm Hom}({\cal O}(-1,-1),{\cal O}(0,-1))=H^0({\cal O}(0,1))={\bf C}^2\cr
         &{\rm Hom}({\cal O}(0,-1),{\cal O}(0,-2))=H^0({\cal O}(0,1))={\bf C}^2\cr
         &{\rm Ext}^1({\cal O}(-1,0),{\cal O}(0,-2))=H^0({\cal
         O}(1))\otimes H^1({\cal O}(-2))={\bf C}^2\cr
        &{\rm Hom}({\cal O}(-1,-1),{\cal O}(-1,0))=H^0({\cal O}(0,1))={\bf C}^2.}
$$
We reverse the signs of the bundles ${\cal O}(0,-2)$ and ${\cal
O}(-1,0)$ (note that $c_1({\cal O}(-1,-1))+c_1({\cal
O}(0,-1))-c_1({\cal O}(0,-2))- c_1({\cal O}(-1,0))=0$). We check that
the directions of the arrows also agree with those in Figure 14.  Note
that there was an extra arrow reversal for due to the ${\rm Ext}^1$ (i.e.\
the bundle ${\rm Hom}({\cal O}(-1,0),{\cal O}(0,-2))$ has negative index).  One
can also obtain a duality cascade in this case as will be discussed in
the next section.

Now we compute the Yukawa couplings.  Classes $t\in{\rm Ext}^1({\cal
O}(-1,0),{\cal O}(0,-2))$ describe extensions $V_t$ of ${\cal O}(-1,0)$ by
${\cal O}(0,-2)$ fitting into an exact sequence
\eqn\exten{0\to {\cal O}(0,-2)\to V_t\to {\cal O}(-1,0)\to 0}
where the extension class $t$ is the obstruction to \exten\ defining
$V_t$ as a direct sum.  We compute Yukawa couplings between $V_t$ and
the remaining two bundles ${\cal O}(-1,-1)$ and ${\cal O}(0,-1)$, then
differentiate with respect to the components of $t$ to obtain the
Yukawa couplings.  As expected, the results will agree with the
superpotential we found in \suupp .

We do the calculation by using a Cech cohomology representation.  If
$(x_1,x_2)$ and $(y_1,y_2)$ are coordinates on the respective ${\bf
P}^1$ factors, for $i=1,2$ let $U_i$ be the open set on which $y_i
\ne0$.  Then we represent classes in ${\rm Ext}^1({\cal O}(-1,0),{\cal
O}(0,-2))\simeq H^1({\cal O}(1,-2))$ on $U_1\cap U_2$ by the cocycle
\eqn\cocycle{\rho_t={t_1x_1+t_2x_2\over y_1y_2},}
where $t=(t_1,t_2)$ is the Ext class.  We define $V_t$ on each of the $U_i$
as the direct sum ${\cal O}(-1,0)\oplus{\cal O}(0,-2)$.  To define $V_t$
globally, sections must be glued via the matrix
\eqn\trans{
\pmatrix{1 & 0\cr \rho_t & 1 \cr}}
We now compute the Hom's between ${\cal O}(0,-1),\ V_t,\ {\cal
O}(-1,-1)$.  We first have
$${\rm Hom}({\cal O}(-1,-1),{\cal
O}(0,-1))=H^0({\cal O}(1,0)),$$
spanned by $x_1$ and $x_2$.

We also
have ${\rm Hom}({\cal O}(0,-1),V_t)= H^0(V_t\otimes{\cal O}(0,1))$.
This can be computed from the transition matrix \trans, if the local
sections are viewed as sections of the bundles ${\cal O}(-1,0)\otimes
{\cal O}(0,1)={\cal O}(-1,1)$ and ${\cal O}(0,-2)\otimes {\cal
O}(0,1)={\cal O}(0,-1)$.  A basis for the holomorphic sections
(expressed in the $U_1$ representation) may be taken to be
\eqn\vtbas{V_1=
\pmatrix{-(t_1x_1+t_2x_2)/y_1\cr y_2\cr},V_2=\pmatrix{
0\cr y_1\cr}}
Note that $\rho_t V_i$ is holomorphic in the $U_2$ variables so we do
have global sections.

Finally, we have ${\rm Hom}(V_t,{\cal O}(-1,-1))
=H^0(V_t^*\otimes{\cal O}(-1,-1))$.  By dualizing \exten\ and tensoring with
${\cal O}(-1,-1)$, we get
$$0\to{\cal O}(0,-1)\to (V_t^*\otimes{\cal O}(-1,-1))\to {\cal O}(-1,1)\to 0.$$
This is almost the same as \exten; we just have to tensor \exten\ by
${\cal O}(0,1)$ and change the extension class from $t$ to $-t$ when we
dualize.  A basis for sections is then given by

\eqn\wtbas{W_1=
\pmatrix{(t_1x_1+t_2x_2)/y_1\cr y_2\cr},W_2=\pmatrix{
0\cr y_1\cr}}

Using \vtbas\ and \wtbas, we compute the products of the Homs as
$$V_1W_2=-t_1x_1-t_2x_2,\ V_2W_1=t_1x_1+t_2x_2,\ V_1W_1=V_2W_2=0.$$
Now introducing variables $X_1,X_2$ for the sections $x_1,x_2$ of
${\rm Hom}({\cal O}(-1,-1),{\cal O}(0,-1))$, we get
$$V_1X_iW_2=-t_i,\ V_2X_iW_1=t_i,\ V_iX_jW_i=0.$$
Introducing $T_i$ for the ${\rm Ext}$s, we get be differentiation
$$V_1X_iW_2T_j=-\delta_{ij},\ V_2X_iW_1T_j=\delta_{ij},\ V_iX_jW_iT_k=0.$$
This agrees with what we found earlier \suupp\ by the identification
$A_i' =\epsilon_{ij}V_j$, $B_j'=X_j$,${\tilde B}_i=W_i$,${\tilde A}_j=T_j$.

\newsec{Large $N$ Dualities Involving Vanishing 4-Cycles}

It is well known that there are transitions in CY 3-folds where
a 4-cycle shrinks and some three cycles grow.  It was suggested
in \vaaug\ that transitions of this kind
may also realize some large $N$ duality in the context of wrapped
branes.  Here we will show that this is indeed the case.
Start with the $A_1$ theory, i.e. with $N_1$ D5 branes wrapping
${\bf P}^1$.  This gives the pure ${\cal N}=1$ theory with $G=U(N_1)$.
It is dual to the conifold with $N_1$ units of $H$ flux through $S^3$.
Now mod out both sides by $Z_2$ acting on the fiber direction of ${\cal O}(-1)
\oplus {\cal O}(-1)\rightarrow {\bf P}^1$.  Note that this corresponds to an
$A_1$ singularity over ${\bf P}^1$ which if we blow up, gives
${\bf P}^1\times {\bf P}^1$.  When they both shrink, it is an example
of a shrunk 4-cycle.
By following this through the
transition this acts as $S^3/Z_2$ with no fixed points (the $S^3$
can be viewed as a sphere bundle over ${\bf P}^1$ which leads to this action).
Thus we already have a concrete realization of these types
of transitions.  Note that
on the quiver side the gauge theory this corresponds to is simply
$U(N_1)\times U(N_1)$ with no matter fields (this is the special
case of the quiver theory we studied with $N_0=0$), shown
in Figure 14.  Just as before $N_1$
controls the size of the $S^3$ and the gaugino condensate field is related
to the modulus field of $S^3$.  Note also that there are now two domain
walls wrapping the $S^3/Z_2$ as discussed in \refs{\gopvaii,\av},
which is consistent with the fact that we have two inequivalent
domain walls, one for each gauge factor.

There is also the duality cascade, just as we discussed in the context
of the ${\widehat A}_1$: Namely as the coupling gets weaker,
as we go back in the UV (by taking $t$ to be a pure imaginary
$B$ field and varying it)
the geometry of the critical points on the $W$ plane change as shown
in Fig. 15 and the gauge factor undergoes Seiberg dualities/mutations
of the type we discussed and we end up in the UV with $U(kN_0)^2\times
U((k+1)N_0)^2$ after $k$ cascades backward.  Note that in this case
the picture we had presented in section 10 for the mechanical
analogy of the particle in the Coxeter box comes to life:  The projection
of the branes on the W-plane are like the strings attached to the point
inside the Coxeter box, which are
related to the coupling of the gauge
theories.

\bigskip
\centerline{\epsfxsize=0.95\hsize\epsfbox{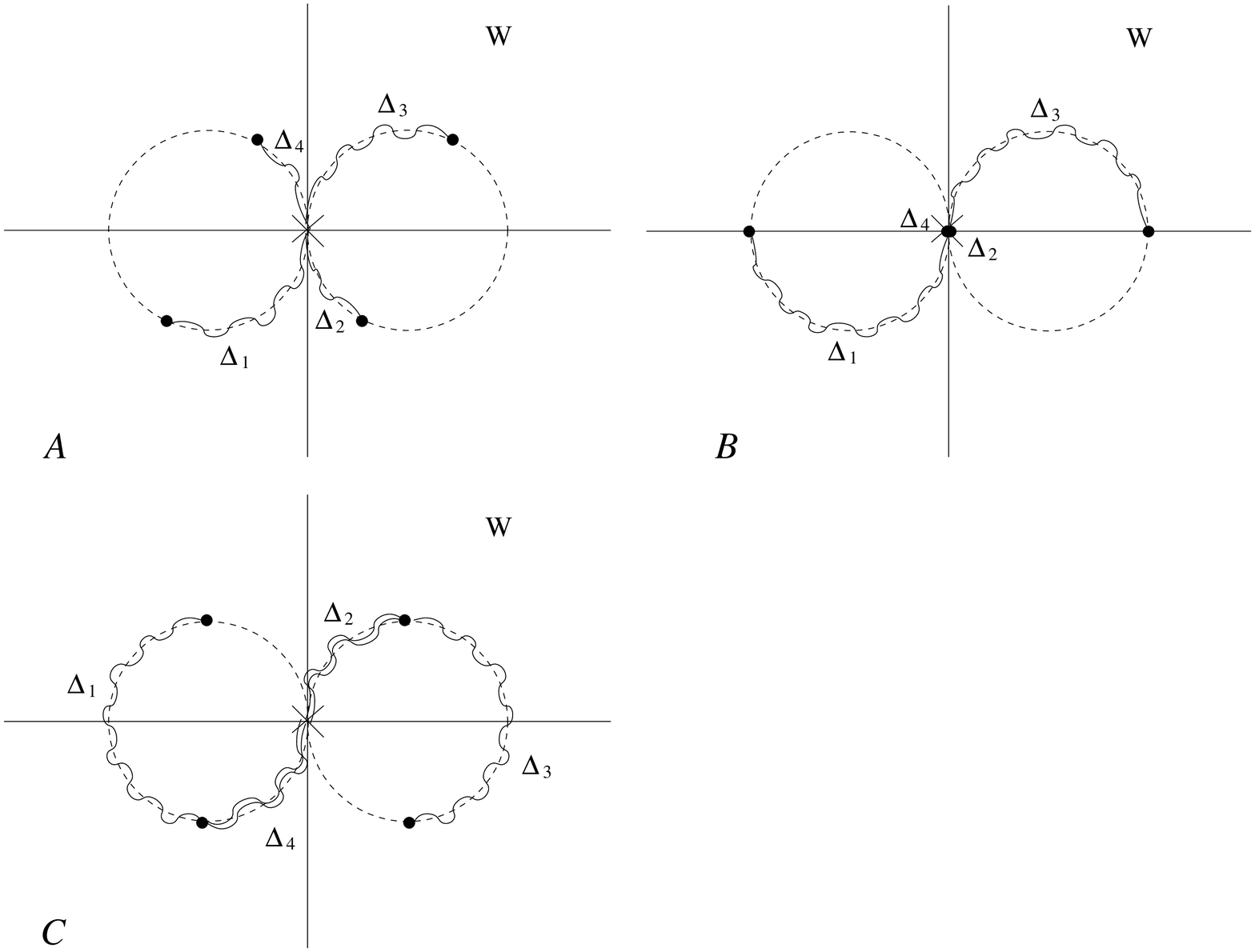}}
\noindent{\ninepoint\sl \baselineskip=8pt {\bf Figure 15}:{\sl
 Realization
of duality cascade in the context of ${\bf P}^1\times {\bf P}^1$ in the
orbifold limit.  The graph
above corresponds to $t$ pure imaginary.  The movement of the particle in the Coxeter
Box, becomes physical in this context, and is identified with a
semi-circle of either of the two circles above.}}
\bigskip

\subsec{Examples of Other 4-cycle/3-cycle transitions as large $N$ dualities}
The above example confirms the validity of the general idea.  Here we
will show other examples of this in the context of quiver theories
associated with del Pezzos.  We will first restrict ourselves to toric del Pezzos
which are given, in addition to the ${\bf P}^1 \times {\bf P}^1$ already
studied, by ${\bf P}^2$ and its blowups with up to 3 points.
Let us denote by $B_k$ the $P^2$ blown up at $k$ points.
It is known that there are no transitions with $k=0,1$, and that
for $B_2$ there is a transition controlled by one
parameter and for $B_3$ there are two inequivalent transitions one
involving 1 parameter and the other involving 2 parameters
(these moduli spaces meet similar to how the $z$-axis meets the
$xy$ plane).  These results have been obtained in
\alt (see also \gros\ for generalities on smoothing other del Pezzo
contractions).  We will now see how to
realize these transitions in the quiver theories arising from del Pezzos.

The corresponding quiver theories have already
been constructed in \refs{\pless,\hananpa}
 or they can be constructed with
the methods discussed here
(which are similar to those discussed in \ih ).
As already discussed we expect that $B_k$ will have
a quiver theory with $r=k+3$  nodes with $k+1$ inequivalent integers
labeling the number of 3-brane and $k$ classes of 5-branes.
Note that the classes of 5-brane charges is what affect the
$H$ flux and thus can possibly control the moduli on 3-cycles
on the other side.  So for $B_k$ we may expect $k$ deformations.
Indeed it is known \gros\ that there are $k$ infinitesimal
deformations for $B_k$ with $k=1,2,3$.  But, as noted above some are obstructed.

The basic idea from physics is that the unobstructed
ones should correspond to gaugino condensates in some
pure ${\cal N}=1$ Yang-Mills theory, as was seen in all the
cases encountered.  Thus we look for possibilities of having
such gauge theories in the corresponding quiver theory.  We have
already encountered exactly this structure in the context of
${\bf P}^1\times {\bf P}^1$, where in the general class of the quiver
theory we could write one which has only pure ${\cal N}=1$ Yang-Mills.
We got $U(N_1)^2$ with no matter, and the single factor $N_1$ controls
the size of the $S^3$.  We now look for this in the other cases.  For
${\bf P}^2$ this is not possible as already anticipated.  So we move
on to the other $B_k$ with $k=1,2,3$.
\vglue 1cm

\centerline{{\bf $B_1$ and transition}}
\vglue 1cm

The quiver theory for this case is given in \bohe\
and shown in Fig. 16 below.  We can identify the nodes with the sheaves
given by
$$ \eqalign{1:&\quad {\cal O}(H)\cr
            2:&\quad {\cal O}\cr
            3:&\quad {\cal O}(H-e)\cr
            4:&\quad {\cal O}(e)}
$$
where $H$ corresponds to the hyperplane class in ${\bf P}^2$ and $e$
is the class corresponding to the blow up.  Also we have to reorient
the nodes $3$ and $4$.  The arrow from 4 to 3 corresponds to an Ext
as in the ${\bf P}^1\times {\bf P}^1$ example.
It is easy to see that if we want
 $U(N)$ factors with no matter fields, compatible with allowed 5-brane
charge labeled by $N_1$ this is not possible.  Note
that the allowed ranks are given by
$$U_1(N_0)\times U_2(N_0-N_1)\times U_3( N_0+N_1)\times U_4( N_0-2N_1)$$
and to get a disconnected figure we need three of the nodes
be zero without the fourth being zero and this is impossible.
Thus this agrees with the geometric anticipation that there are no
deformations in this case.
Note also that the class given by the above $N_0$ and $N_1$ give
$$N_0 H+(N_0+N_1)(e-H)-(N_0-2N_1)(e)=N_1(3e-H)$$
which is consistent with the fact that $c_1\cdot (3e-H)=
(3H-e)\cdot (3e-H)=0$ (using $H^2=-e^2=1, H\cdot e=0$).

\bigskip
\centerline{\epsfxsize=0.30\hsize\epsfbox{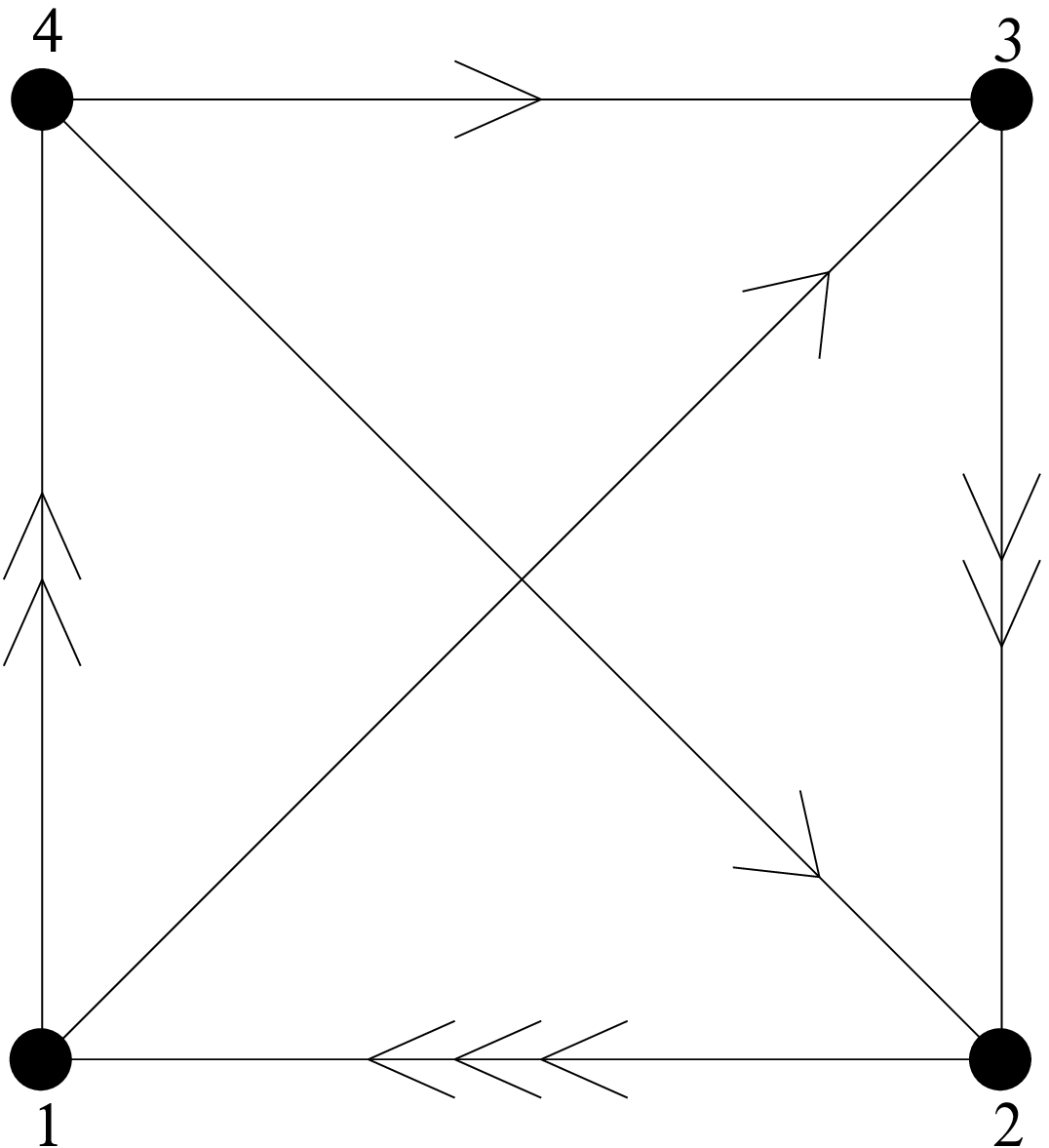}}
\noindent{\ninepoint\sl \baselineskip=8pt {\bf Figure 16}:{\sl
 Quiver
diagram for the field theory in the $B_1$ case.}}
\bigskip

\centerline{\bf {$B_2$ and transition}}
\vglue 1cm

The quiver for this case is given in \bohe.  There are some inequivalent
choices (related by dualities).  We have chosen one  in Fig 17.  To obtain
pure ${\cal N}=1$ Yang-Mills
would require setting at least three ranks to zero, which is the number
of degrees of freedom we have for the ranks. It turns out that this is
possible.  In particular set $N_2=N_5=N$ and $N_1=N_3=N_4=0$. Thus
we expect a one parameter moduli of deformation and this is indeed the
case.  The geometry of the deformed space
admits one compact 3-cycle and is given as follows.

\bigskip
\centerline{\epsfxsize=0.35\hsize\epsfbox{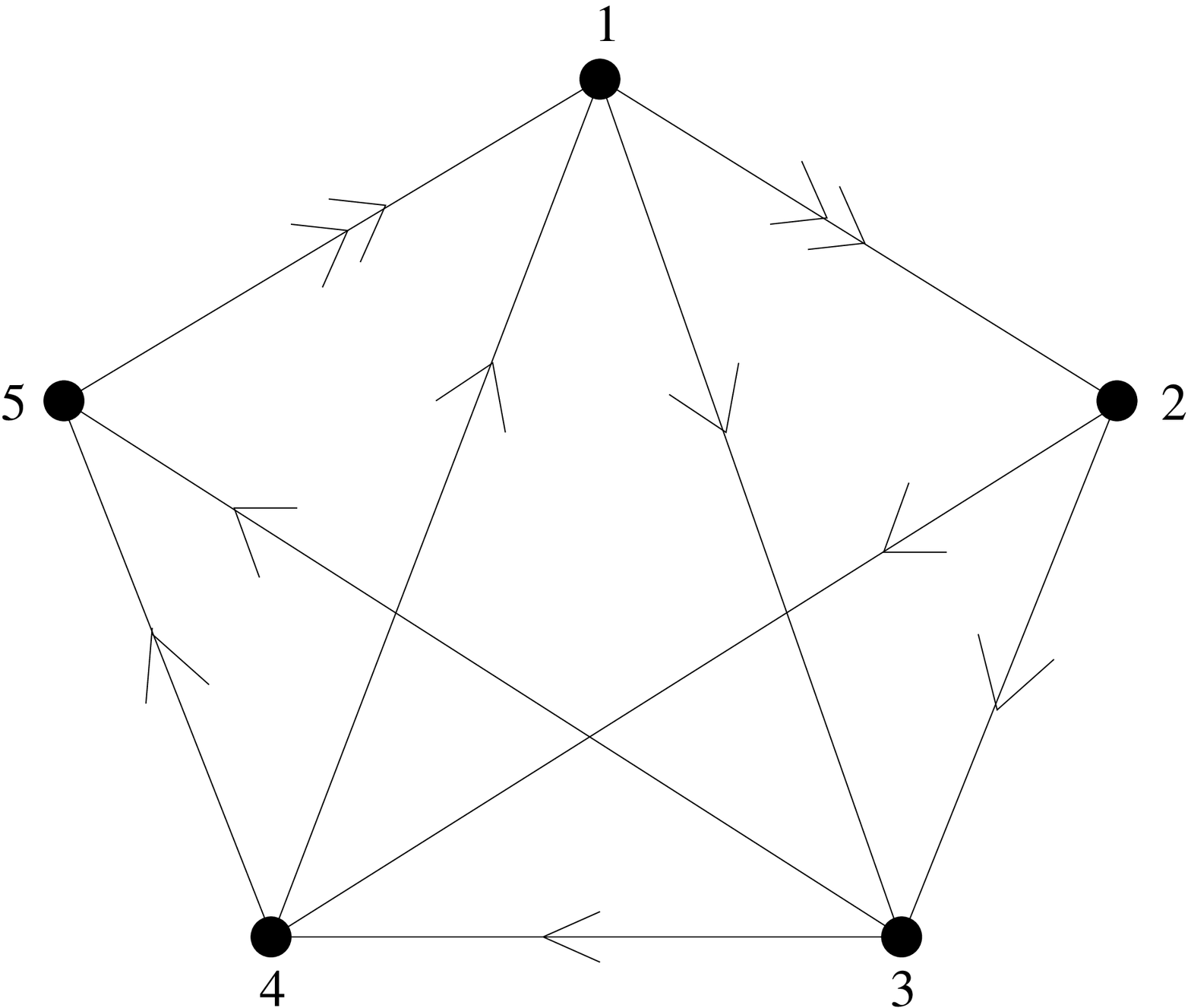}}
\noindent{\ninepoint\sl \baselineskip=8pt {\bf Figure 17}:{\sl Quiver
diagram for the field theory in the $B_2$ case.}}
\bigskip

Here and later we will use a general model for the blown-down geometry
of $B_k$: for $k\le 6$ there is an (anticanonical) embedding of $B_k$
into ${\bf P}^{9-k}$.  Thinking of ${\bf P}^{9-k}$ as the
projectivization of ${\bf C}^{10-k}$, the embedded $B_k$ is the
projectivization of its associated cone, which is a three dimensional
variety in ${\bf C}^{10-k}$, singular at the vertex of the cone.  This
is a local model for the geometry, the del Pezzo being contracted to
obtain the singular vertex.

The idea used in constructing the deformation is to find a
bigger cone $C$ such that the above local model can be constructed from
$C$ by imposing the vanishing of certain linear equations vanishing at the
vertex.  The deformation is obtained by simply deforming these linear
equations away from the vertex.  We now illustrate with $k=2$, where
the local model is a cone in ${\bf C}^8$.

We start by noticing that $B_2$ can be described as the blowup of
${\bf P}^1\times {\bf P}^1$ at one point, and that ${\bf P}^1\times
{\bf P}^1$ is a degree 2 hypersurface in ${\bf P}^3$.  It follows that
$B_2$ is a hypersurface in the blowup of ${\bf P}^3$ at a point, which
we may as well take to be $(1,0,0,0)$.  The hypersurface has degree 2
in the variables of ${\bf P}^3$ and contains $(1,0,0,0)$.  Such a
quadratic polynomial is a linear combination $\sum a_im_i$ of the 9 monomials
$$x_1x_2,\ x_1x_3,\ x_1x_4,\ x_2^2,\ x_2x_3,\ x_2x_4,\ x_3^2,\
x_3x_4,\ x_4^2.$$ Said differently, these 9 monomials define the
coordinates of a mapping $\phi$ from the blowup of ${\bf P}^3$ to
${\bf P}^8$, and $B_2$ is a hyperplane section of the image of $\phi$.
In the coordinates $(y_1,\ldots,y_9)$ of ${\bf P}^8$, the image of
$\phi$ is defined by quadratic equations such as $y_1y_5=y_2y_4,\
y_1y_6=y_3y_4,\ldots$.  If these equations are viewed as equations in
${\bf C}^9$, the result is a 4 dimensional variety in ${\bf C}^9$ which
is the cone over the image of $\phi$.  Taking the hyperplane section
$\sum a_i y_i=0$
containing the vertex, we get the local model inside the hyperplane
which is identified with ${\bf C}^8={\bf C}^{10-k}$ and we have
recovered the general construction concretely.  The vertex of the cone is
identified with the singular point after contracting $B_2$.  The
transition is completed by smoothing this singularity.
The deformation is given by
$$\sum_i a_iy_i=t,$$
where $t$ is a deformation parameter.  For $t=0$, this is the blown-down
geometry; for general $t$, this is a smooth threefold.
The size of $t$ is determined, using the minimization of the
superpotential, as discussed before.  It will change
depending on the single integer $N$ controlling the rank
of the disconnected gauge groups.

\vglue 1cm

\centerline{\bf $B_3$ and transition}
This is the case which was studied in \refs{\plb,\bohe}.
In particular four inequivalent models related to each other
by Seiberg duality were studied there.  In the notation of \plb\
consider these theories, shown in Fig. 18 below.

\bigskip
\centerline{\epsfxsize=0.65\hsize\epsfbox{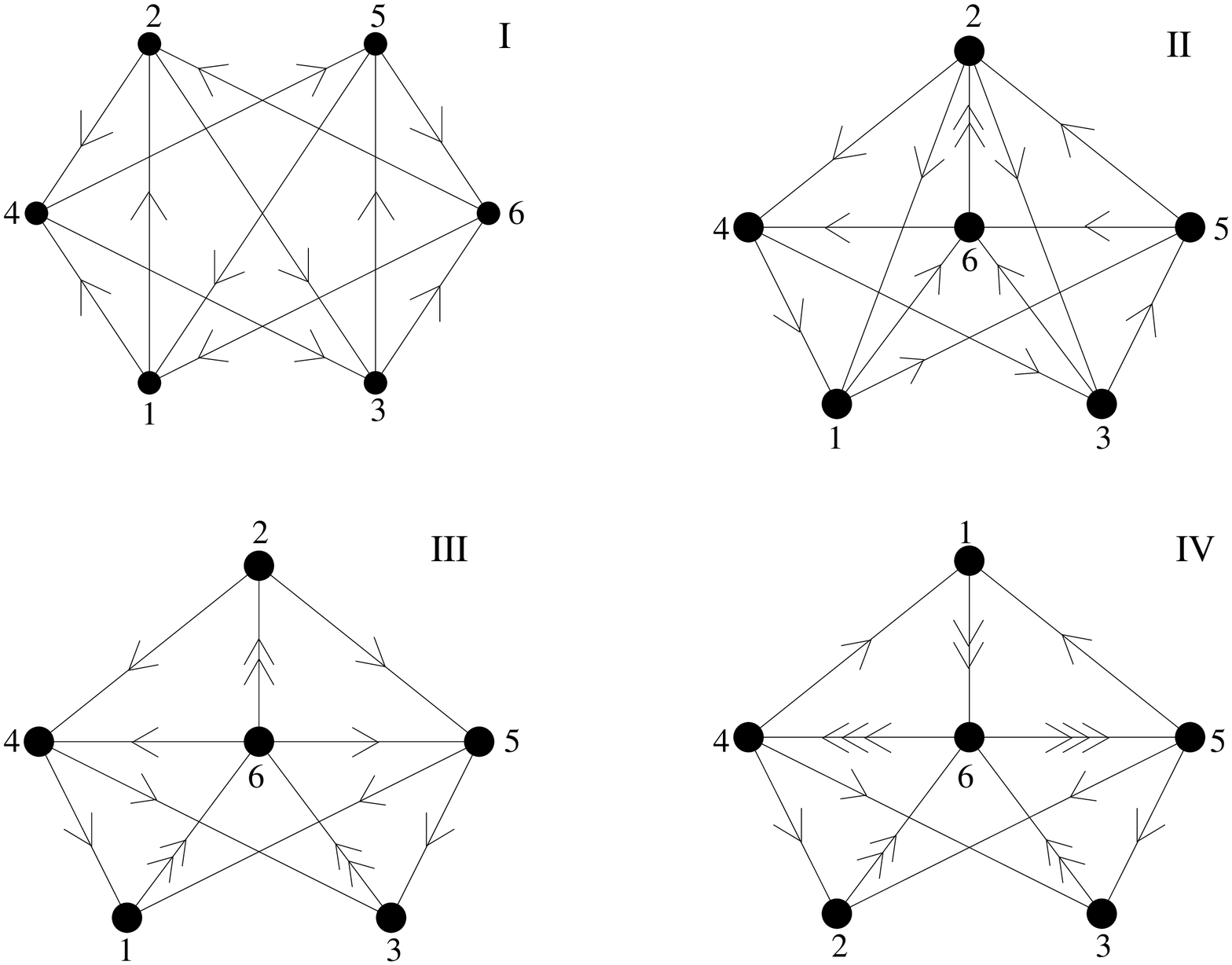}}
\noindent{\ninepoint\sl \baselineskip=8pt {\bf Figure 18}:{\sl
Quiver
diagrams of the four different field theory realizations in the $B_3$
case. These are related by Seiberg duality.}}
\bigskip

The conditions for anomaly cancellation can be written for these
models.  As expected  there are four independent integers which
satisfy anomaly free condition as expected.  We can write the
two conditions, for example, as
$$ Model\ I: N_5=N_1+N_2-N_3, \quad N_6=N_3+N_4-N_1.$$
$$Model\ II: N_5=2N_2+N_4-N_1-N_3,\quad N_6=N_1+N_3-N_2$$
$$Model\ III: N_5=2(N_1+N_3-N_2)-N_4,\quad N_6=N_1+N_3-N_2$$
$$Model\ IV: N_5={2\over 3}(N_1+N_2+N_3)-N_4,\quad N_6={1\over 3}(N_1+N_2+N_3).$$

These theories can be represented by sheaves as discussed before.
For example model III corresponds to
$$\eqalign{ 1:& \quad {\cal O}(e_3) \cr
            2:&\quad {\cal O}(e_1)\cr
            3:&\quad {\cal O}(e_2)\cr
            4:&\quad {\cal O}(H-e_1)\cr
            5:&\quad {\cal O}(H-e_2)\cr
            6:&\quad {\cal O}(H)}
$$
where $e_i$ denotes the class of the three blowups and
$H$ is the hyperplane class.  Also we have to reverse
the orientation of 2,4 and 5.

For the case studied in \refs{\plb, \ih}
 where the $N_i$ are all equal, it
was shown that one can
go through dualizing node 1 of model I to get model
II, then dualizing node 5 to get model III and then
dualizing node 2 to get model IV.  Of course these
dualities all immediately generalize to the case
where the $N_i$ are not equal with suitable
superpotentials, as discussed in full generality before.

As an example, consider model III.  Dualizing node 1 takes
one back to model III, with a change conjugation and 
relabeling of nodes $2\leftrightarrow 3$.
The resulting theory has $N_1'=N_4+N_5-N_1$, $N_2'=N_3$, $N_3'=N_2$
and $N_4 \dots N_6$ unchanged.  In all of the models, one could also dualize
the nodes having more than two arrows in and two out, and this would
generate duals with quivers not appearing in fig. 18.

For general $N_i$'s, we'll have an analog of the cascading flow to the
IR. For some of the possible endpoints of these flows we could get
e.g. model I with only the disconnected nodes $N_4=N_6$ non-zero (or
only $N_1=N_3$ non-zero, or only $N_2=N_5$ non-zero).  For this end
point we have a product of pure Yang-Mills, leading to gaugino
condensation, controlled by one integer.  Another example of an
endpoint is model III with only the disconnected nodes $N_1,N_2, N_3$
non-zero with $N_2=N_1+N_3$, which has a two parameter choice.  This
corresponds to the class given by
$$N_1(e_3)-N_2(e_1)+N_3(e_2)=N_1(e_3-e_1)+N_3(e_2-e_1)$$
which is orthogonal to $c_1=3H-e_1-e_2-e_3$.
Thus in this case we would expect to have
a transition involving two independent blown up 3-cycles,
controlled by two complex parameters.  Thus from
the quiver analysis we are led to expect transitions
involving either one or 2 dimensional moduli space.
This is indeed the case!

We can embed $B_3$ in ${\bf C}^{10-k}=
{\bf C}^7$ as before.  But this can be realized
as a hyperplane section of a 4 dimensional variety in ${\bf C}^8$
as well as the intersection of two hyperplanes with a 5 dimensional
variety in ${\bf C}^9$.   The deformations are obtained by simply
deforming the hyperplanes to avoid the vertex.  There are as many
deformation parameters as there are hyperplanes.

One-dimensional component: Embed ${\bf P}^1 \times {\bf P}^1
\times {\bf P}^1$ in ${\bf P}^7$ using the 8 multilinear polynomials
on ${\bf P}^1\times {\bf P}^1\times {\bf P}^1$.  A general hyperplane
section is identified with $B_3$ in ${\bf P}^6$.
Then take the cone over the 3 dimensional image  to get a 4 dimensional
singular variety in ${\bf C}^8$.  A hyperplane section through the
vertex is identified with the blown-down geometry in ${\bf C}^7=
{\bf C}^{10-k}$.
The deformation is $f(x_1,...,x_8)=t$, where $f$ is a fixed homogeneous linear
polynomial.

Two-dimensional component: Embed ${\bf P}^2 \times {\bf P}^2$ in ${\bf
P}^8$ using the 9 bilinear polynomials.  Intersecting with 2
hyperplanes yields $B_3$.  Take the cone to get a singular 5
dimensional variety in ${\bf C}^9$.  Now the deformation is given by
taking 2 linear polynomials:
$$g(x_1,...,x_9)=t_1\qquad h(x_1,...,x_9)=t_2,$$
a deformation
in ${\bf C}^7$.

Of course in all these cases we can put the fluxes
on the dual geometry to find the exact sizes for the blown up
$S^3$'s, as is by now rather familiar.  This yields exact
results for such ${\cal N}=1$ quiver theories, for which
no other method is known.  Also we can cascade backwards
as in the ${\bf P}^1\times {\bf P}^1$ example already discussed,
where as we go up in the cascade the number of 3-branes
continues to increase.  This class of examples
illustrates that not only can these chiral
theories undergo gaugino condensate, but that they
can also be viewed as arising from an infinite tower of cascades.

\subsec{The general del Pezzo case and affine $E_k$ symmetry}

Now let us analyze the more general case of $P^2$ blown up at
$k$ points with $k=4,...,8$. Even though these
are not toric, one can construct their quiver
diagram and the superpotential
terms from the study of the associated exceptional collection,
as we have discussed (the
quiver diagrams for these
classes has been proposed in \ih ) .
 Our comments
also apply to the cases with $k<4$ (with suitable
definitions for $E_k$).  The transitions
in this case have also been considered \gros\
and the result is that there are $C_2-1$ deformations
where $C_2$ is the dual coxeter number of $E_k$ groups with
$E_4=A_4, E_5=D_5$.  In particular the $E_k$ Weyl group
is quite relevant for such geometries as we will explain below.

Let $x$ denote an $H_2$ class of $B_k$ and let $c_1$
denote a class in $H_2$ dual to the chern class of $B_k$.
We know that the total allowed charge we have is given by
an element $x$ such that
$$x\cdot c_1=0$$
This yields a $k$ dimensional integral space, which
gives all the allowed 5-brane charges.  This space,
which is $k$-dimensional, turns out to be
given by the root lattice of $E_k$.

Let us denote the blow up classes by $e_i$ and the hyperplane class
by $H$.  These have a Lorentzian self-intersection
$$H\cdot H=1 \quad H\cdot e_i=0 \quad e_i\cdot e_j=-\delta_{ij}$$
and $c_1=3H-\sum_i e_i$.
A basis for the simple roots of $E_k$ can be chosen to be
$H-e_1-e_2-e_3$ and $e_i-e_{i+1}$ as $i=1,...,{k-1}$.  One
can see that these span the root lattice of $E_k$ with the canonical
inner product (up to an overall sign).
More is in fact true. The automorphism of del Pezzo cohomology is given
by the Weyl group of $E_k$. In particular for any such
root $x$ the action on another class is given by
$$y\rightarrow y+(y\cdot x)x$$
That is very much in the spirit of what we studied in the context
of A-D-E spaces.

There is also a natural set of spherical bundles, which are related
to the weight lattice of the $E_k$. Namely, consider the classes $x$
such that
\eqn\sph{x\cdot c_1=p}
for some fixed $p$.
Note that the difference of any such $x$'s is on the root lattice.
Moreover $x\cdot (root\ lattice)\in {\bf Z}$ which implies that $x$ is in
the weight lattice of $E_k$.  For $p=1$  the corresponding $x$ are given
by the weights of the fundamental
representation of the $E_k$ (except for $E_8$ which gives the $240$ roots).
Note that all the line bundles corresponding
to ${\cal O}(x)$ correspond to spherical bundles, and so the corresponding
brane would yield an ${\cal N}=1$ Yang-Mills theory with no adjoint fields.
If we consider branes corresponding to such sheaves, the Weyl reflection
by the root vectors
acts on the underlying quiver theory, as that is related to the automorphism
of the del Pezzo, very much like how it arose in the A-D-E quiver theory.
Also adding the 3-branes to the story should promote the $E_k$ to the affine
version.

Suppose we now want to study generally what are the allowed
phases for these theories yielding pure
$\N=1$ gauge theory leading to
gaugino condensates.  We would have to find the allowed
branches of these theories and look for spherical
bundles with no matter between them.
Note that if $R$ is a root, then the bundle ${\cal O}(R)$ has no
cohomology.  Since $-R$ is also a root, then the dual bundle ${\cal
O}(-R)= {\cal O}(R)^*$ has no cohomology either.  So given a
collection of bundles $V_\alpha$ such that each $V_\alpha^*\otimes
V_\beta$ is a root for all $\alpha\ne\beta$, then the corresponding
nodes in the quiver are totally disconnected.

We now ask: what is the cardinality of the maximal set of bundles such that
each pair differs by a root in the above sense?  Or more generally, differs
by something with only 1 cohomology?  It is not difficult to see that there
are at most $k+1$ such choices.
For example, in the $B_4$ case we have $C_2=5$ bundles which we write
symmetrically as
$$H-e_1,\ H-e_2,\ H-e_3,\ H-e_4,\ 2H-e_1-e_2-e_3-e_4.$$
This somewhat cumbersome form reflects our method for obtaining
these bundles: by finding geometric curve classes $x$ satisfying
\sph\ with $p=2$.
The difference are all of the form $e_i-e_j$ or $\pm(H-e_i-e_j-e_k)$, all
of which are roots.  So in this case, we have 5=$C_2$ disjoint bundles.
The number drops to 4 after requiring orthogonality to
$c_1$, which is the expected number of deformations
of $B_4$ namely $C_2-1=4$.  This also works in the $B_3,B_2$ and $
{\bf P}^1\times {\bf P}^1$
cases that we already discussed. For example in the $B_3$ case
we have $E_3=A_1+A_2$ and the two branches that we found correspond to
weights which differ either by the root of $A_1$ (the one dimensional
branch) or two roots of $A_2$ (the two dimensional branch).
The ${\bf P}^1\times {\bf P}^1$ case corresponds to $E_1=A_1$
and that is the case where the relevant class corresponds to the difference
of the two ${\bf P}^1$'s as we have discussed. Note that in this case
the affine $E_1=A_1$ action is indeed realized as was discussed
before.  In particular the $A_1$ Weyl reflection is generated
by the class given by the difference of the two ${\bf P}^1$'s.

However we encounter the puzzle
for higher $k$'s that there are less pure
$\N=1$ branches that we have identified,
than predicted from the deformations.  In particular $C_2-1 >k$
for $E_k$ with $k>4$.
  Note that these would be the cases
where on the A-D-E quiver theories we would also have the
non-abelian branches.  It would be interesting to understand how these extra
branches would appear
in the cases of these del Pezzos.  One would naturally expect
them to correspond to higher rank bundles.

\vskip 1cm

\centerline{\bf Acknowledgements}

We would like to thank M. Aganagic, O. Aharony, M. Berkooz, D.-E. Diaconescu,
M.R. Douglas, A. Elezi, K. Hori, A. Iqbal, R. Leigh, E. Sharpe and E. Zaslow
for valuable discussions. BF would also like to thank Harvard University for
hospitality during part of this project.

The research of FC and CV is supported in part by NSF grants PHY-9802709
and DMS-0074329. The research of BF is supported by a Marie Curie fellowship,
and also by the IRF Centers of Excellence program, by the European RTN
network HPRN-CT-2000-00122, and by Minerva. The research of KI is supported
by DOE-FG03-97ER40546. The research of SK is supported in part by NSF grant
DMS-073657 and NSA grant MDA904-00-1-052.

\listrefs

\end